\documentclass[a4paper,fleqn]{cas-dc}

\usepackage[authoryear,longnamesfirst]{natbib}
\usepackage{xcolor}
\usepackage{subfig}
\usepackage{tikz}

\def\tsc#1{\csdef{#1}{\textsc{\lowercase{#1}}\xspace}}
\tsc{WGM}
\tsc{QE}

\newcommand{\traite}{\rule[0.05cm]{7.5mm}{0.5mm}}

\definecolor{purple}{rgb}{0.5, 0.0, 0.5}
\definecolor{orange}{rgb}{1.0, 0.5, 0.0}
\definecolor{dark_blue}{rgb}{0,0.4470,0.7410}
\definecolor{dark_orange}{rgb}{0.8500,0.3250,0.0980}
\definecolor{dark_yellow}{rgb}{0.9290,0.6940,0.1250}
\definecolor{dark_purple}{rgb}{0.4940,0.1840,0.5560}
\definecolor{medium_green}{rgb}{0.4660,0.6740,0.1880}
\definecolor{forest_green}{rgb}{0.1953,0.8008,0.1953}
\definecolor{light_blue}{rgb}{0.3010,0.7450,0.9330}
\definecolor{dark_red}{rgb}{0.6350,0.0780,0.1840}
\definecolor{gris}{rgb}{0.5,0.5,0.5}
\definecolor{cyan_blue}{rgb}{0.,1.,1.}
\definecolor{mymagenta}{rgb}{1.,0.,1.}
\definecolor{yellg}{rgb}{0.8,0.8,0.}

\begin{document}
\let\WriteBookmarks\relax
\def\floatpagepagefraction{1}
\def\textpagefraction{.001}

\shorttitle{}    

\shortauthors{}  

\title [mode = title]{A high-fidelity numerical database for free-stream-turbulence-induced transition
under varying pressure gradients}  


\author[1,2]{L. Zemmour}
\cormark[1]
\ead{louenas.zemmour@sorbonne-universite.fr}
\ead[url]{}
\affiliation[1]{organization={Sorbonne Université, CNRS, Institut Jean Le Rond D'Alembert},
  city={Paris},
  postcode={75005}, 
  country={France}}

\author[2]{X. Gloerfelt}
\ead{xavier.gloerfelt@ensam.eu}
\ead[url]{}
\affiliation[2]{organization={DynFluid Laboratory, Arts et Metiers Institute of Technology},
  city={Paris},
  postcode={75013}, 
  country={France}} 

\author[1]{P. Cinnella}
\ead{paola.cinnella@sorbonne-universite.fr}

\begin{abstract}
  Free-stream-turbulence-induced transition under streamwise-varying pressure gradients, as encountered on turbine blades, has received much less attention in high-fidelity simulations than its zero-pressure-gradient counterpart. The ERCOFTAC T3 experiments provide a carefully documented reference for this problem and remain central to transition-model calibration. Complementing them with full-field, three-dimensional data opens new opportunities for the development of physics-based and machine-learning transition models. To this end, wall-resolved implicit large-eddy simulations of the T3A, T3B, and T3C2–C5 flat-plate cases are performed with a high-order compressible solver, using an upstream slip-wall block that lets the synthetic free-stream turbulence develop up to the leading edge. The simulations reproduce well the measured transition locations, integral quantities, and velocity profiles, and predict a separation bubble near the end of the plate for the lowest-Reynolds-number case T3C4. As the Reynolds number decreases and transition moves into the adverse-pressure-gradient region, the mechanism evolves continuously from classical bypass transition to transition governed by the tendency of the laminar boundary layer to separate, up to separation-induced transition. Intermittency, conditioned statistics, and Reynolds-stress budgets document this evolution and the associated non-equilibrium effects, which challenge current RANS transition models.
\end{abstract}

\begin{highlights}
\item High-fidelity LES database of ERCOFTAC T3 transition cases with pressure gradients
\item Transition shifts continuously from bypass to separation-induced as Re decreases
\item Intermittency and Reynolds-stress budgets reveal strong non-equilibrium effects
\item Open full-field data to support physics-based and machine-learning transition modelling
\end{highlights}


\maketitle

\section{Introduction}\label{sec:introduction}

The accurate prediction of laminar-to-turbulent transition remains a major challenge in computational fluid dynamics (CFD), particularly for boundary layers subjected to free-stream turbulence (FST) and non-zero pressure gradients. Transition has a strong impact on skin friction, heat transfer, and flow separation, and therefore plays a central role in the design and performance of aerodynamic and turbomachinery systems. Its prediction remains challenging because of the large variety of transition mechanisms and their sensitivity to Reynolds and Mach numbers, free-stream turbulence characteristics, pressure gradients, surface geometry, and three-dimensional effects \citep{Mayle91,Arnal00}.

Among the experimental databases available for transition modelling, the ERCOFTAC T3 experiments \citep{Roach92,Savill93} have played a particularly important role. The T3 series provides a systematic set of flat-plate boundary-layer experiments subjected to different levels of free-stream turbulence, both under zero pressure gradient (T3A and T3B) and under varying pressure gradients (T3C1--C5 and T3L1--L3). These experiments have consequently been widely used for the development and validation of transition models, from early closures \citep{Westin97,Suzen00} to modern correlation-based and data-driven approaches \citep{Langtry09,Srivastava21}. Their value stems from the controlled variation of the inflow turbulence and pressure-gradient conditions, together with measurements of the streamwise evolution of the boundary layer and transition.

However, the T3 experiments also present limitations from the perspective of high-fidelity numerical simulations and data-driven modelling. The measurements provide only limited spatial and temporal resolution and do not contain the three-dimensional, time-resolved information required to analyse the detailed dynamics of transition. In addition, the characterization of the incoming turbulence and, in particular, its integral length scales and spatial structure, is incomplete in several of the historical experiments. For the pressure-gradient configurations, the exact geometry and development of the external flow are also not always sufficiently documented to allow an unambiguous numerical reproduction. These limitations become particularly restrictive when the objective is to develop data-driven models requiring large and internally consistent datasets.

Several numerical databases and high-fidelity simulations have subsequently been developed to complement the experimental T3 benchmarks, although most of them have focused on zero-pressure-gradient (ZPG) boundary layers. An early numerical database was produced by \cite{Voke95} using large-eddy simulation (LES) to reproduce the T3B experiment. The incoming turbulence was generated from random perturbations imposed on a laminar boundary-layer profile. Despite being pioneering work, the database was obtained with the relatively limited computational resources available at the time, resulting in a comparatively coarse spatial resolution. The database nevertheless provides mean and fluctuating velocity profiles as well as Reynolds-stress budgets at several streamwise locations, making it one of the first numerical counterparts to the T3 experiments.
The first DNS studies of bypass transition considered simplified configurations in which the leading edge was not explicitly included. \cite{Jacobs01} investigated a T3A-like configuration by superimposing Fourier modes on a laminar boundary layer (with Orr--Sommerfeld modes used to provide the appropriate wall-normal decay). A similar approach was adopted by \cite{Brandt04}, who focused on the dynamics of Klebanoff streaks and investigated the influence of the incoming turbulence characteristics on transition. These studies established the importance of the incoming turbulence and its integral length scale in the transition process, but the databases are not publicly available.
The treatment of the leading edge constitutes another important distinction between numerical databases. \cite{Nagarajan07} and \cite{Ovchinnikov08}, for example, explicitly included a finite-thickness leading edge, using a super-elliptic geometry. \cite{Nagarajan07} investigated the influence of leading-edge bluntness on bypass transition, whereas \cite{Ovchinnikov08} considered a T3B-like configuration with a high turbulence intensity ($Tu=6.7\%$) and a large integral length scale. Two configurations were considered, one with the full leading edge and another using a symmetry condition at the centre plane to reduce the computational cost. Interestingly, the latter configuration provided a closer representation of the experimental T3B skin-friction evolution, whereas the full leading-edge simulation showed discrepancies in the transitional region. The study also identified the possible formation of $\Delta$-shaped structures associated with high-amplitude, large-scale free-stream turbulence, providing an alternative transition pathway to the classical secondary instability of Klebanoff streaks. \cite{Lee18a} produced a highly resolved DNS of the T3A configuration, including a super-elliptic leading edge and a synthetic free-stream turbulence field with an intensity of approximately 3\%. The corresponding database contains both time-averaged statistics and thousands of instantaneous three-dimensional fields and is publicly available through the Johns Hopkins University Turbulence Database (JHTDB) \citep{Lee18b}.
More recent studies have further diversified the available numerical datasets. \cite{Perrin23} compared DNS and implicit LES (iLES) resolutions for the T3A case, and also insvestigated the effect of including or not the leading edge in their iLES. \cite{Durovic24} performed highly resolved DNS using synthetic free-stream turbulence ahead of a smooth chamfered leading edge and investigated the influence of the incoming integral length scale on transition. In parallel, \cite{Bienner24d} performed iLES for a wider range of turbulence intensities, considering both an integral length scale comparable to the T3A conditions and a second length scale approximately ten times larger. Their database additionally covers different Mach numbers and both air and dense-gas configurations. The publicly available dataset contains streamwise evolutions of averaged quantities, boundary-layer profiles, and spectra for the different configurations, together with Python tools and documentation for accessing the data. These simulations confirmed the emergence of nonlinear $\Delta$-shaped structures for sufficiently large integral length scales, in agreement with the observations of \cite{Ovchinnikov08} and \cite{Wu23}.
Finally, \cite{Wu26} recently extended the range of ZPG DNS data to free-stream turbulence intensities between $0.75\%$ and $6\%$. The incoming turbulence is generated from a spatially decaying isotropic-turbulence simulation (SDIT). This approach allows the spatial decay of the free-stream turbulence to be controlled more consistently and provides a systematic dataset covering five different turbulence intensities. The simulations use a fine spatial resolution, and the database provides boundary-layer profiles at multiple streamwise locations. Overall, these recent studies have considerably improved the availability and quality of numerical reference data for bypass transition.

Despite this substantial progress, the existing numerical databases remain heterogeneous in terms of their inflow treatment, leading-edge representation, resolution, and available statistical quantities. In particular, the choice of whether or not to explicitly include the leading edge is not merely a numerical detail. The leading edge modifies the receptivity of the boundary layer to free-stream turbulence \citep{Nagarajan07} and can alter the local pressure gradient and the initial development of the laminar boundary layer \citep{Mamidala22b,Perrin23}. It may consequently affect the turbulence level experienced by the boundary layer downstream and, ultimately, the transition location. Furthermore, several numerical studies report a substantially faster decay of free-stream turbulence than observed experimentally \citep{Jacobs01,Brandt04,Zaki13}, demonstrating that reproducing the experimentally relevant inflow turbulence remains a non-trivial issue.
A further limitation is that almost all publicly available high-fidelity databases concern zero-pressure-gradient boundary layers. Numerical investigations of free-stream-turbulence-induced transition under pressure gradients are considerably less numerous. One of the most systematic numerical studies is that of \cite{Nolan13}, who performed four DNS of a flat-plate boundary layer subjected to different constant pressure-gradient conditions, ranging from a strong adverse pressure gradient (APG) to a favorable pressure gradient (FPG). The authors analysed conditional statistics for the laminar and turbulent regions, making the database particularly useful for investigating the modification of transition dynamics by pressure gradients. However, these data are not, to our knowledge, available through an open online database. \cite{Brinkerhoff15} investigated transition under a combined favorable- and adverse-pressure-gradient distribution using a commercial CFD solver, with a relatively high free-stream turbulence intensity of $Tu\simeq5.3\%$. Their study illustrates the strong interaction between FST, pressure-gradient history and transition.
The situation is similar for experimental databases. Several studies have investigated transition under favourable or adverse pressure gradients, including the experiments of \cite{AbuGhannam80}, \cite{Narasimha84}, \cite{Blair92}, \cite{Roberts03}, and the series of experiments performed by Gostelow and co-workers \citep{Gostelow89,Gostelow94}. The recent review by \cite{Davis26} highlights that comprehensive and systematically documented datasets with pressure gradient remain scarce.
More recently, \cite{Simoni19} established a comprehensive experimental database combining controlled free-stream turbulence with systematically varied adverse pressure gradients. Subsequent studies used these data to investigate laminar separation bubbles and their bursting mechanisms \citep{Dellacasagrande20,Dellacasagrande24}. However, most of these experiments involve separated flows, whereas the present work focuses on attached boundary layers undergoing free-stream-turbulence-induced transition. Moreover, these datasets are not as readily accessible as the major public numerical databases, and the legacy ERCOFTAC T3 experiments therefore remain widely used for transition-model development and assessment.

This review highlights a gap between the requirements of modern transition modelling and the currently available reference data. Public high-fidelity databases provide detailed three-dimensional information for ZPG bypass transition, but no equivalent database is readily available for systematically studying the combined effects of free-stream turbulence and pressure gradients in attached boundary layers.
The present work addresses this gap by introducing a high-fidelity numerical database designed to reproduce selected ERCOFTAC T3 transitional boundary-layer experiments under both zero and varying pressure gradients. The database is generated using wall-resolved implicit large-eddy simulations with a high-order compressible Navier--Stokes solver. The numerical configurations are designed to control the free-stream turbulence intensity, integral length scale, inflow spectrum, and external pressure-gradient distribution. In addition to mean-flow quantities, the database provides time-resolved three-dimensional velocity and pressure fields together with Reynolds stresses, velocity gradients, turbulence statistics, and other quantities relevant to transition modelling.
Beyond providing a high-fidelity numerical counterpart to the T3 experiments, the database is intended as a common framework for analysing and modelling free-stream-turbulence-induced transition. The use of a unified numerical methodology across different turbulence and pressure-gradient conditions reduces ambiguities associated with changes in geometry, boundary conditions, and numerical treatment, thereby facilitating systematic comparisons between transition regimes. The resulting dataset is particularly suited to the development and assessment of physics-based and data-driven transition models, including transition indicators, intermittency models, and data-driven corrections to Reynolds-Averaged Navier--Stokes (RANS) closures.

The remainder of the paper is organized as follows. Section~\ref{sec:methodology} describes the numerical methodology, including the flow solver and the turbulence injection approach. Section~\ref{sec:database_description} presents the target experiments and the numerical setup of the simulated configurations. Section~\ref{sec:validation} validates the simulations for the zero-pressure-gradient cases T3A and T3B against the experiments and previous DNS and LES. Section~\ref{sec:T3C} is devoted to the T3C cases with a varying pressure gradient, including the evolution of mean-flow quantities, the velocity and Reynolds-stress profiles, and an analysis of intermittency and conditioned statistics based on a laminar--turbulent discrimination. Section~\ref{sec:budgets} analyses the Reynolds-stress budgets to characterize the modification of the transition mechanisms by the pressure gradient. Finally, Section~\ref{sec:conclusion} summarizes the main findings and discusses their implications for transition modelling. The re-analysis of the experimental database, the sensitivity of the T3C cases to the inflow turbulence, and the laminar--turbulent discrimination algorithm are detailed in Appendices~\ref{app:expe}, \ref{app:T3C_sens}, and \ref{app:lam_turb}, respectively.

\section{Methodology}\label{sec:methodology}

\subsection{Flow solver}

The simulations in this study are conducted using the in-house, high-order compressible Navier--Stokes solver MUSICAA.
This solver evaluates the Eulerian fluxes using tenth-order centered finite differences and computes the viscous fluxes using
fourth-order central differences. Time advancement is achieved via a four-step, fourth-order Runge--Kutta integration scheme.
To extend the stability limits and enable simulations at typical CFL numbers of 4 to 5, a fourth-order implicit residual smoothing (IRS)
technique is employed \citep{Cinnella16}. This high-order IRS approach, modified for enhanced robustness \citep{Bienner24a},
maintains an accuracy comparable to explicit methods while significantly improving computational efficiency.
Additionally, a tenth-order selective filter is applied to damp unresolved high-wavenumber fluctuations, effectively acting as an
implicit LES (iLES) subgrid-scale model. Such a strategy has proven highly effective in previous studies \citep{Gloerfelt19}.
Further details regarding the numerical methods and their validation can be found in \cite{Bienner24a}.

\subsection{Turbulence injection}\label{subsection:turbulence_injection}

Freestream turbulence is introduced into the computational domain using a synthetic flow field generation technique based
on random Fourier modes (RFM) \citep{Bechara94}. A homogeneous isotropic turbulent velocity field is generated
as the superposition of $N$ independent RFMs:
\begin{equation}
  \mathbf{u}_{in}'(\mathbf{x}, t)=\sum_{n=1}^{N}\hat{u}_n\cos(\mathbf{k}_n\cdot(\mathbf{x}-\mathbf{\overline{u}} t)+\psi_n)\mathbf{a}_n
  \label{eq:uprime_rfm}
\end{equation}
where the velocity direction $\mathbf{a}_n$, the phase $\psi_n$, and the wavenumber orientation $\mathbf{k}_n/k_n$
are random variables assigned according to specified probability density functions \citep{Bienner24c}.
The wavenumber modulus $k_n$=$|\mathbf{k}_n|$ is discretized using a logarithmic distribution, $\exp[\log k_{\min}+(n-1)\Delta k_n]$ with
$\Delta k_n$=  $(\log k_{\max}-\log k_{\min})/(N-1)$, $k_{\min}$ and $k_{\max}$ denoting the minimum and maximum resolved wavenumbers, respectively.
The amplitude of the RFM modes is defined by:
\begin{equation}
  \hat{u}_n=\sqrt{2E(k_n)\Delta k_n}
  \label{eq:amp_rmf}
\end{equation}
and is prescribed following a von Kármán spectrum, supplemented by a Saffman viscous dissipation function and a bottleneck correction:
\begin{flalign}
& E(k)= 1.453\frac{(u_{rms})^2 k^4/k_e^5}{\exp\left[\frac{17}{6}\log(1+(k/k_e)^2)\right]}
  \exp\left(-1.5 c_{k} (k\eta)^2\right) \notag \\
&\times \left[1+0.522\left(\frac{1}{\pi}\arctan\left(10\log_{10}(k\eta)+12.58\right)+\tfrac{1}{2}\right)\right] &&
  \label{eq:von_karman_spectrum}
\end{flalign}
where $k_e$=$0.747/L_{f,in}$, $c_k$=1.613, $\eta$ represents the Kolmogorov viscous scale,
and $L_{f, in}$ the target integral length scale of the inlet turbulence.
The turbulent fluctuating velocity  $\mathbf{u}_{in}'$ is superimposed to the inlet velocity and
injected through the boundary conditions. The number of modes is set as $N$=100.
For each case, the prescribed parameters are $u_{rms}$=$T_u U_{ref}$, $L_{f, in}$ and the
bounds $k_{\min}$ and $k_{\max}$, which depend on the computational grid.

\section{Database description}\label{sec:database_description}

This section introduces the numerical database and details the simulated cases.
First, the target experimental setup and corresponding flow conditions are outlined, followed by a comprehensive
description of the numerical framework and boundary conditions.

\subsection{Experimental setup}

\begin{figure*}[t]
\centering
\includegraphics[width=\linewidth]{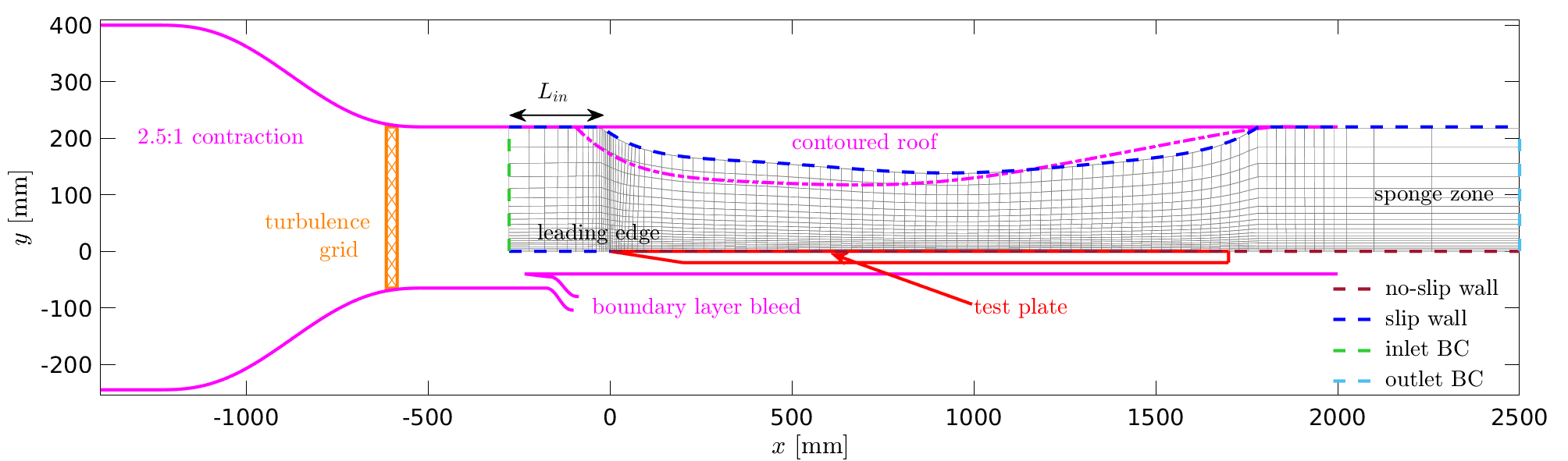}
\caption{Sketch of the experimental and numerical setups for the T3 ERCOFTAC configurations. The experimental
  contours are adapted from \cite{Savill93} (not to scale), and the computational grid corresponds to the T3C4
  simulation (every 12 points). The profiled top wall for the T3C cases follows Eq.~(\ref{eq:t3cr_upper_wall})
  from \cite{Suluksna09}, and is designed to reproduce, with slip-wall conditions, the pressure
  distribution obtained with the experimental configuration (magenta dash-dotted line).}
\label{fig:setup}
\end{figure*}

The reference experimental database consists of a series of flat-plate boundary layers subjected to freestream
turbulence (FST) intensities ranging from 0.9\% to 6.6\%. These test cases are designated as T3A, T3B, and T3C1--C5.
Cases T3A and T3B represent zero-pressure-gradient (ZPG) boundary layers, whereas cases T3C1--C5 feature varying pressure gradients,
representative of an aft-loaded turbine blade. The experimental setup is documented in \cite{Roach92} and \cite{Savill93} and
is reproduced schematically in Fig.~\ref{fig:setup}. A flat plate of 1700~mm length, 710~mm width, and 20~mm thickness is
placed in the low-speed wind-tunnel section. The upper surface of the plate is the test wall in the experiments.
The origin of the coordinate system is taken at the leading edge, formed by a 5-degree chamfer on the side opposite
to the test surface, with a tip radius of 0.75~mm. For the T3C cases, variations in the streamwise pressure gradient are produced
with a contoured wall attached to the wall opposite the test plate, located 200~mm from the plate. The geometry of the contoured wall
is not given in the literature and is sketched in the figure from \cite{Savill93}. The contraction first induces an accelerating
flow (favorable pressure gradient, FPG), followed by a decelerating flow (adverse pressure gradient, APG). A turbulence
grid is installed at the exit of the wind-tunnel contraction, 610~mm upstream of the leading edge. The distance from the
upper surface of the plate to the bottom wall is only 40~mm, and a boundary-layer bleed is used to minimize
interaction with the boundary layer developing on the bottom wall, as depicted in Fig.~\ref{fig:setup}.
The inflow conditions for the reproduced cases are summarized in Table~\ref{tab:inlet_exp_ercoftac}.

\begin{table}[pos=h]
\centering
\caption{\fontsize{8}{8}{Summary of experimental inlet conditions for the T3 ERCOFTAC cases \citep{Savill93}.}}
\label{tab:inlet_exp_ercoftac}
\begin{tabular}{|c|c|c|c|}
\hline
Case   & $U_{in}$ (m/s) & $Tu_{exp}^{x=0}$ (\%)  & Pressure gradient   \\
\hline
T3A    &      5.4      &       3         &       zero            \\
T3B    &      9.4      &       6         &       zero            \\
T3C2   &      5.0      &       3         &       variable        \\
T3C3   &      3.8      &       3         &       variable        \\
T3C4   &      1.2      &       3         &       variable        \\
T3C5   &      8.6      &       3         &       variable        \\
\hline
\end{tabular}
\end{table}

\subsection{Numerical setup}\label{subsec:numerical_setup}

As discussed in the introduction, various numerical setups have been used to simulate bypass transition,
including setups with or without the leading edge (LE). Simpler settings start with a finite-thickness Blasius profile.
Turbulent fluctuations are then superimposed at the inlet but should be damped to mimic the boundary-layer
receptivity (also referred to as the shear-sheltering effect). Either continuous modes of the Orr--Sommerfeld--Squires
operator in the wall-normal direction \citep{Jacobs01,Brandt04} or an analytical damping function
\citep{Pinto19,Perrin23} can be used. The exact location of the transition then depends on the chosen
injection height (see discussions in \cite{Pinto19}, and Appendix B of \cite{Bienner24d}).
On the other hand, some studies have included the LE in the computational setup \citep{Nagarajan07,Ovchinnikov08,Lee18a,Perrin23,Durovic24}.
These authors used a super-elliptic leading edge to reproduce T3-like boundary-layer transitions, except
\cite{Durovic24}, who meshed the experimental LE geometry to compare with the experiments of \cite{Fransson20}.
\cite{Perrin23} compared iLES simulations with and without the LE and showed a better quantitative agreement with T3A experiments
when the LE is present, although at a higher computational cost. In the present study, as sketched in Fig.~\ref{fig:setup},
an inlet domain of length $L_{in}$ with a slip wall on its lower side is used upstream of the plate to allow
the inlet turbulence to develop, thereby avoiding ambiguities associated with the injection height.
Furthermore, whereas most previous studies used DNS-like grid resolutions, iLES is employed here to further reduce
the computational cost and allow the exploration of various cases. A similar strategy is described in \citep{Perrin23},
where careful validation against DNS is conducted. The implicit regularization of the unresolved scales is different
in the present solver, which is based on solution filtering, and the wall-resolved LES resolution has been selected based on
a previous study \citep{Gloerfelt23b}, in which iLES and DNS were compared for natural transition. The IRS smoothing,
which allows the use of time steps corresponding to CFL numbers between 3 and 4, is another important ingredient
for reducing the computational cost and was also validated in \cite{Gloerfelt23b}.
Given the compressible formulation of the MUSICAA solver, the simulations are conducted at moderate Mach numbers,
whereas the reference experiments correspond to essentially incompressible conditions. Reynolds-number similarity
is maintained by appropriately rescaling the flat-plate dimensions. For the ZPG cases, an inlet Mach number of
$M_a$=0.5 is specified (as in \cite{Pinto19} and \cite{Bienner24d}), while for the pressure-gradient configurations
T3C, a lower value of $M_a$=0.2 is used. It was shown in \cite{Bienner24d} that ZPG flat-plate bypass transition
is weakly affected by compressibility effects in the subsonic regime, with very similar results for $M_a$=0.1
and $M_a$=0.9 boundary layers.

\begin{table*}[t]
\centering
\caption{Summary of computational parameters and mesh resolutions for the T3 configurations.
$L_y$ and $L_z$ are the vertical and spanwise lengths; $L_{f, in}$ the inlet integral length scale; 
$\delta_{out}$ and $\delta_{onset}$ are the 99\% boundary-layer thickness near the outlet and at the
transition onset, respectively. $L_{in}$ is the distance from the inlet to the leading edge ($x=0$).
The mesh column reports $\Delta x^+ \times \Delta y^+ \times \Delta z^+$ and the grid size in each direction.
$Tu^{x=0}_x$ is the logitudinal turbulence intensity at the plate LE.}
\label{tab:geometry_mesh_ercoftac}
\resizebox{\linewidth}{!}{
\begin{tabular}{|c|c|c|c|c|c|c|c|c|c|c|}
\hline
Case & $L_y/\delta_{out}$ & $L_z/\delta_{out}$ & $L_z/\delta_{onset}$ & $L_z/L_{f,in}$
& $L_{in}/L_p$ & $\Delta x^+ \times \Delta y^+ \times \Delta z^+$ & Grid Size &
$\text{Re}_{L_{f,in}}$ & $Tu^{x=0}_x$ (\%)\\
      \hline
      T3A  &  4.5 & 2.2 & 6.7 &  7 & 0.05 & 20$\times$ 0.9$\times$ 12 & 1000$\times$ 200$\times$  60 & 2261 & 3.2 \\
      T3B  & 11.5 & 6.2 &  37 &  7 & 0.05 & 23$\times$ 1.0$\times$ 13 & 1000$\times$ 200$\times$ 200 & 7914 & 6.2 \\
      T3C5 & 10.4 & 2.3 &  16 & 17 & 0.08 & 14$\times$ 1.4$\times$ 14 & 2800$\times$ 600$\times$ 100 & 2319 & 3.1 \\
      T3C2 &  9.5 & 2.3 & 8.0 & 18 & 0.08 & 16$\times$ 1.5$\times$ 10 & 1750$\times$ 450$\times$ 100 & 1249 & 2.7 \\
      T3C3 &  9.2 & 2.9 & 7.1 & 28 & 0.12 & 19$\times$ 1.5$\times$ 11 & 1200$\times$ 350$\times$  90 &  713 & 2.7 \\
      T3C4 &  2.4 & 3.1 & 8.5 & 44 & 0.16 & 17$\times$ 0.8$\times$ 12 & 1000$\times$ 250$\times$  90 &  460 & 3.2 \\
      \hline
    \end{tabular}
  }
\end{table*}

The grid characteristics for the present T3-like campaign are summarized in Table~\ref{tab:geometry_mesh_ercoftac}.
For the ZPG cases, the computational domain is a rectangular prism, with non-reflective boundary conditions \citep{Tam96a}
imposed at the inlet, outlet, and upper surface. As sketched in Fig.~\ref{fig:setup}, a slip-wall condition is applied
to the bottom wall upstream of the leading edge, while an adiabatic no-slip condition is enforced along the plate itself.
The experimental configuration for the pressure-gradient (T3C) cases is more complex and lacks an exhaustive description
in the literature, which complicates its numerical replication. Generally, a streamlined slip wall is used for the top boundary.
For instance, \cite{Menter15} recalibrated the top-wall geometry for each T3C2--5 case from the experimental velocity
profiles. In the present study, the top-wall shape proposed by \cite{Suluksna09} is used to approximate the desired
pressure gradients, with a slip-wall boundary condition prescribed along this contoured boundary. Denoting $D$ the
inlet domain height (at $x$=0) and $h$ the local height, the geometry is given by
\begin{align}
\frac{h}{D}=&\min\biggl(1.231x^6-6.705x^5+14.061x^4 \notag\\
            &-14.113x^3+7.109x^2-1.9x+0.95; 1 \biggr)\,,
\'e\'e\'e\label{eq:t3cr_upper_wall}
\end{align}
where $x$ is the dimensional streamwise coordinate.
The bottom-wall boundary conditions remain identical to those of the ZPG cases. Characteristic boundary conditions
are employed at the inlet, and a backpressure condition at the outlet is combined with a sponge layer.

To check the absence of artificial confinement effects, the ratio of the domain height and width to the maximum
boundary-layer thickness (at the outlet) is reported in Table~\ref{tab:geometry_mesh_ercoftac} and
ensured to be higher than two \citep{Poggie15}. For the T3C cases, $L_y$ corresponds to the domain
height at the inlet ($L_y$=$D$ of Eq.(\ref{eq:t3cr_upper_wall})). In the spanwise direction, the width $L_z$ is at least
seven times the inlet integral length scale $L_{f,in}$.
The mesh resolution is sufficient to resolve the transition mechanisms, near-wall coherent structures, and outer boundary-layer
turbulence. The corresponding mesh spacings, evaluated at the maximum boundary-layer thickness, are provided in wall units
in Table~\ref{tab:geometry_mesh_ercoftac}.
As previously noted, experimental information on the inlet turbulence integral length scale is limited.
Consequently, the numerical inlet turbulent intensity and integral length scale were iteratively calibrated to
approximate the available experimental target data. The resulting numerical inflow conditions are also reported
in Table~\ref{tab:geometry_mesh_ercoftac}.

\begin{figure*}[t]
\centering
\includegraphics[width=\linewidth]{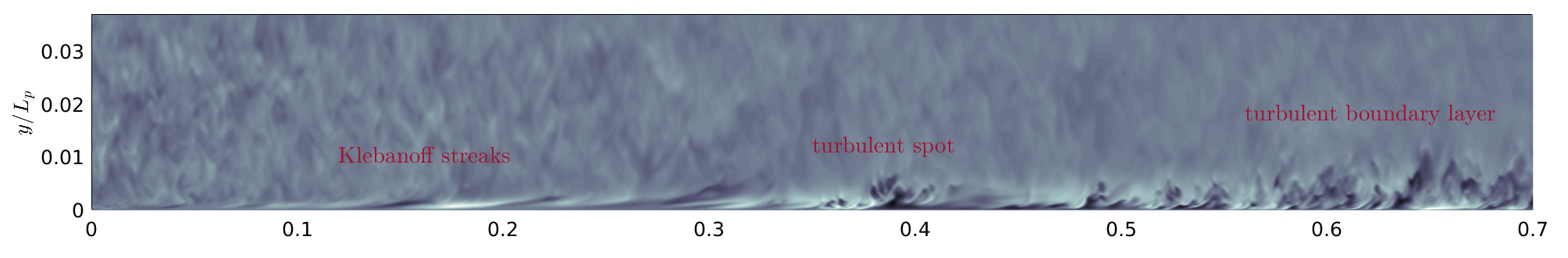}\\
\includegraphics[width=\linewidth]{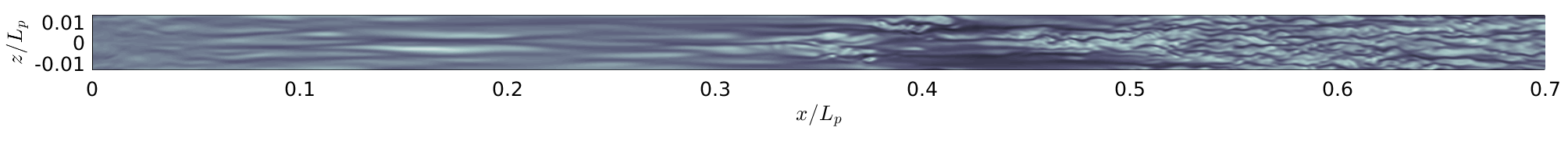}
\caption{Instantaneous side and top views of the streamwise velocity fluctuations for the T3A case.}
\label{fig:view_t3a}
\end{figure*}

\section{Validation for zero-pressure-gradient FST-induced transition}\label{sec:validation}

In this section, the T3A and T3B cases, corresponding to ZPG bypass transition under moderate and high freestream turbulence (FST) levels, are validated against Coupland's experiments and published DNS and LES results. Although these cases have already been investigated using scale-resolving simulations, this comparison provides an opportunity to validate both the present numerical setup, which includes an upstream block with a slip wall, and the adopted LES grid resolution. It also allows us to assess the ability of the simulations to reproduce the experimental results and to identify possible limitations of the measurements.

\subsection{ZPG case T3A}

First, the T3A case, with $Tu$=3\%, corresponds to the canonical scenario of FST-induced bypass transition, referred to as "bypass transition in the narrow sense" by \cite{Wu23}. The instantaneous views in Fig.~\ref{fig:view_t3a} show the development of streaky structures within the laminar boundary layer in response to freestream turbulent buffeting. These so-called Klebanoff streaks are amplified through a transient-growth mechanism, corresponding to the optimal response of the Blasius boundary layer to finite-amplitude streaks. Their lift-up by the mean shear is clearly visible in the side view. Secondary instabilities and interactions between neighbouring streaks subsequently lead to turbulent eruptions, which develop into turbulent spots. Their irregular inception and merging produce an unsteady turbulent front before the boundary layer reaches a fully turbulent state.

\begin{figure}[htbp]
\centering
\includegraphics[width=0.9\linewidth]{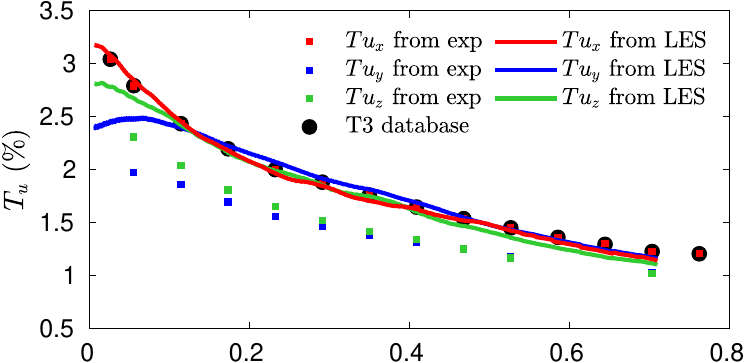}
\includegraphics[width=0.9\linewidth]{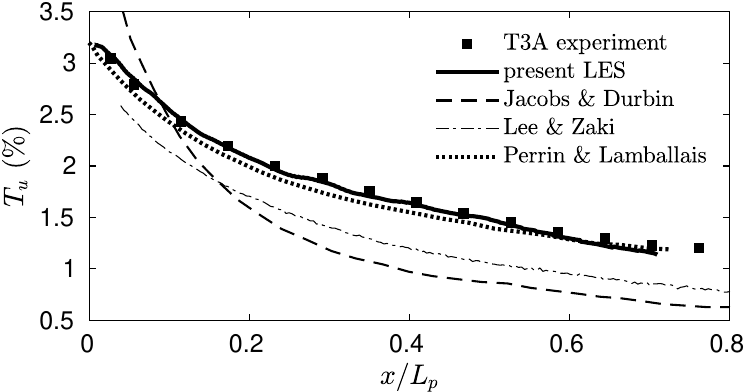}
\caption{Decay of the turbulent intensity along the plate for the T3A case.}
\label{fig:Tu_t3a}
\end{figure}

Figure~\ref{fig:Tu_t3a} shows the decay of the incoming turbulence, which must first be accurately reproduced to obtain a realistic transition location. The decay rate depends on the prescribed integral length scale $L_f$, with smaller values of $L_f$ leading to a more rapid decay. The evolution of the turbulent intensities for the three velocity components is reported in the top panel, with $Tu_x=u_{rms}/U_e$, $Tu_y=v_{rms}/U_e$, and $Tu_z=w_{rms}/U_e$, and is directly compared with the experimental values obtained from hot-wire anemometry at the last measurement points from the wall. An important point is that the $Tu$ reported in the ERCOFTAC database corresponds to $u_{rms}$ alone, as discussed in Appendix \ref{app:expe}. The present $Tu_x$ is in very good agreement with $Tu_{exp}=Tu_{x,exp}$. In contrast, the experimental values of $Tu_y$ and $Tu_z$ are lower than those obtained from the LES. Measurements have confirmed the isotropy of the grid-generated turbulence, and the numerical results also exhibit a good degree of isotropy. This suggests that the measured $v_{rms}$ and $w_{rms}$ may be underestimated. In particular, $u_{rms}$ was measured using a single-wire probe, whereas an X-wire was used for the transverse components, whose sensitivity and calibration may be less reliable, as already noted by previous authors and further illustrated in Appendix \ref{app:expe}.
The bottom panel compares the present results with other simulations of the T3A configuration. As noted by \cite{Wu26}, the decay rate obtained in the DNSs of \cite{Jacobs01} and \cite{Lee18a} is too fast, suggesting that their effective $L_f$ differs from the experimental value. Consequently, although the turbulence levels are close to 3\% at the leading-edge location, the downstream boundary layer is exposed to insufficient turbulent buffeting. The present results are in very good agreement with the iLES of \cite{Perrin23}, which includes a super-elliptic leading edge. This agreement indicates that the present setup is sufficient to reproduce the boundary-layer receptivity in the leading-edge region. It is also worth noting that a noticeable anisotropy is observed in the vicinity of the leading edge in the top panel.
A similar anisotropy is also observed in the T3B measurements, as discussed below.

\begin{figure}[h]
\centering
\includegraphics[width=0.9\linewidth]{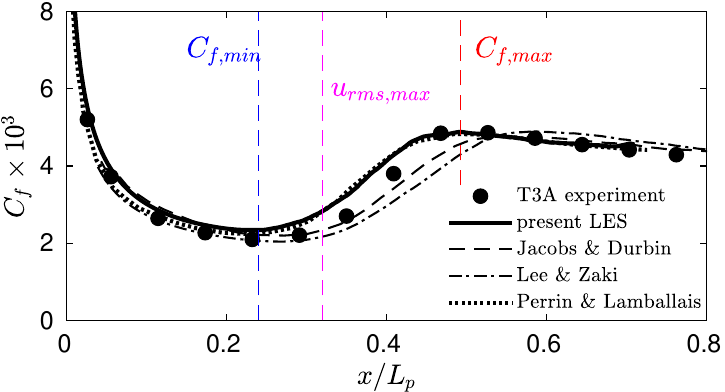}
\includegraphics[width=0.9\linewidth]{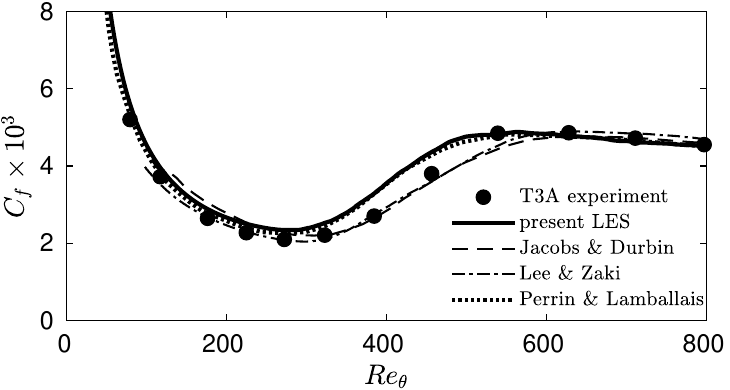}
\caption{Evolution of the skin-friction coefficient for the T3A case.}
\label{fig:Cf_t3a}
\end{figure}

The evolution of the skin-friction coefficient,
$C_f=\tau_w/(\frac{1}{2}\rho_e U_e^2)$, is plotted in Fig.~\ref{fig:Cf_t3a} as a function of $x$ normalized by the plate length $L_p$, and as a function of the Reynolds number based on the edge velocity $U_e$ and the boundary-layer momentum thickness $\delta_\theta$. The first representation provides a direct comparison with the experimental transition location and shows very good agreement with Coupland's measurements and the iLES of \cite{Perrin23}. In contrast, the DNSs of \cite{Jacobs01} and \cite{Lee18a} predict a slightly delayed transition, consistent with their lower turbulence levels. The use of $Re_\theta$ as the abscissa is commonly adopted to compare bypass transition under different flow conditions. In this representation, the increase in $C_f$ associated with the laminar-to-turbulent transition occurs closer to the experimental data in the DNSs of \cite{Jacobs01} and \cite{Lee18a}, whereas the present results remain close to those of \cite{Perrin23}. Since very good agreement is obtained when the results are plotted as a function of $x/L_p$, the origin of this small shift in $Re_\theta$ remains unclear.

\begin{figure}[h]
\centering
\includegraphics[width=0.9\linewidth]{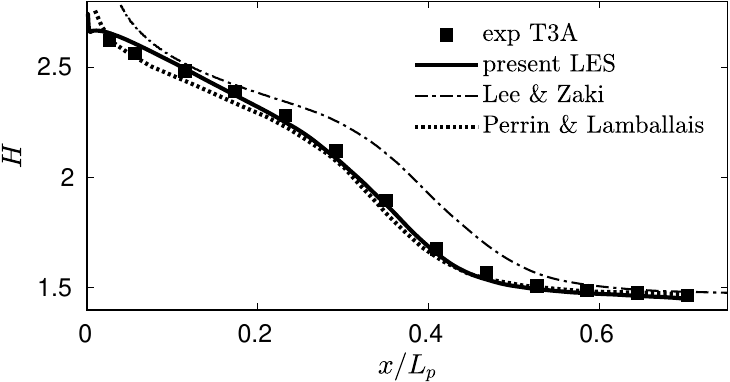}
\caption{Evolution of the shape factor for the T3A case.}
\label{fig:H_t3a}
\end{figure}

The evolution of the shape factor, $H=\delta^*/\delta_\theta$, where $\delta^*$ is the displacement thickness, is reported in Fig.~\ref{fig:H_t3a}. The present results are in good agreement with the boundary-layer measurements and are almost superimposed on those of \cite{Perrin23}. These authors highlighted discrepancies in the initial evolution of $H$ between simulations performed with and without a leading edge, emphasizing the importance of accurately reproducing leading-edge receptivity. Although a slight discrepancy with the measurements remains, the close agreement with the iLES of \cite{Perrin23}, which includes the leading edge, indicates that the present setup provides an adequate representation of the initial development. As expected, the DNS of \cite{Lee18a} exhibits a delayed decrease in $H$ from the laminar value of approximately 2.5 towards the turbulent value of approximately 1.4, consistent with their delayed transition.

\begin{figure}[h]
\centering
\includegraphics[width=0.9\linewidth]{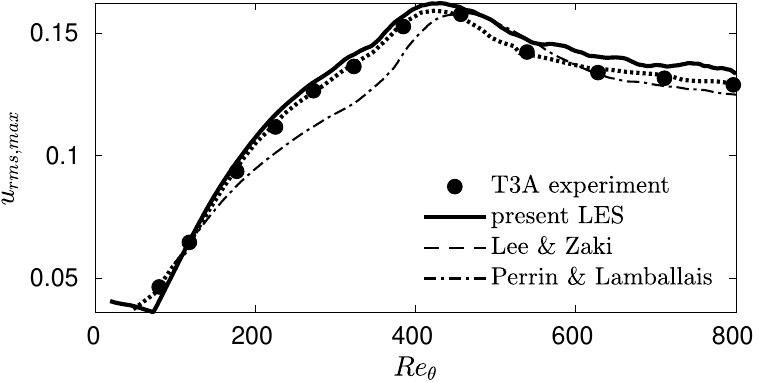}
\caption{Evolution of the maximum streamwise turbulence intensity for the T3A case.}
\label{fig:urms_t3a}
\end{figure}

\begin{figure}
  \centering
  \includegraphics[width=0.8\linewidth]{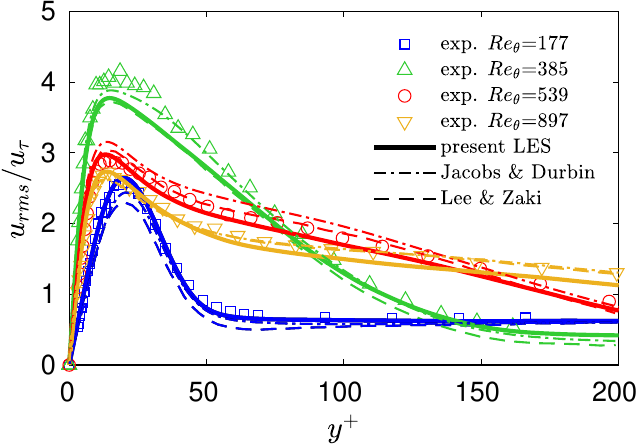}
  \includegraphics[width=0.8\linewidth]{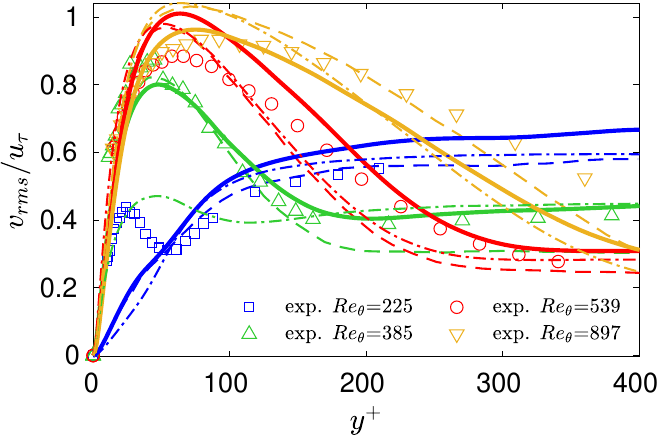}
  \includegraphics[width=0.8\linewidth]{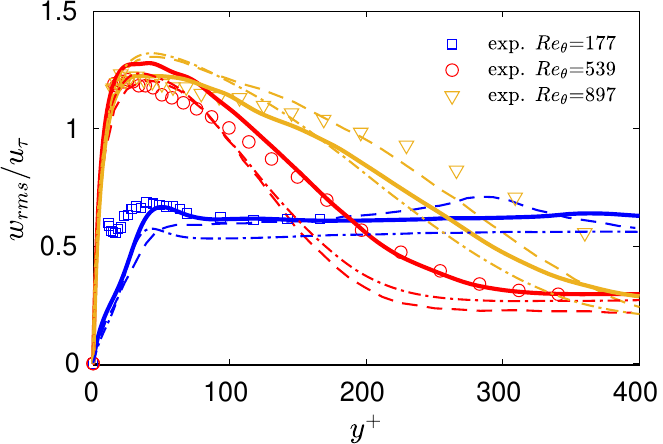}
  \includegraphics[width=0.8\linewidth]{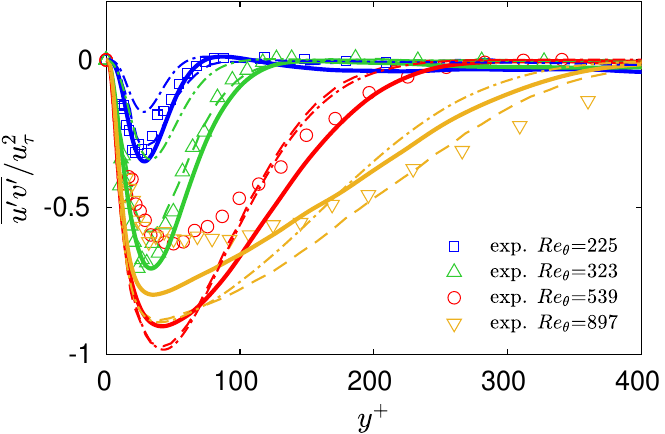}
  \caption{Comparison of Reynolds stress profiles with T3A experiments and DNSs of \cite{Jacobs01} and \cite{Lee18a}}
  \label{fig:turb_t3a}
\end{figure}

Another quantity of interest for identifying the onset of transition is the maximum streamwise turbulence intensity in the wall-normal direction, tracked along the plate (Fig. \ref{fig:urms_t3a}). In particular, this quantity provides a useful indicator of the onset of turbulent spots, at least for ZPG or equilibirum boundary layers \citep{Nolan13}. The location and magnitude of the peak are in good agreement with the experiments and the iLES of \cite{Perrin23}, whereas the peak obtained from the JHTDB database is shifted downstream.
Finally, Fig.~\ref{fig:turb_t3a} compares the Reynolds-stress profiles with the T3A experiments and the DNSs of \cite{Jacobs01} and \cite{Lee18a}. The streamwise stations correspond to those selected by \cite{Jacobs01}. The streamwise component is in good agreement with the single-wire measurements in the laminar, transitional, and turbulent regions for all simulations. Larger discrepancies are observed for the transverse components measured with the X-wire. In particular, the measured levels for $v^+_{rms}$ and $\overline{u'v'}^+$ appear to be too low at the fully turbulent stations, since all simulations predict similar values. Some differences with the reference DNSs are also observed, although it is difficult to relate them unambiguously to either the LES resolution or the slight mismatch in the predicted transition locations. A near-wall peak in $v^+_{rms}$ is observed at the first station, $Re_\theta=225$, but is not reproduced by any of the simulations, as first noted by \cite{Jacobs01}. The peak is sufficiently well defined to suggest that it may result from a measurement error, although its origin remains unexplained. Comparison with other cases suggests that the crossflow unsteadiness near the leading edge may be affected by the proximity of the lower wall of the wind tunnel and the boundary-layer bleed device.

\subsection{ZPG case T3B}

\begin{figure*}[t]
\centering
\includegraphics[width=0.95\linewidth]{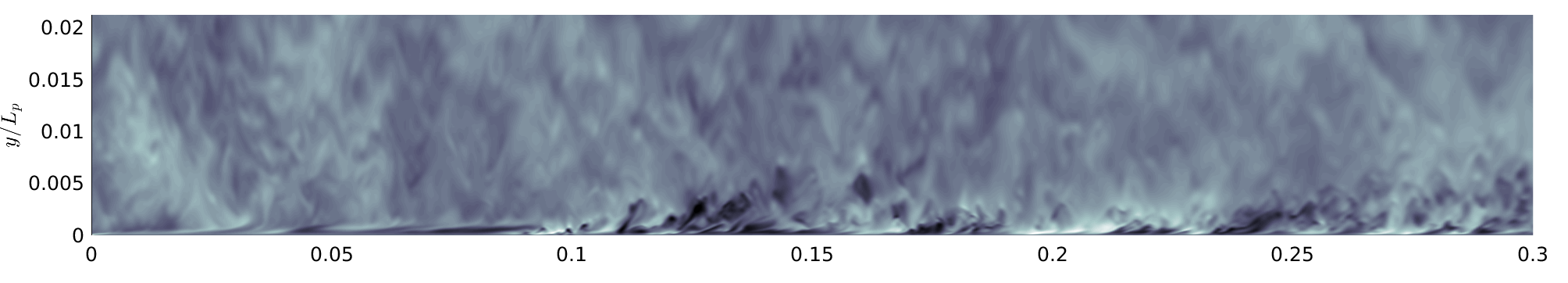}\\
\includegraphics[width=0.94\linewidth]{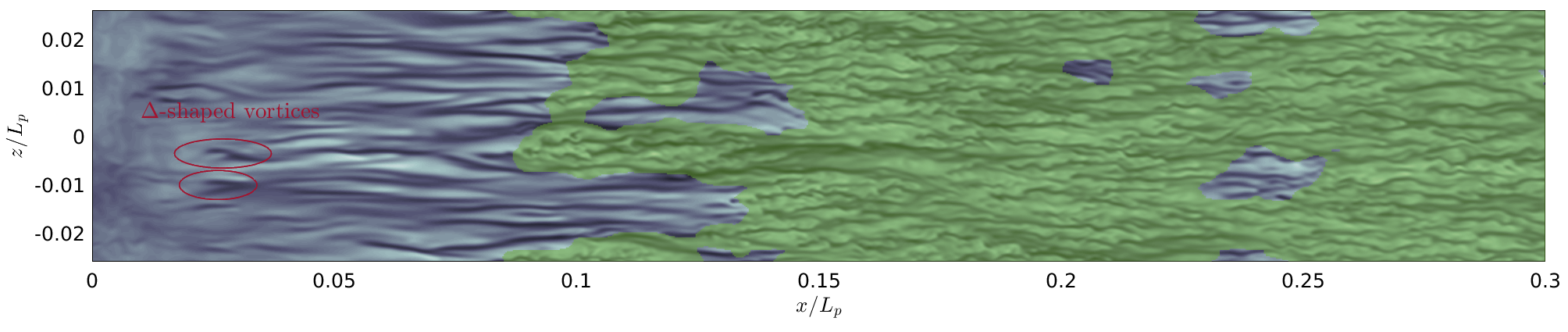}
\caption{Instantaneous side and top views of the streamwise velocity fluctuations for the T3B case.
The turbulent region from binary segmentation is displayed in semi-transparent green.}
\label{fig:view_t3b}
\end{figure*}

\begin{figure}
\centering
\includegraphics[width=0.9\linewidth]{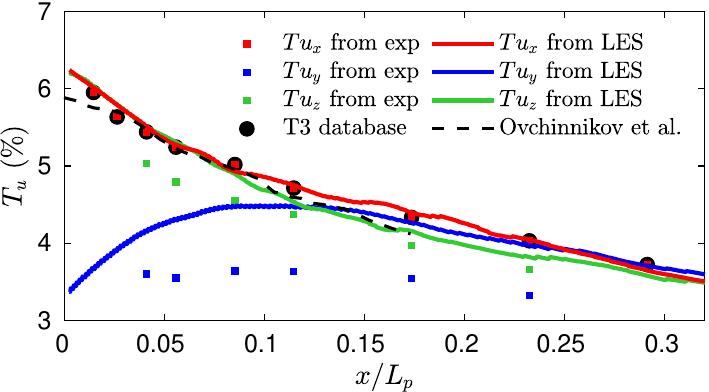}
\caption{Decay of the turbulent intensity along the plate for the T3B case.}
\label{fig:Tu_t3b}
\end{figure}

Instantaneous snaphots for T3B are displayed in Fig. \ref{fig:view_t3b}. The significantly
higher $Tu$ level (6\%) and larger integral scale are clearly visible in the side view.
A wider computational domain has been used to include several $L_f$ and its longitudinal extent
is shorter ($\approx 0.3 L_p$) since the transition appears earlier. This case has
been less often simulated. \cite{Voke95} performed an LES study and \cite{Ovchinnikov08}
used a DNS-like resolution including a super-elliptic LE (the results reported here are
those obtained with a symmetry plane). The first observations of $\Delta$-shaped vortices
previous to the streak breakdown was done on this configuration by \cite{Ovchinnikov08}
and confirmed later by \cite{Wu23} in their bypass-transition simulations using $Tu$ levels greater
than 4\%. \cite{Bienner24d} also show that a large integral lengthscale in conjunction with
high $Tu$ levels can trigger these $\Delta$-shaped vortices, that can breakdown before the
streaks. A competition is then at work between the scenario of bypass in the narrow sense
and the direct inception of spot infants by nonlinear receptivity. In the present LES,
an examination of instantaneous snapshots does not
clearly indicate spot inceptions due to $\Delta$-shaped vortices, even if strong disturbances
appear just downstream of the LE. The presence of $\Delta$-shaped vortices is examplified
in the chosen snapshot of Fig. \ref{fig:view_t3b}.
The decay of freestream turbulence is reported in Fig. \ref{fig:Tu_t3b}. As explained before,
the $Tu$ reported in Ercoftac database corresponds to the longitudinal component and is
good agreement with our results or those of \cite{Ovchinnikov08}. Here again, turbulent
fluctuations measured by X-wires seem underestimated. An interesting feature, even more
marked than for T3A case, is the lower $Tu_y$ levels in the first 10\% of the plate,
before a satisfying isotropy is recovered. The low levels of $Tu_y$ in the experiments
may also be related to this flow distortion in the vicinity of the leading edge.

\begin{figure}[h]
\centering
\includegraphics[width=0.9\linewidth]{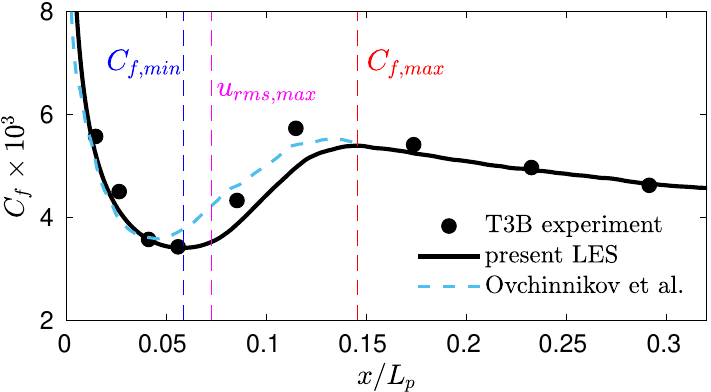}
\caption{Evolution of the skin-friction coefficient for the T3B case.}
\label{fig:Cf_t3b}
\end{figure}

\begin{figure}
\centering
\includegraphics[width=0.9\linewidth]{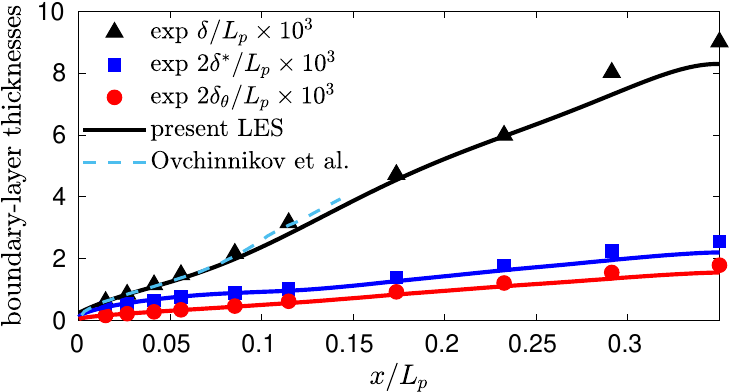}
\includegraphics[width=0.9\linewidth]{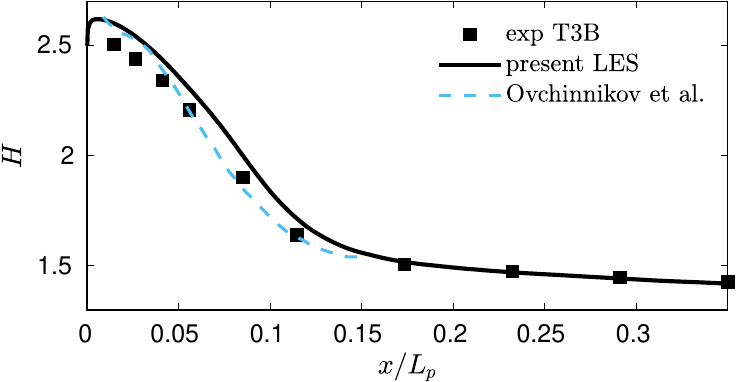}
\caption{Boundary layer development for the T3B case.}
\label{fig:delta_t3b}
\end{figure}

The skin-friction distributions along the plate in Fig. \ref{fig:Cf_t3b} indicate that transition occurs a little earlier in the simulation of \cite{Ovchinnikov08} than in the present LES, both results being in fair agreement with the measurements. The growth of the boundary-layer thicknesses, $\delta$, $\delta^*$, and $\delta_\theta$, also follows the experimental data closely (Fig. \ref{fig:delta_t3b}). The initial 99\%-thickness is reproduced slightly more accurately by the DNS of \cite{Ovchinnikov08}, suggesting that their simulation better captures the effective origin of the turbulent boundary layer and, consequently, the precise transition location. This difference is also apparent in the shape factor (bottom panel), which is nevertheless well reproduced by both simulations.

\begin{figure}[h]
\centering
\includegraphics[width=0.9\linewidth]{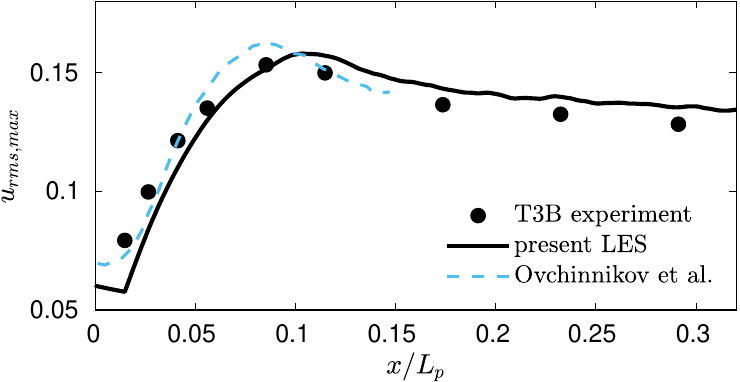}
\caption{Evolution of the maximum streamwise turbulence intensity for the T3B case.}
\label{fig:urms_t3b}
\end{figure}

The evolution of the maximum streamwise velocity fluctuation shown in Fig. \ref{fig:urms_t3b} exhibits the same overall trends. In the high-$Tu$ and high-$L_f$ bypass-transition cases investigated by \cite{Bienner24d} and \cite{Wu26}, where $\Delta$-shaped vortices either triggered or contributed to the breakdown to turbulence, a first maximum was observed in the evolution of $u_{rms,max}$ and was associated with these precursor nonlinear breakdown events. Such an early local maximum is not observed in the present simulation, nor in the simulation of \cite{Ovchinnikov08}, despite the clear identification of $\Delta$-shaped structures in their flow field. The strong buffeting of the initially laminar boundary layer by the external disturbances is nevertheless clearly evident.

\begin{figure}
  \centering
  \includegraphics[width=0.8\linewidth]{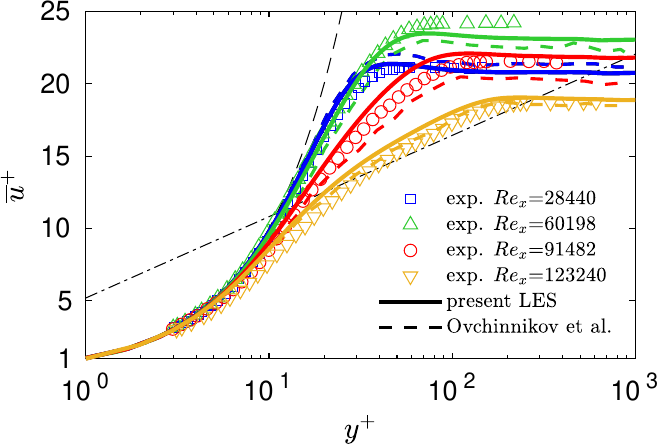}
  \includegraphics[width=0.8\linewidth]{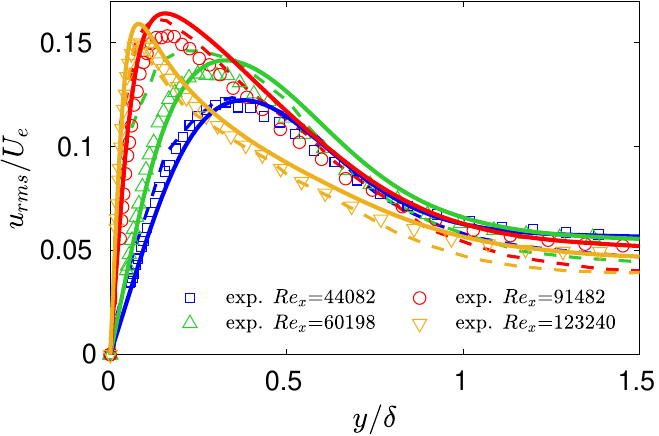}
  \includegraphics[width=0.8\linewidth]{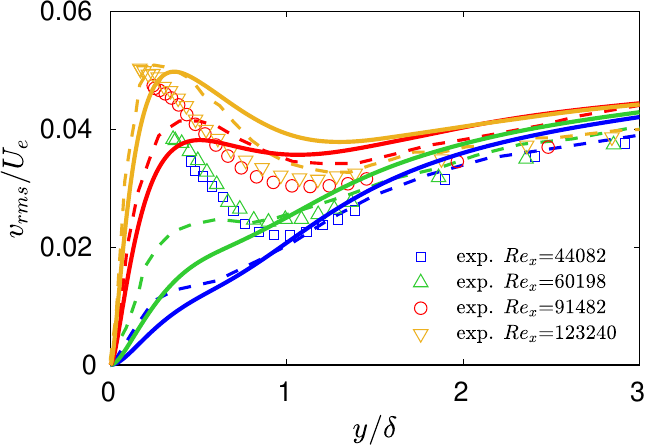}
  \includegraphics[width=0.8\linewidth]{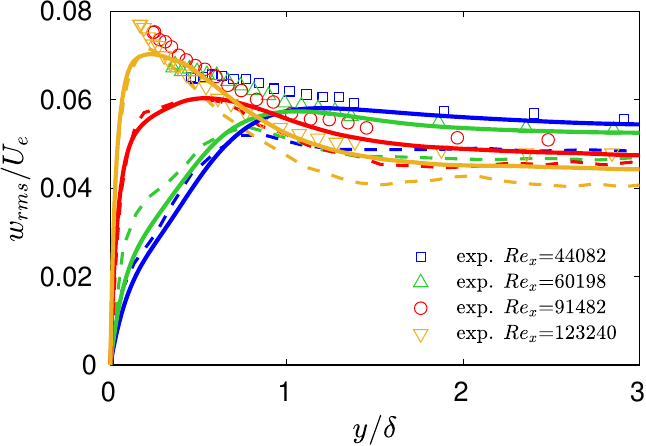}
  \caption{Comparison of mean streamwise velocity and Reynolds stress profiles with T3B experiments and DNS of \cite{Ovchinnikov08}}
  \label{fig:prof_t3b}
\end{figure}

The mean velocity and turbulent intensity profiles, normalized by the edge velocity, are assessed in Fig. \ref{fig:prof_t3b} at the $Re_x$ stations selected by \cite{Ovchinnikov08}. The transition from the laminar to the turbulent streamwise velocity profile, including the development of a logarithmic layer, is well reproduced. The longitudinal fluctuation profiles are also well captured, particularly at low Reynolds numbers, where they reflect the activity of the Klebanoff streaks. The spanwise fluctuation profiles are in fair agreement with the measurements, although the experimental values are too high below $\delta$ at the first stations. The discrepancies at these two stations are also observed in the crossflow component, which exhibits a strong near-wall peak that is not present in the simulations. A similar increase is also observed in the profiles of \cite{Ovchinnikov08}, who used a symmetry condition in the present region, and it is even more pronounced when the entire leading edge is included in the simulation. These differences in the region close to the leading edge therefore appear to result from the specific flow around the experimental plate, which is positioned close to the lower wall of the wind tunnel.

\section{Pressure gradient cases T3C}\label{sec:T3C}

\begin{figure*}
  \begin{center}
    \subfloat[][Case T3C5, $Re_{L_p}$=952k]
    {\includegraphics[width=.4\textwidth]{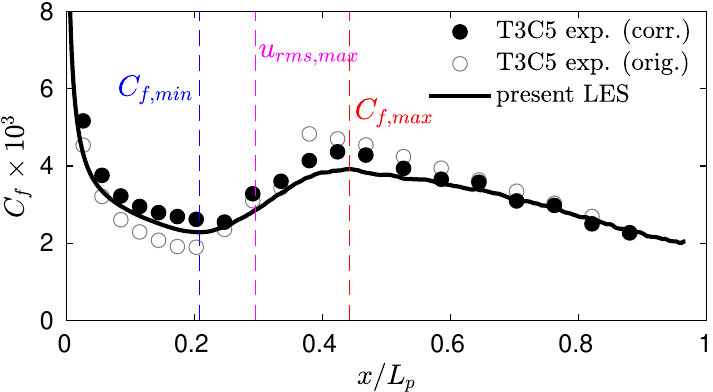}}
    \hspace{0.5cm}
    \subfloat[][Case T3C2, $Re_{L_p}$=570k]
    {\includegraphics[width=.4\textwidth]{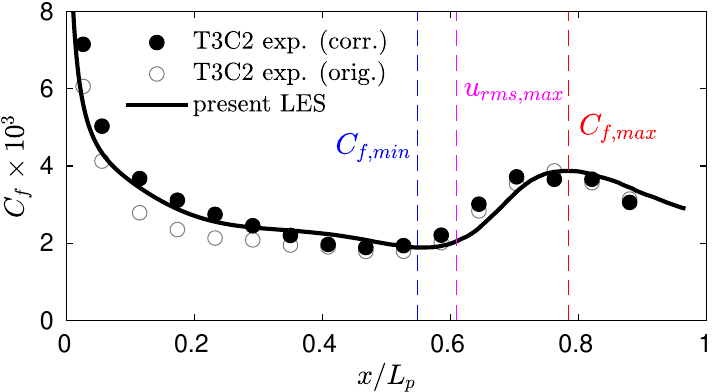}}
    
    \subfloat[][Case T3C3, $Re_{L_p}$=420k]
    {\includegraphics[width=.4\textwidth]{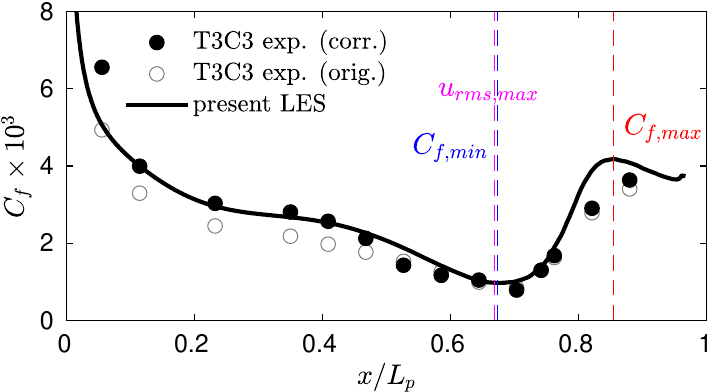}}
    \hspace{0.5cm}
    \subfloat[][Case T3C4, $Re_{L_p}$=136k]
    {\includegraphics[width=.4\textwidth]{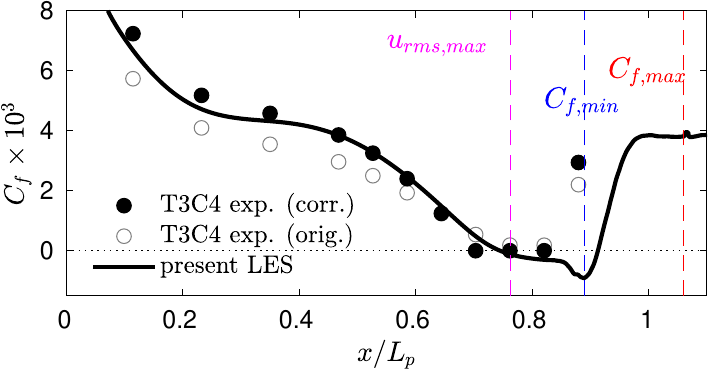}} 
    \caption{Skin friction coefficient distributions for T3C configurations.}
    \label{fig:Cf_T3C}  
  \end{center}
\end{figure*} 

\subsection{Evolution of mean flow quantities}

The T3C cases have never been investigated using wall-resolved LES, at least to the best of our knowledge, although there are almost as many RANS simulations as papers dealing with transition in RANS models. There is also one attempt using a hybrid RANS/LES approach by \cite{Yin21}. The T3C2--C5 series corresponds to the same experimental conditions, with $Tu\approx$3\%, but with different flow velocities. The parameter that therefore varies is the Reynolds number. Based on the plate length, the corresponding values are $Re_{L_p}$=$136\,000$ for T3C4, $420\,000$ for T3C3, $570\,000$ for T3C2, and $952\,000$ for T3C5. Other experiments have also systematically investigated the effect of an external pressure gradient on transition by deforming the upper wall of the wind tunnel, while varying the three parameters consisting of the Reynolds number, the incoming turbulence intensity, and the strength of the pressure gradient (variable or APG) \citep{Volino01,Roberts03,Dellacasagrande21}. A common finding from these studies is that the boundary layer may separate over certain ranges of these parameters. These experimental configurations were subsequently used by these authors to characterize separation-bubble-induced transition. In the T3C experiments, the boundary layer remains attached, with the mean $C_f$ remaining positive as seen in Fig.~\ref{fig:Cf_T3C} (symbols). Several authors have nevertheless suspected separation near the end of the plate for T3C4, and some RANS transition models predict separation for T3C2--C3. It is therefore of interest to compare these conditions with those of similar experiments reported in the literature.
Based on the aforementioned experiments, maintaining an attached flow requires either a sufficiently high Reynolds number, a sufficiently high turbulence intensity, or a sufficiently weak adverse pressure gradient. For example, the experimental data of \cite{Volino01} cover two turbulence intensities, $Tu$=0.2 and 7\%, and four Reynolds numbers (based on the nominal exit velocity and the wetted length of the suction surface) ranging from 50k to 300k. All cases exhibit separation except for the high-$Tu$ cases and the two highest Reynolds numbers, 200k and 300k.
In the experiments of \cite{Roberts03}, the test matrix covers $Tu$ values between 0.5 and 9\% and four Reynolds numbers (based on the freestream velocity and plate length), namely 350k, 470k, 650k, and 930k. Two ceiling geometries are considered: a convergent-divergent configuration and a weakly divergent one. Their 2003 proceedings paper mainly describes attached-flow cases. With the convergent-divergent ceiling, most cases exhibit separation, except at the two highest Reynolds numbers and for high turbulence intensities ($>$4.5--6\%). At $Tu$=6.4\% and $Re$=650k, a slight modification of the contoured wall ($C_{p_1}$ and $C_{p_2}$ distributions in their paper), the boundary layer can be attached or can separate.
The database of \cite{Simoni19,Dellacasagrande21} covers four values of $Tu$ (=1.5, 2.5, 3.5, and 5\%), three values of $Re$ (=70k, 150k, and 220k, based on the plate length), and four different APG strengths obtained by varying the angle of a divergent ceiling, $\alpha$=3, 5, 9, and 12$^\circ$. This database therefore encompasses a wide range of separated-flow cases. The few attached-flow cases, corresponding to high turbulence levels ($Tu>$2.5\%), relatively high Reynolds numbers ($Re$=150k or 250k), and weak pressure gradients ($\alpha$=3 or $5^\circ$), were investigated by \cite{Dellacasagrande21} to assess the impact of these parameters on intermittency and spot nucleation rates. Given the moderate turbulence intensity maintained at 3\% throughout the T3C2--C5 experiments, these results indicate that the low-Reynolds-number cases are theoretically close to separation. Another important observation is that, even when the same ceiling geometry is used for all cases, blockage effects may become significant at the lowest Reynolds number (T3C4). This has in fact led some authors to propose a different streamlined geometry for this case \citep{Suluksna09,Menter15}. We will show that the configurations C2 and C3 cases are consequently particularly sensitive to the prescribed turbulence intensity (see appendix \ref{app:T3C_sens}).
It should also be noted, as pointed out by \cite{Roberts03}, that even if the ceiling geometry is fixed, the displacement of the transition location modifies the pressure gradient experienced by the boundary layer at the onset of transition. Consequently, the pressure gradient acting during the breakdown process is not necessarily the same from one case to another.

\begin{figure}
\centering
\includegraphics[width=0.9\linewidth]{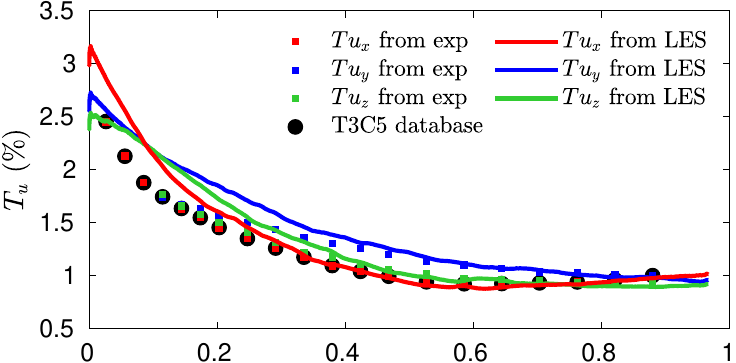}
\includegraphics[width=0.9\linewidth]{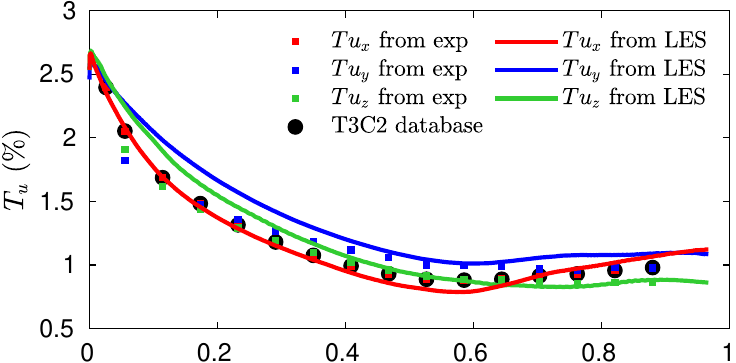}
\includegraphics[width=0.9\linewidth]{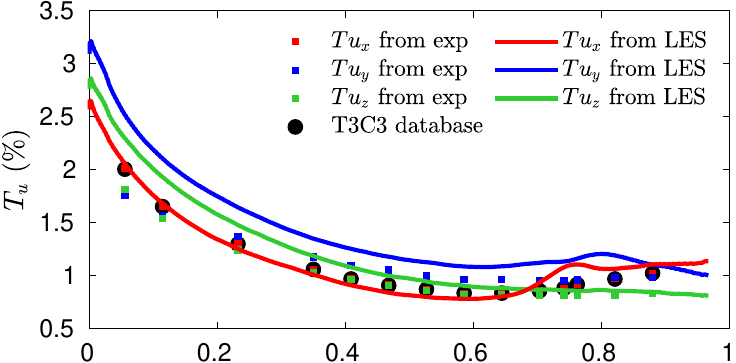}
\includegraphics[width=0.9\linewidth]{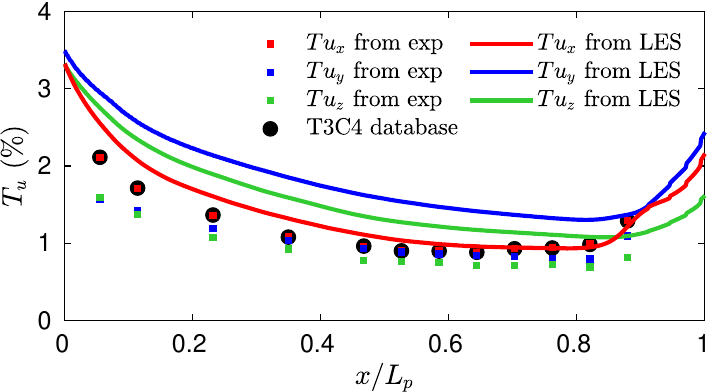}
\caption{Decay of the turbulent intensity along the plate for the T3c cases.}
\label{fig:Tu_t3c}
\end{figure}

\begin{figure*}
  \centering
  \begin{minipage}{0.4\textwidth}
    \includegraphics[width=\textwidth]{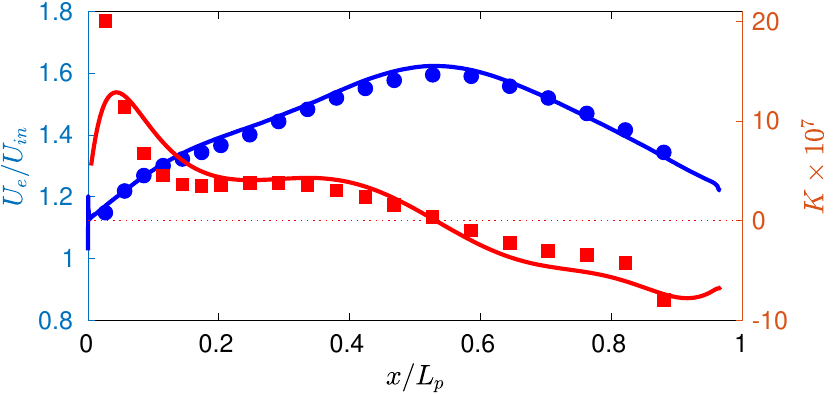}
  \end{minipage}
  \begin{minipage}{0.4\textwidth}
    \includegraphics[width=\textwidth]{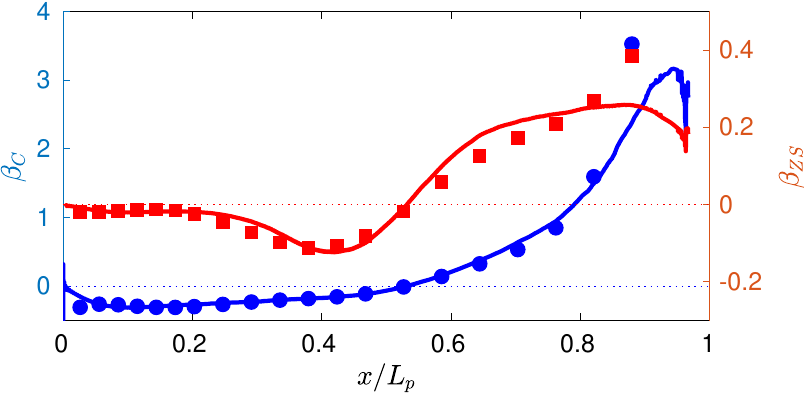}
  \end{minipage}\\
  \begin{minipage}{0.4\textwidth}
    \includegraphics[width=\textwidth]{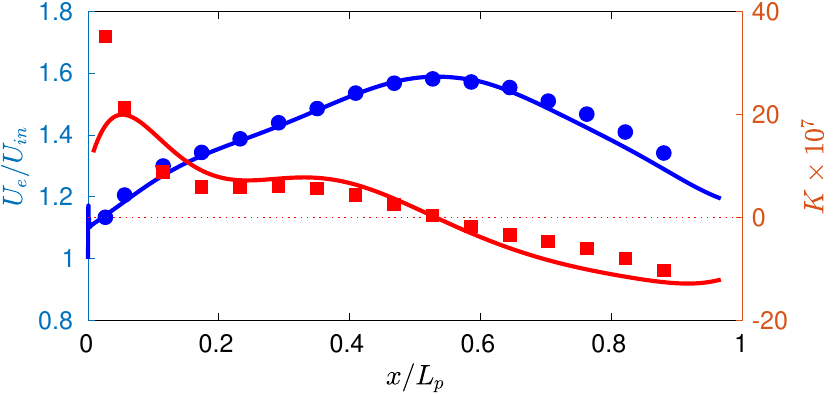}
  \end{minipage}
  \begin{minipage}{0.4\textwidth}
    \includegraphics[width=\textwidth]{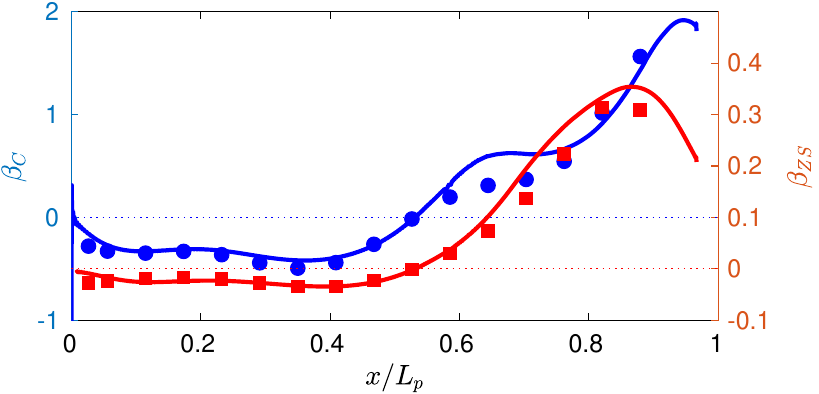}
  \end{minipage}\\
  \begin{minipage}{0.4\textwidth}
    \includegraphics[width=\textwidth]{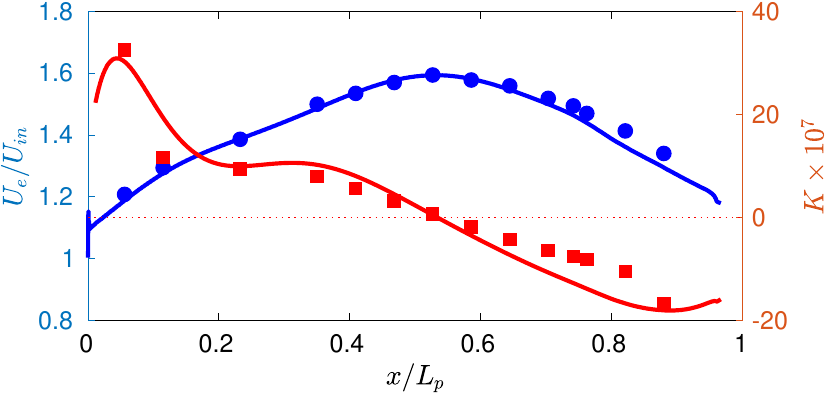}
  \end{minipage}
  \begin{minipage}{0.4\textwidth}
    \includegraphics[width=\textwidth]{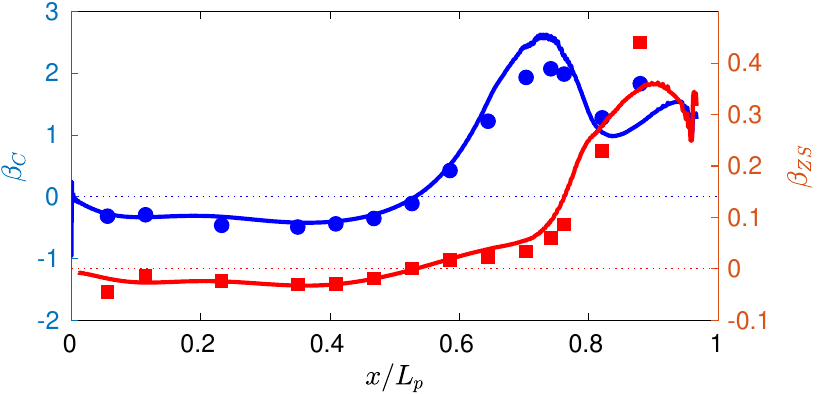}
  \end{minipage}\\
  \begin{minipage}{0.4\textwidth}
    \includegraphics[width=\textwidth]{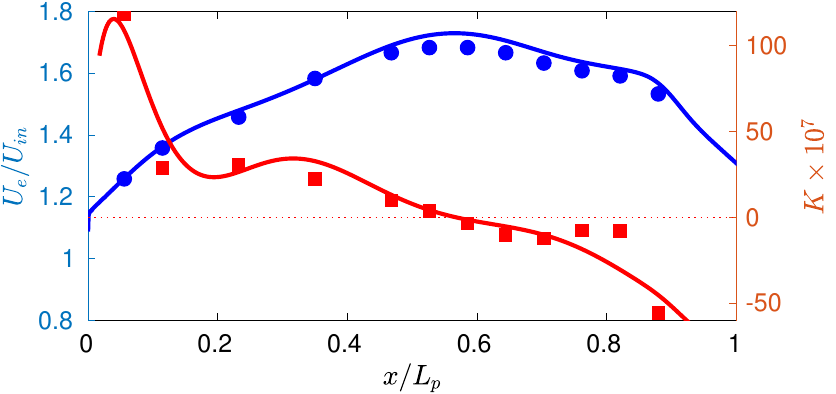}
  \end{minipage}
  \begin{minipage}{0.4\textwidth}
    \includegraphics[width=\textwidth]{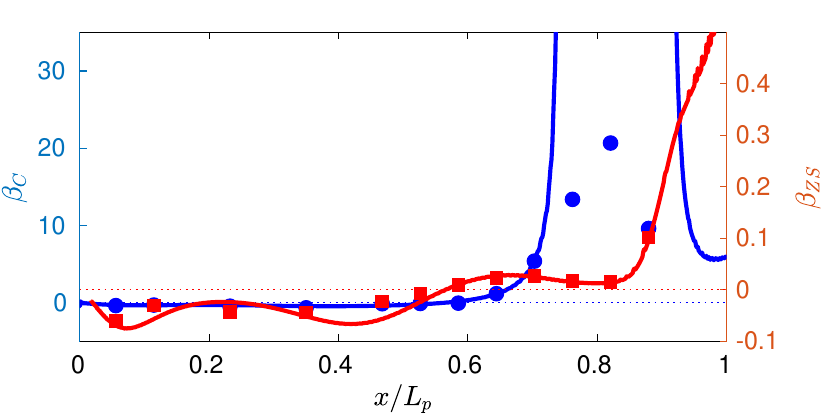}
  \end{minipage}
  \caption{Freestream velocity and gradient parameters for T3C cases.}
  \label{fig:pgrad_T3C}  
\end{figure*}

The comparison of the skin-friction coefficient evolution along the plate between simulations and experiments in Fig. \ref{fig:Cf_T3C} shows that the simulations correctly reproduce the influence of the varying pressure gradient and Reynolds number on transition. The quantitative agreement is improved when using the corrected $C_f$ values (see Appendix~\ref{app:expe}). Indeed, the values reported in the experimental database tend to underestimate the laminar skin friction, as confirmed by the momentum-balance re-evaluation presented here, while a direct evaluation of $C_f$ (based on $\partial\overline{u}/\partial y$) is preferable in the transitional region, particularly for T3C5, since the measured velocity profiles after transition onset do not yet exhibit the characteristics of a fully developed turbulent flow.
For the T3C4 configuration, the simulation exhibits a separated region ($C_f<0$) that is not observed experimentally, resulting in a delayed transition. Numerous tests were performed by varying the characteristics of the incoming turbulence ($Tu$ and $L_f$). However, except for an unrealistically high $Tu$, all simulated boundary layers separate near the end of the plate (see Appendix~\ref{app:T3C_sens}).
The evolution of the freestream turbulence intensity for the parameters selected in the final simulations is reported in Fig. \ref{fig:Tu_t3c}. As mentioned previously and discussed in Appendix~\ref{app:expe}, the measured $Tu$ corresponds to the streamwise component and is in good agreement with the present simulations. A slight overestimation is observed near the beginning of the plate for the T3C5 and T3C4 cases, but the $Tu_x$ level at the transition location is consistent with the measurements. It is also interesting to note the anisotropy generated by the external pressure gradient, which is similar for all cases but tends to increase as the Reynolds number decreases. This anisotropy is also present in the experiments and is visible for the T3C5 and T3C2 cases, which have the highest Reynolds numbers. The systematic underestimation of the transverse velocity components by the X-wire measurements, already discussed above, is more pronounced at lower Reynolds numbers, as will also be observed in the profiles of the Reynolds-stress components.

The distributions of the edge velocity and of the pressure-gradient parameters are reported in Fig. \ref{fig:pgrad_T3C}. The accelerating and then decelerating external velocity is fairly well reproduced compared with the measurements. A slight overestimation near the end of the plate may be related to the shape of the streamlined slip wall, which does not necessarily reproduce the contoured ceiling used in the experiments. Several parameters are considered to characterize the external pressure gradient, namely the acceleration parameter $K$, the Clauser pressure-gradient parameter $\beta_C$, and the Zagarola--Smits version of the Clauser parameter $\beta_{ZS}$, defined as
\begin{equation}
K=\frac{\mu_e}{\rho_e U_e^2}\frac{\partial U_e}{\partial x}\;,\;
\beta_C=\frac{\delta^*}{\rho_w u_\tau^2}\frac{\partial P_e}{\partial x}\;,\;
\beta_{ZS}=\frac{\delta}{\rho_e U_{ZS}^2}\frac{\partial P_e}{\partial x}\;,
\label{PG_param}
\end{equation}
where $U_{ZS}=U_e\delta^*/\delta$ is the Zagarola--Smits velocity. The Kline parameter is commonly used to characterize the effect of a favorable pressure gradient on turbulent boundary layers. Values below approximately $10^{-6}$ correspond to weak acceleration, whereas values of about $3\times10^{-6}$ are commonly associated with the onset of strong modifications of the turbulent boundary layer and possible relaminarization. These values should nevertheless be regarded as approximate indicators rather than universal thresholds, since the response also depends on the Reynolds number and the flow history. For the present T3C cases, the values of $K$ are increasing as the Reynolds number diminishes, and the acceleration felt by the laminar boundary layer after the leading is very high. $K$ changes sign as the external flow first accelerates and then decelerates, with the latter corresponding to the adverse pressure-gradient region. The Clauser parameter $\beta_C$ is the most used measure of the APG strength based on the wall shear stress. However, its use becomes questionable for boundary layers experiencing a large velocity defect or separation. Following \cite{Maciel18} and the subsequent analysis of \cite{Gungor22}, $\beta_{ZS}$ is therefore also considered. The Zagarola--Smits velocity, $U_{ZS}=U_e\delta^*/\delta$, decreases as the velocity defect increases and provides an outer velocity scale that accounts for the development of the boundary-layer thickness and displacement thickness. Consequently, $\beta_{ZS}$ provides a more appropriate measure of the local pressure-gradient effect when the boundary layer departs significantly from a canonical equilibrium state. In particular, \cite{Gungor22} showed that $\beta_C$ can increase to very large values and subsequently decrease in large-defect APG boundary layers, whereas $\beta_{ZS}$ remains representative of the local pressure-gradient forcing. For instance, $\beta_C$ becomes undefined for T3C4 in the separation bubble. 

\begin{figure*}
  \centering
  \begin{minipage}{0.32\textwidth}
    \includegraphics[width=\textwidth]{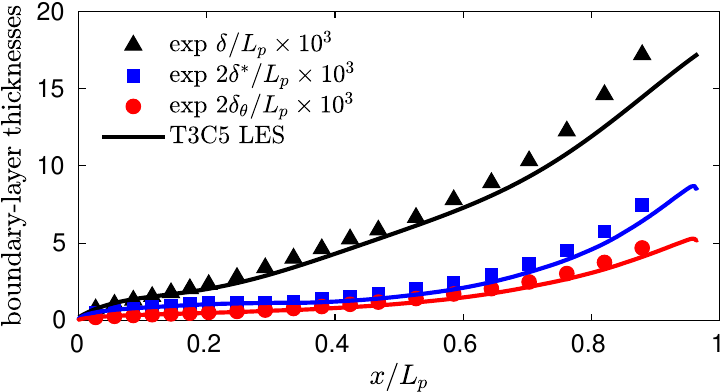}
  \end{minipage}
  \begin{minipage}{0.32\textwidth}
    \includegraphics[width=\textwidth]{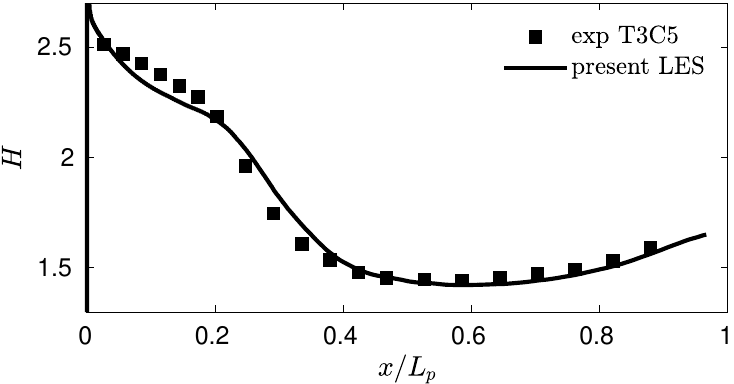}
  \end{minipage}
  \begin{minipage}{0.32\textwidth}
    \includegraphics[width=\textwidth]{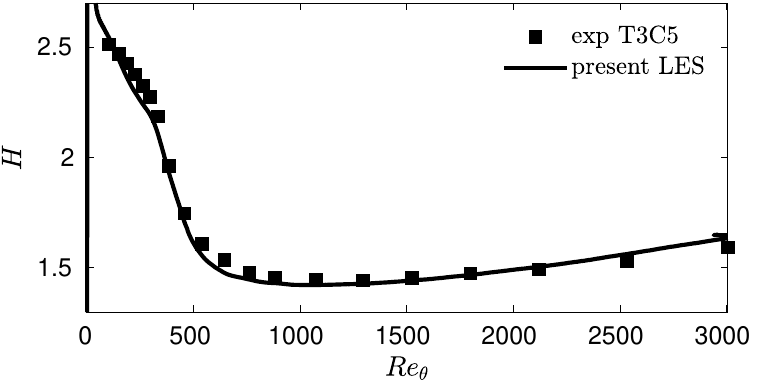}
  \end{minipage}\\
  \begin{minipage}{0.32\textwidth}
    \includegraphics[width=\textwidth]{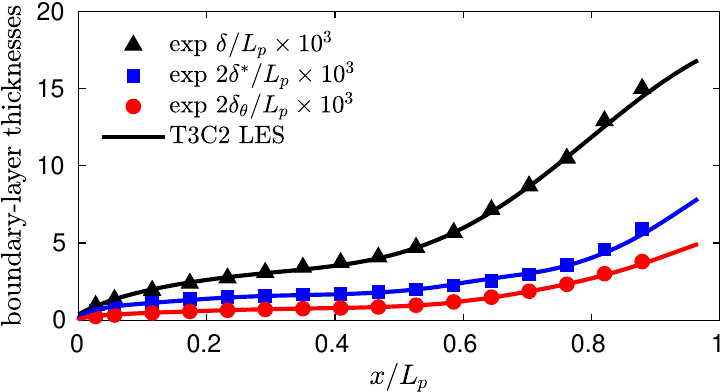}
  \end{minipage}
  \begin{minipage}{0.32\textwidth}
    \includegraphics[width=\textwidth]{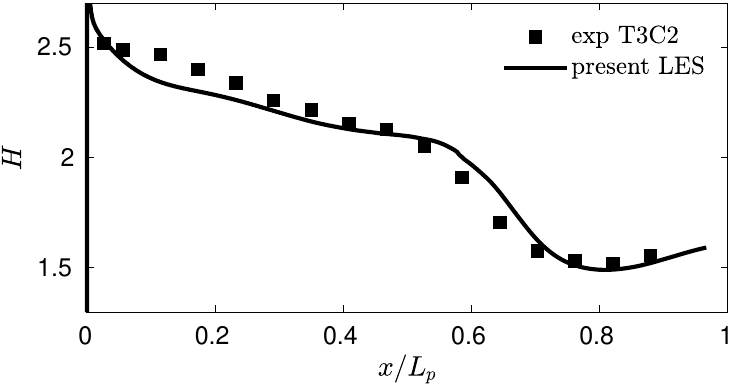}
  \end{minipage}
  \begin{minipage}{0.32\textwidth}
    \includegraphics[width=\textwidth]{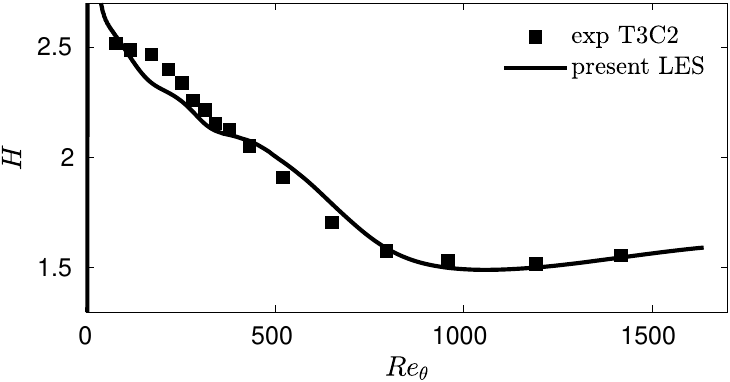}
  \end{minipage}\\
  \begin{minipage}{0.32\textwidth}
    \includegraphics[width=\textwidth]{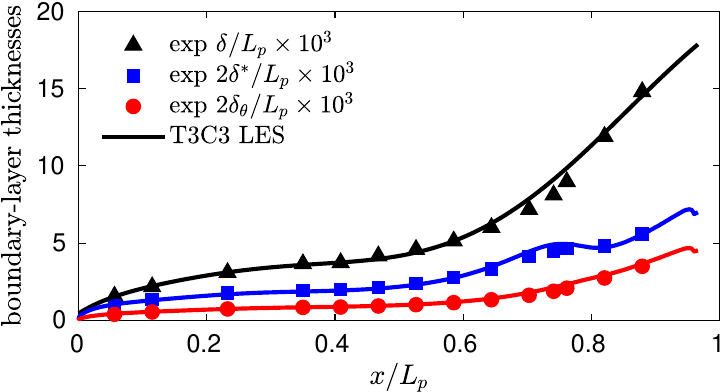}
  \end{minipage}
  \begin{minipage}{0.32\textwidth}
    \includegraphics[width=\textwidth]{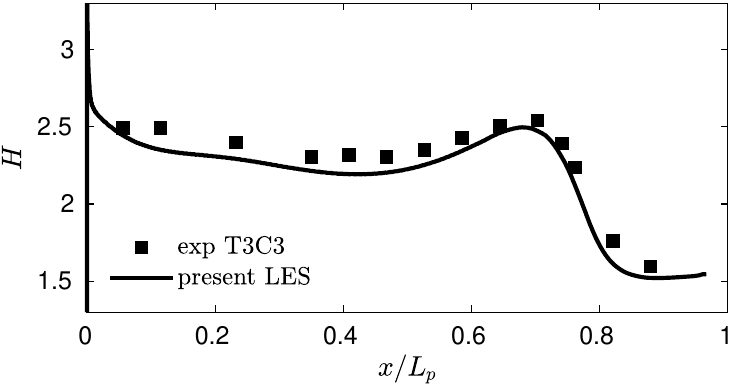}
  \end{minipage}
  \begin{minipage}{0.32\textwidth}
    \includegraphics[width=\textwidth]{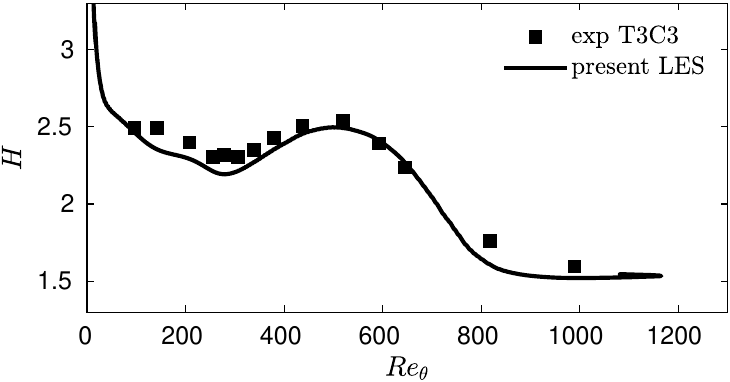}
  \end{minipage}\\
  \begin{minipage}{0.32\textwidth}
    \includegraphics[width=\textwidth]{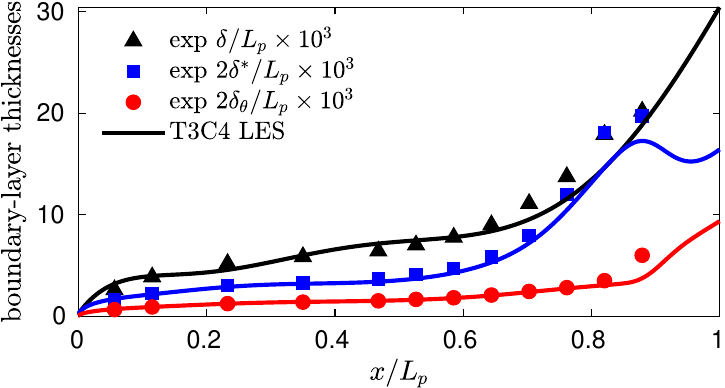}
  \end{minipage}
  \begin{minipage}{0.32\textwidth}
    \includegraphics[width=\textwidth]{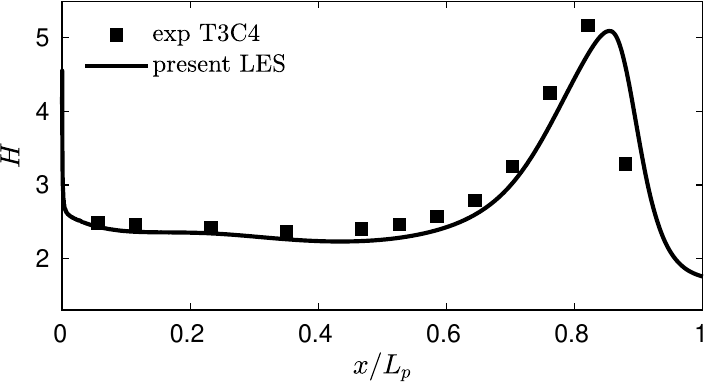}
  \end{minipage}
  \begin{minipage}{0.32\textwidth}
    \includegraphics[width=\textwidth]{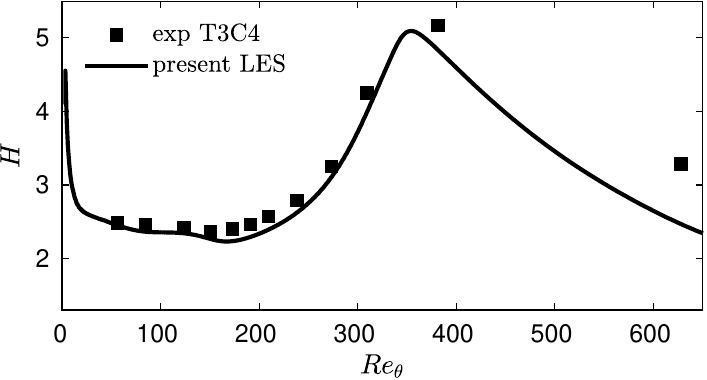}
  \end{minipage}
  \caption{Boundary-layer thicknesses and shape factor for T3C cases.}
  \label{fig:delta_T3C}  
\end{figure*}

The evolution of the boundary-layer thicknesses and shape factor is shown in Fig.~\ref{fig:delta_T3C}. For T3C5, the change in the growth rate of $\delta$ occurs around $0.22L_p$, consistently with the evolutions of $C_f$ and $H$. Unlike in the ZPG cases, the subsequent turbulent growth is not linear but accelerates as the pressure gradient becomes adverse. The slight underestimation of $\delta$ near the end of the plate may indicate that the adverse pressure gradient generated by the streamlined upper wall is slightly too weak, or that the effective origin of the turbulent boundary layer lies slightly too far downstream. The transition location is more readily identified from the shape factor and is in very good agreement with the experiment. Starting from the Blasius value ($H\approx2.5$), $H$ first decreases under the strong favorable pressure gradient near the leading edge (see $K$ in Fig.~\ref{fig:pgrad_T3C}). This decrease is slightly faster than in the measurements, a trend observed for all T3C cases that reflects the sensitivity of the leading-edge region, but a similar value of about 2 is reached at transition onset. $H$ then drops rapidly to about 1.5 and, instead of reaching a plateau, increases again as the adverse pressure gradient accelerates the growth of $\delta^*$. When plotted against $Re_\theta$, transition onset is found around $Re_\theta=500$, while the pressure gradient is still favorable. T3C5 can therefore be regarded primarily as a case of transition under a favorable pressure gradient, whereas, for the other cases, transition moves towards the end of the plate and occurs under an increasingly strong adverse pressure gradient as the Reynolds number decreases.

For T3C2--C4, the evolution of $\delta$ is qualitatively similar, with a marked change in growth rate at transition and a growing difference between $\delta^*$ and $\delta_\theta$. When transition occurs late, the tendency of the laminar boundary layer to separate under the adverse pressure gradient produces a local maximum of $\delta^*$, which is absent from $\delta_\theta$ and therefore results in a peak of the shape factor. For T3C3, this maximum occurs around $0.7L_p$, as in the experimental data, and the resulting peak in $H$ ($\approx2.7$) is well reproduced, both qualitatively and quantitatively. After a slow decrease under the favorable pressure gradient, $H$ increases ever more rapidly as the boundary layer approaches separation, until transition brings it down to about 1.5. This peak reflects the competition between the tendency of the laminar boundary layer to separate and the development of an attached turbulent flow, which is highly sensitive to the transition location and hence to the free-stream turbulence level (Appendix~\ref{app:T3C_sens}). For T3C2, no increase in $H$ is observed, indicating that transition occurs just before the adverse pressure gradient can significantly affect the laminar boundary layer. A slightly lower turbulence level shifts transition only marginally downstream but is sufficient for the increase in $H$ to appear, so that T3C2 lies very close to the boundary between the two regimes. Conversely, this behaviour is amplified in T3C4: the peak in $\delta^*$ is pronounced and located near the end of the plate, and $H$ reaches about 5.5 before dropping sharply with the rapid transition. The adverse pressure gradient is strong enough to generate a shear layer, at least instantaneously, which becomes unstable through a Kelvin--Helmholtz-type mechanism and accelerates transition \citep{Istvan18}. In the simulation, the boundary layer clearly separates, and artificially advancing transition by increasing $Tu$ to mitigate this separation results in a smaller peak in $H$ (Appendix~\ref{app:T3C_sens}).

\begin{figure*}[t]
  \centering
  \subfloat[][Case T3C5, $Re_{L_p}$=952k]
  {\includegraphics[width=0.245\linewidth]{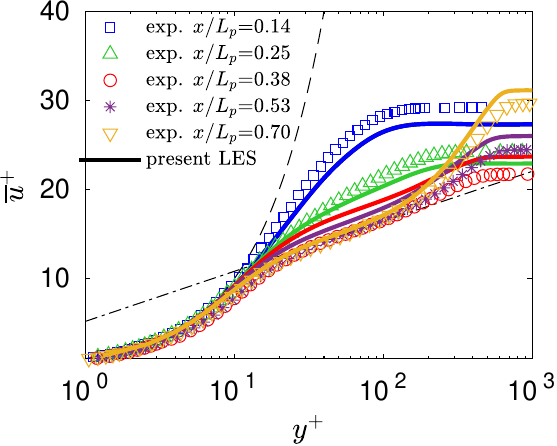}
   \includegraphics[width=0.245\linewidth]{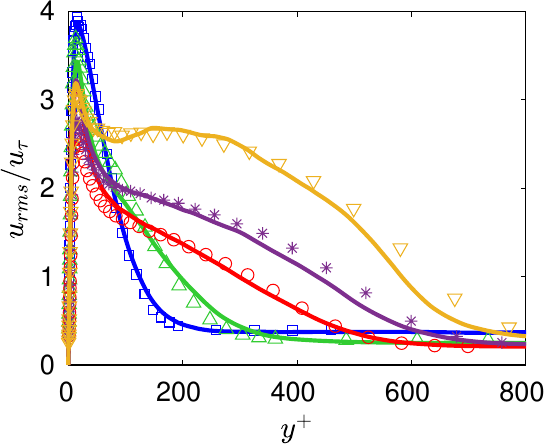}
   \includegraphics[width=0.245\linewidth]{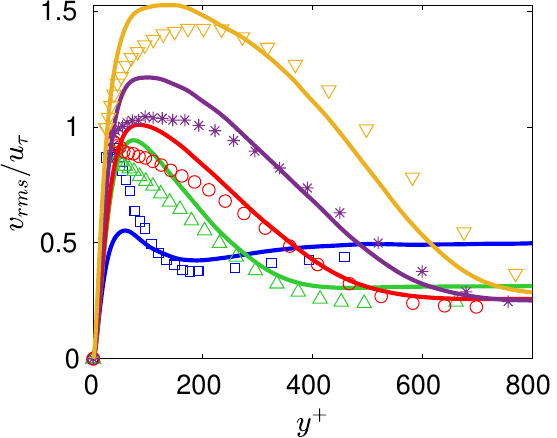}
   \includegraphics[width=0.245\linewidth]{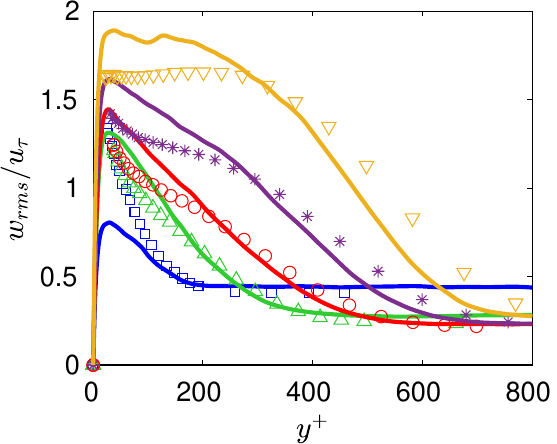}}\\
  \subfloat[][Case T3C2, $Re_{L_p}$=570k]
  {\includegraphics[width=0.245\linewidth]{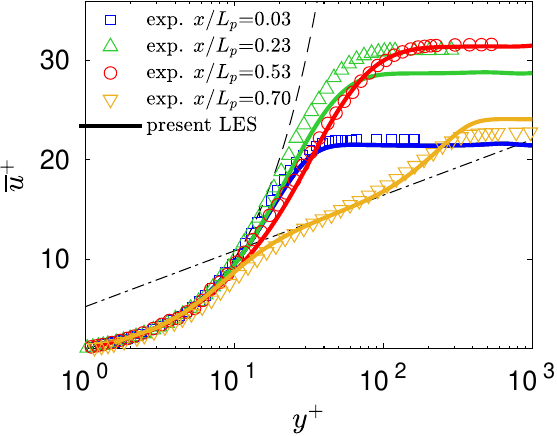}
   \includegraphics[width=0.245\linewidth]{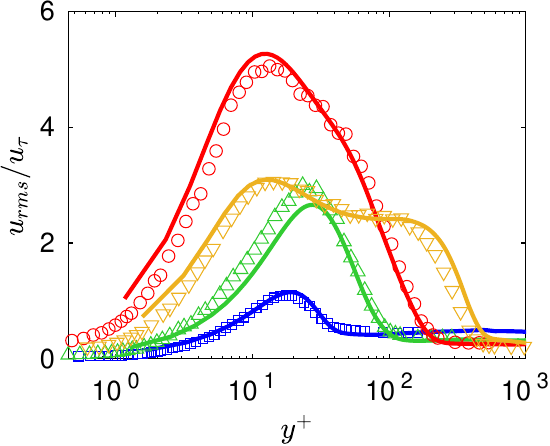}
   \includegraphics[width=0.245\linewidth]{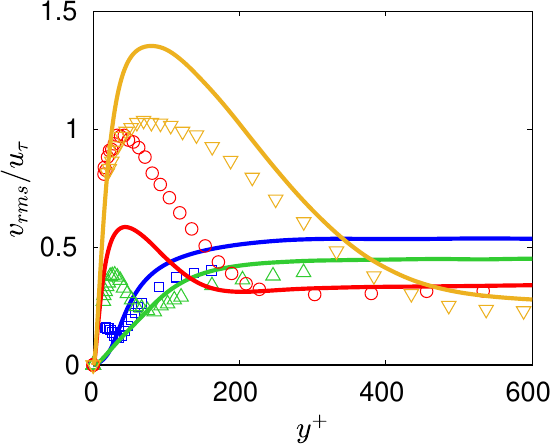}
   \includegraphics[width=0.245\linewidth]{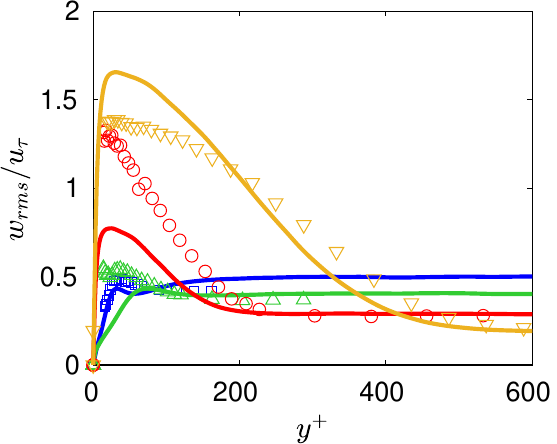}}\\
  \subfloat[][Case T3C3, $Re_{L_p}$=420k]
  {\includegraphics[width=0.245\linewidth]{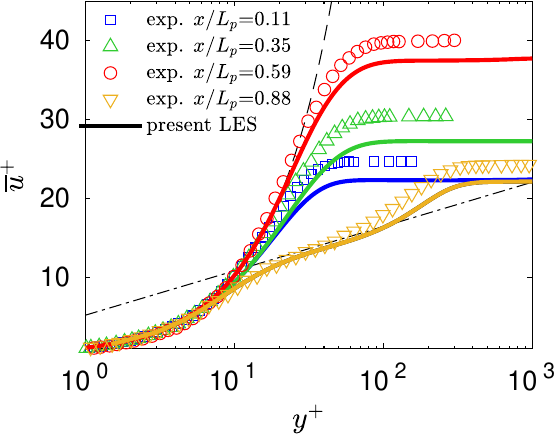}
   \includegraphics[width=0.245\linewidth]{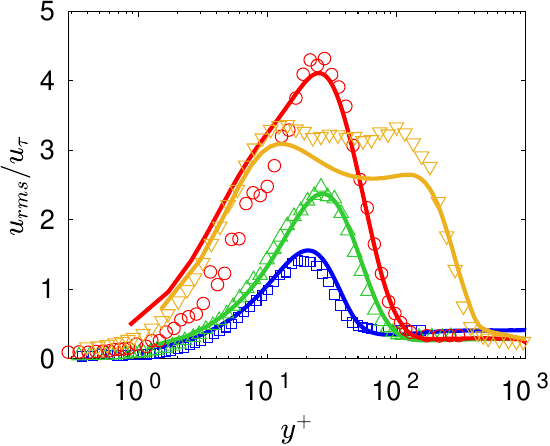}
   \includegraphics[width=0.245\linewidth]{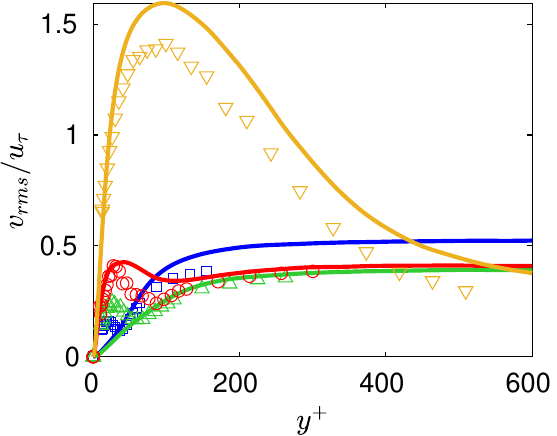}
   \includegraphics[width=0.245\linewidth]{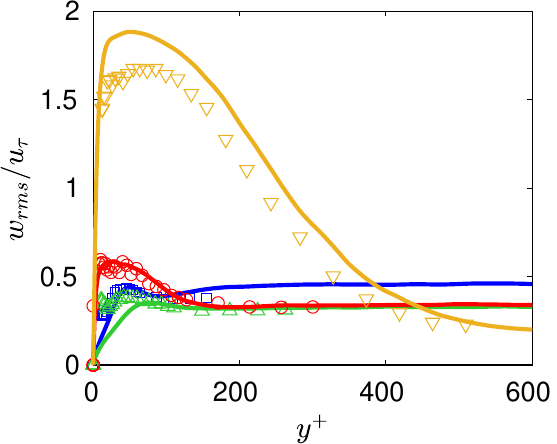}}\\
  \subfloat[][Case T3C4, $Re_{L_p}$=136k]
  {\includegraphics[width=0.245\linewidth]{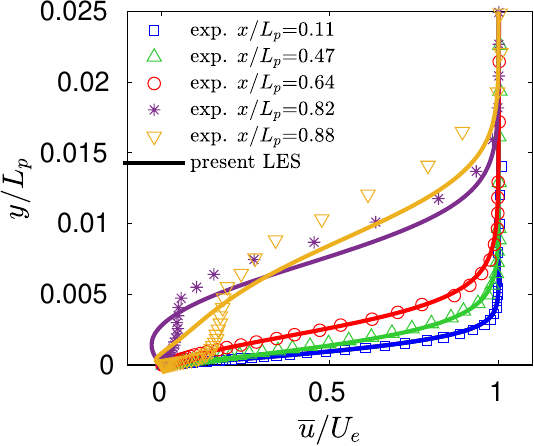}
   \includegraphics[width=0.245\linewidth]{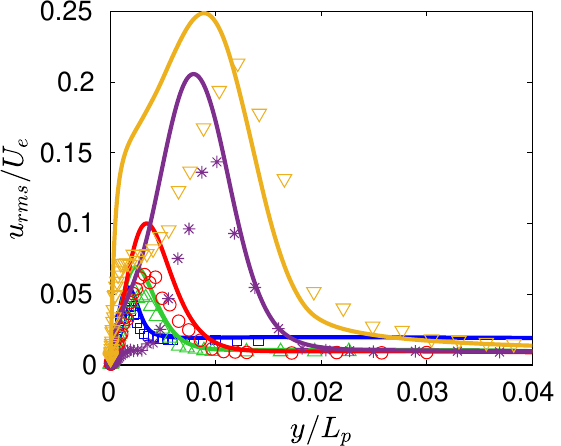}
   \includegraphics[width=0.245\linewidth]{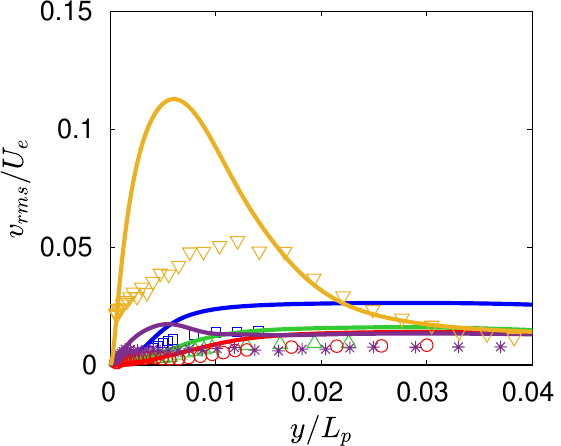}
   \includegraphics[width=0.245\linewidth]{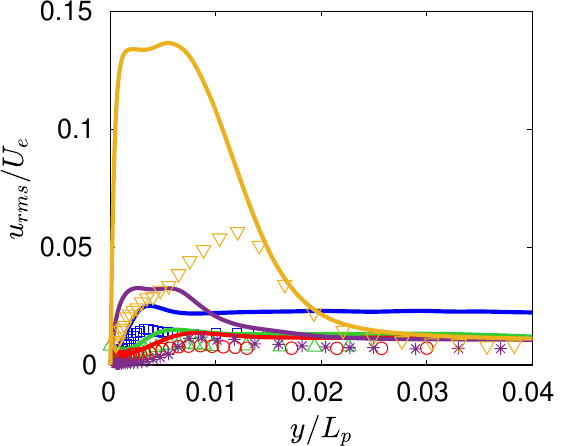}}
  \caption{Mean streamwise velocity and Reynolds stress profiles for various T3C configurations.}
  \label{fig:prof_t3c}
\end{figure*}

\subsection{Velocity and turbulent intensity profiles}
We now compare the mean streamwise velocity profiles and the three components of the Reynolds stresses for the T3C cases, ordered by decreasing Reynolds number: T3C5, T3C2, T3C3, and T3C4.

The profiles for T3C5 are shown at five successive stations in Fig. \ref{fig:prof_t3c}(a). The first two streamwise velocity profiles still correspond to laminar and transitional boundary layers and are in reasonable agreement with the experiments, given the sensitivity of this representation to the value of the friction velocity. At the two intermediate local Reynolds numbers, the profile is close to a turbulent profile, but with a higher intercept of the logarithmic law. This feature is not reproduced by the experiments, since the classical logarithmic law, $u^+=(1/\kappa)\log(y^+)+C$, with $\kappa=0.41$ and $C=5.2$, is used in the Clauser-chart method to determine the friction velocity $u_\tau$. The last profile, corresponding to a fully turbulent boundary layer, is in good agreement with the measurements and exhibits a typical wake profile for an APG boundary layer, with a large wake parameter. The development of the streamwise velocity fluctuations during transition is also very well reproduced. Again, the last profile is characteristic of a turbulent boundary layer under an adverse pressure gradient. For the transverse fluctuations, the progressive development of the fluctuations is correctly captured, although a systematic overestimation is observed. Based on the previous cases, this discrepancy is more likely due to an underestimation of the measurements obtained with X-wires. The pronounced near-wall peak observed at the first station, in both $v$ and $w$, is also absent from the simulations and probably corresponds to a specific flow feature immediately upstream of the plate, as already noted for the ZPG cases.

\begin{figure}
\centering
\includegraphics[width=0.9\linewidth]{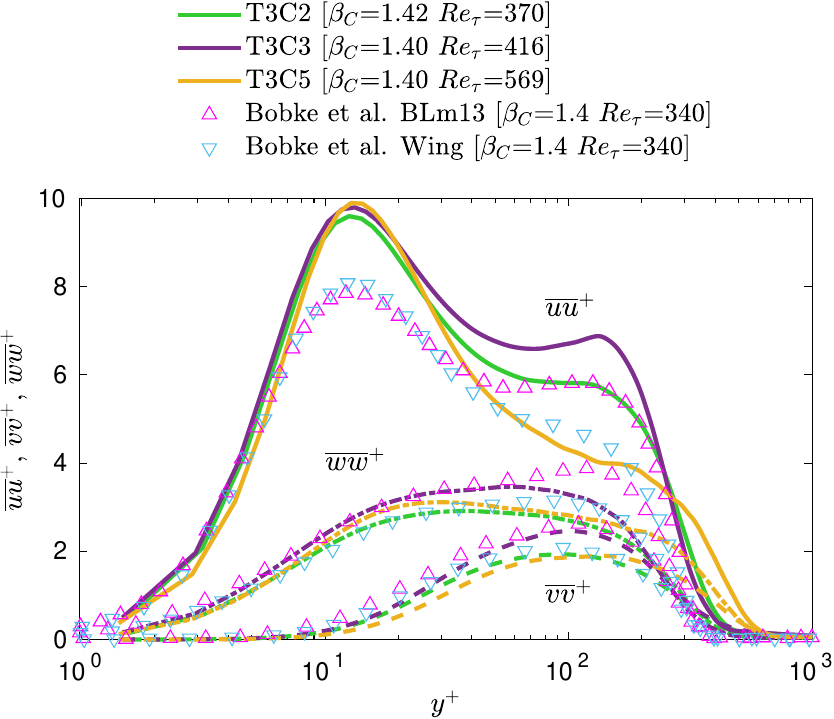}
\caption{Inner-scaled turbulent intensities at a location where the adverse pressure gradient
  corresponds to $\beta_C$=1.4.}
\label{fig:APG}
\end{figure}

The profiles for T3C2 are shown in Fig. \ref{fig:prof_t3c}(b). Although transition occurs later, the evolution of the velocity profiles remains typical of bypass transition and follows the measurements reasonably well. The $u_{rms}$ profiles also show very good agreement, considering the sensitivity to the flow conditions ($Tu$ and pressure-gradient location). The logarithmic scale clearly highlights the outer peak in the streamwise fluctuations, which is characteristic of the presence of an adverse pressure gradient. The comparison is less favorable for $v_{rms}$ and $w_{rms}$. For the first transitional profiles, only the increase in the fluctuation levels in the outer part of the boundary layer, towards the freestream levels, is consistent with the measurements. The experiments, however, exhibit a near-wall peak that increases up to the mid-length of the plate, which is completely absent in the iLES. Moreover, the fluctuation levels in the turbulent profile are significantly higher than the experimental values, further confirming the uncertainty associated with the X-wire measurements.
For T3C3, the same overall observations can be made in Fig. \ref{fig:prof_t3c}(c). The increase in the fluctuation levels at the last station is more pronounced, owing to the thickening of the laminar boundary layer induced by the adverse pressure gradient immediately before transition.
A comparison of the turbulent-intensity profiles at the same value of the Clauser APG parameter, $\beta_C=1.4$, is provided in Fig. \ref{fig:APG}. Because of the early transition, this condition is reached at a significantly higher Reynolds number in T3C5. The T3C2 and T3C4 cases can instead be compared with the profiles of \cite{Bobke17} at $Re_\tau=340$. The simulations of \cite{Bobke17} consider two flow conditions: an equilibrium boundary layer with a power-law external velocity distribution (case BLm13) and a boundary layer developing over a NACA profile under a variable pressure gradient (case Wing). A common feature of these cases is an increase in the fluctuation levels in the wake region, resulting in a secondary maximum in the $\overline{u'u'}$ profiles.
This effect is not solely determined by the local pressure-gradient strength, but also depends on the flow history, i.e., on the way in which the pressure gradient has evolved upstream. This is particularly apparent for the T3C cases: although the same local value of $\beta_C$ is considered, the profiles differ because this value is reached at different streamwise locations, $x/L_p$, and following different pressure-gradient histories. Moreover, the T3C profiles are obtained shortly after transition, which explains the higher amplitude of the inner peak. The flow history therefore combines the effects of the variable pressure gradient and the transition process.
Finally, for T3C4, the profiles in Fig. \ref{fig:prof_t3c}(d) are normalized by $U_e$ rather than by $u_\tau$, which approaches zero as the boundary layer separates. Most of the plate remains laminar, and the velocity profiles at the first stations are in good agreement with the experiments, with the same discrepancies in the velocity fluctuations as observed for the other cases.
The profile at $x/L_p=0.82$ lies within the separated bubble in the simulation and exhibits negative streamwise velocities near the wall, whereas the experimental profile shows a steep near-wall gradient connected to a thick boundary layer. It is unclear whether the single-hotwire measurements can reliably distinguish positive and negative streamwise velocities in this region, or whether the profile instead reflects some degree of flow recovery. The fact that the last experimental profile, at $x/L_p=0.88$, exhibits a similar near-wall distribution is also puzzling. At this location, the LES indicates that the boundary layer is already reattaching. The rapid increase in the magnitude of $u_{rms}$ at the last stations is qualitatively reproduced by the LES, although the near-wall behaviour of the measured profiles remains difficult to interpret.

\begin{figure*}[htbp]
\centering
\includegraphics[width=\linewidth]{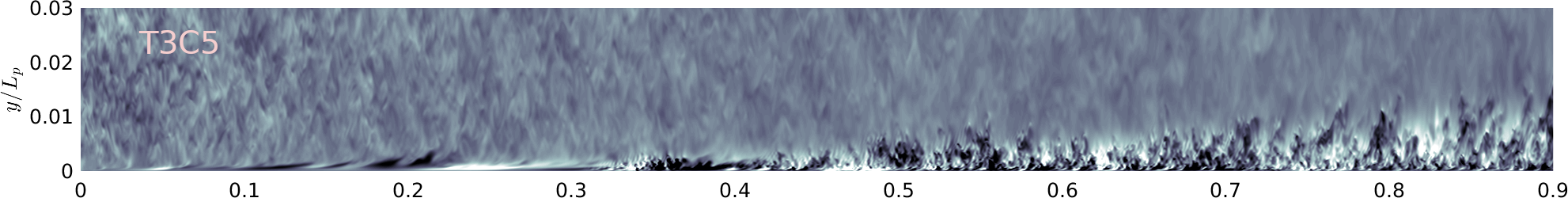}\\
\includegraphics[width=\linewidth]{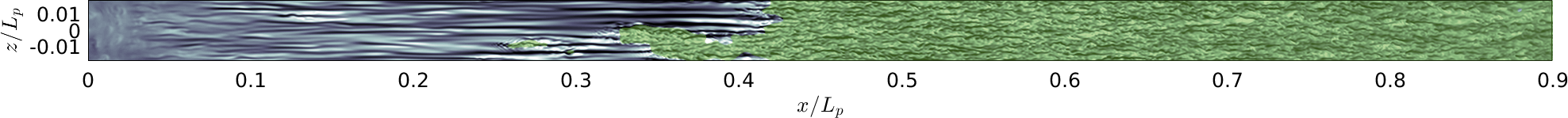}\\
\includegraphics[width=\linewidth]{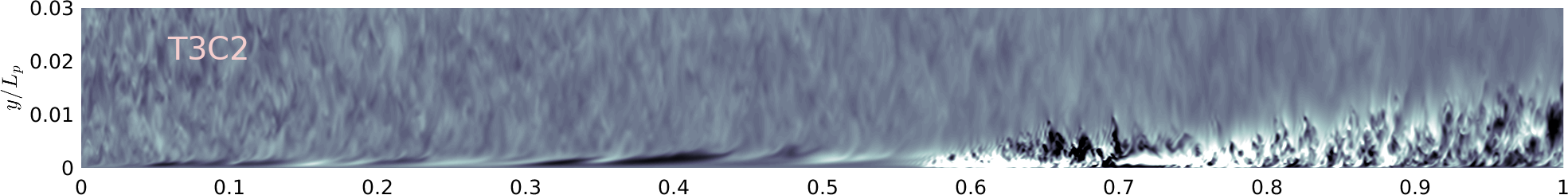}\\
\includegraphics[width=\linewidth]{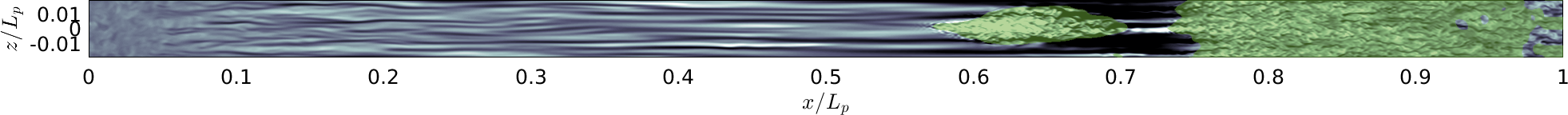}\\
\includegraphics[width=\linewidth]{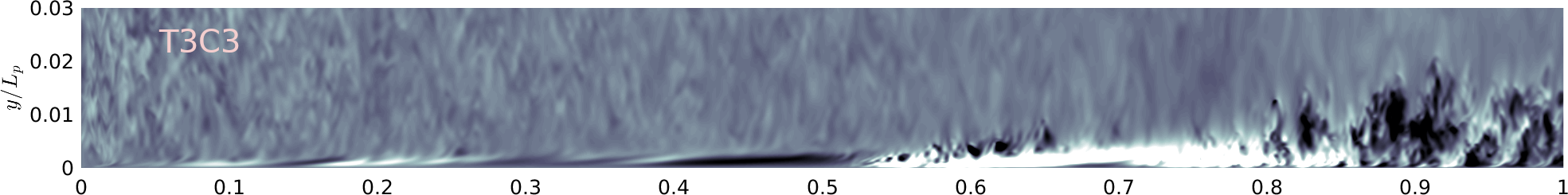}\\
\includegraphics[width=\linewidth]{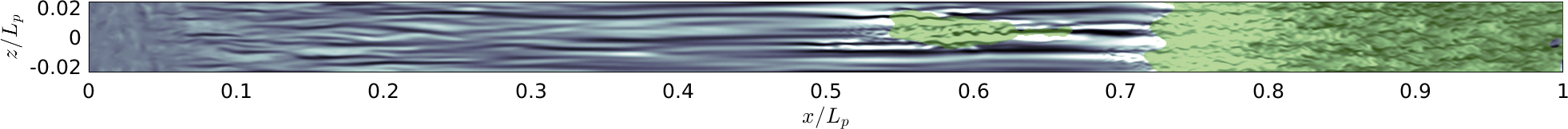}\\
\includegraphics[width=\linewidth]{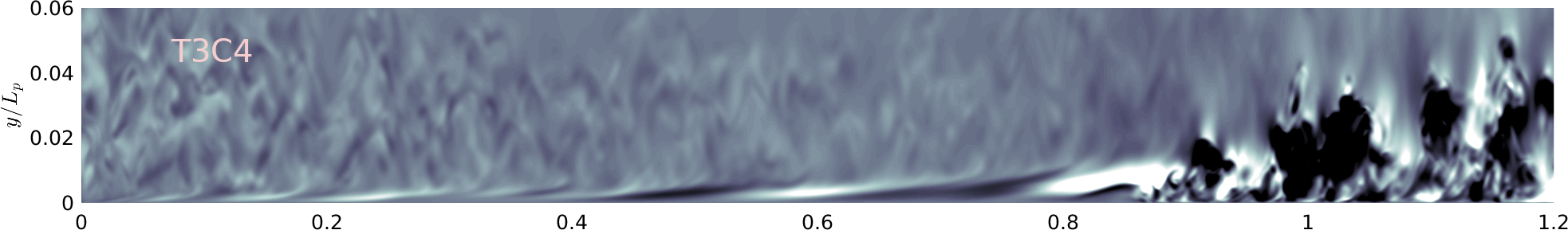}\\
\includegraphics[width=\linewidth]{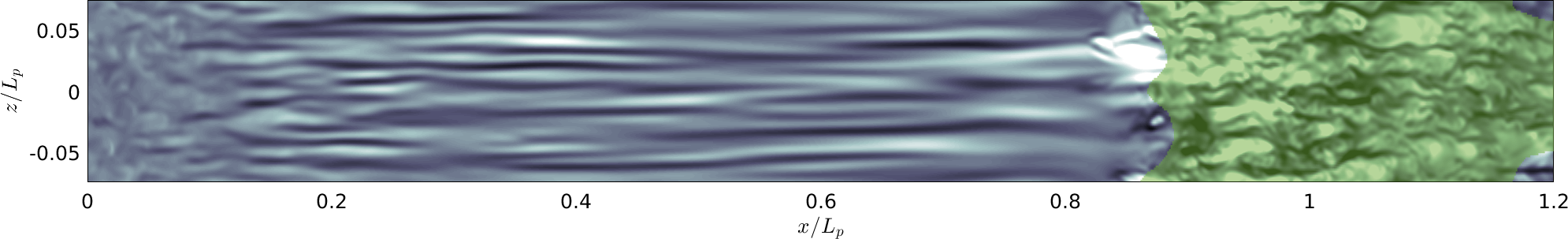}
\caption{Instantaneous side and top views of the streamwise velocity fluctuations for the T3C cases.
The turbulent region from binary segmentation is displayed in semi-transparent green.}
\label{fig:view_t3c}
\end{figure*}

\subsection{Laminar-turbulent discrimination} \label{ssec:lam_turb}

Most transition models use a transport equation for intermittency or laminar kinetic energy. It is therefore of interest to examine how a varying pressure gradient affects the intermittency distribution or the discriminated statistics, in which the laminar and turbulent contributions are artificially separated using an intermittency indicator taking values of 0 or 1. A detailed discussion is provided by \cite{Nolan13} for zero external pressure gradients and for constant favorable or adverse pressure gradients. In the present configurations, the pressure-gradient variation representative of an aft-loaded turbine blade leads to a non-equilibrium boundary layer.
Intermittency is a quantity that is frequently reported in experimental studies, as long time-series measurements are available, for example from hotwires. However, experimental intermittency measurements are not free from ambiguity, since their determination depends on the appropriate choice of a threshold. In simulations, the main difficulty is to store the instantaneous volumetric quantities over a sufficiently long time interval to obtain a converged estimate of this quantity. The laminar-turbulent discrimination algorithm used here is summarized and discussed in Appendix \ref{app:lam_turb}.
An instantaneous view of the different T3C configurations is shown in Fig. \ref{fig:view_t3c}. The decrease in Reynolds number from T3C5 to T3C2--C3--C4 can be appreciated from the size of the structures, in particular the Klebanoff streaks. The laminar-turbulent discrimination criterion is shown in transparent green for the region identified as fully turbulent by our algorithm. The earlier transition in T3C5 can thus be observed, as well as the formation of turbulent spots that merge into an irregular turbulent front for T3C2 and T3C3. The separation-induced transition in T3C4 is more abrupt, with a more regular front near the end of the plate.

\begin{figure}[t]
\centering
\includegraphics[width=0.9\linewidth]{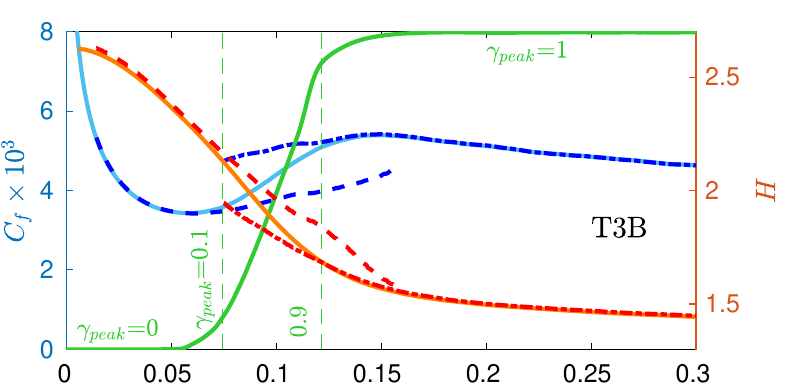}
\includegraphics[width=0.9\linewidth]{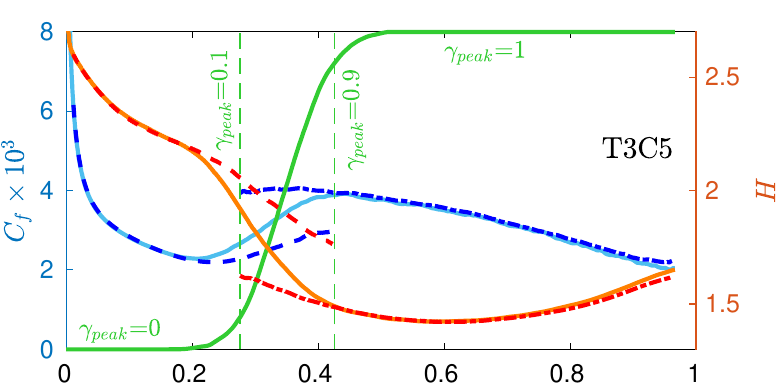}
\includegraphics[width=0.9\linewidth]{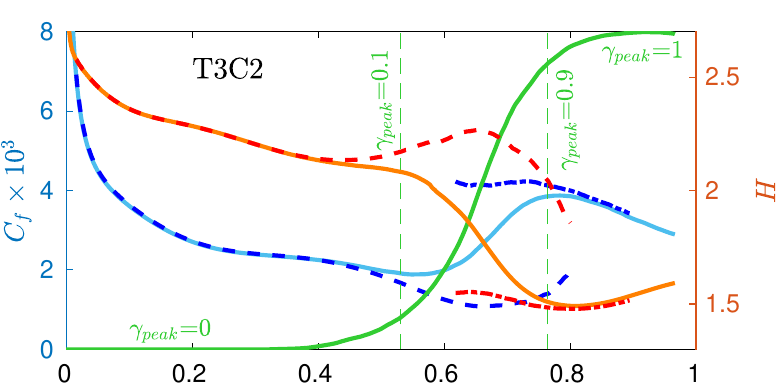}
\includegraphics[width=0.9\linewidth]{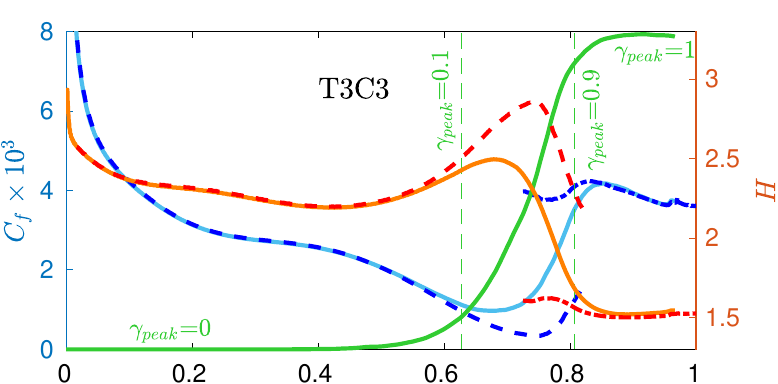}
\includegraphics[width=0.9\linewidth]{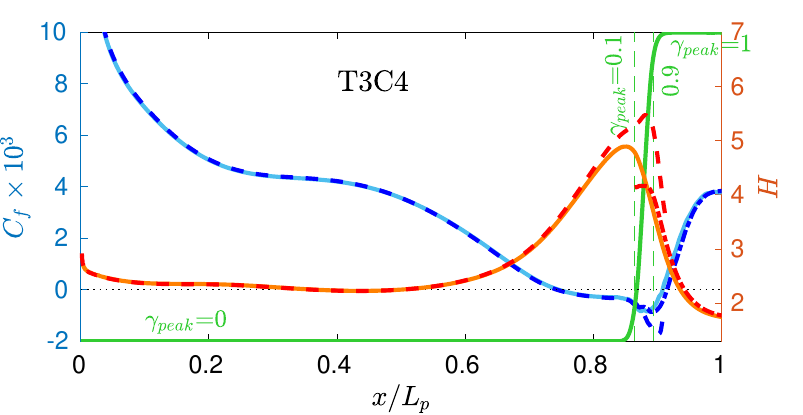}
\caption{Evolutions of intermittency (green), friction (blue) and shape factor (red) for T3C and T3C cases.
  For $C_f$ and $H$, solid lines denote the non-discriminated statistics, the dashed ones the laminar-conditioned
  and the dash-dotted the turbulent-conditioned statistics.}
\label{fig:intermittency}
\end{figure}

By averaging this laminar-turbulent indicator in time and along the spanwise direction $z$, we obtain the intermittency distribution $\gamma$ in the $x$--$y$ plane. We then consider the evolution of its maximum value, $\gamma_{peak}$, in each wall-normal section as a function of $x$. Figure \ref{fig:intermittency} shows the results for the T3C cases together with the T3B case, which provides a reference without a pressure gradient. The increase from 0 to 1 of the intermittency provides a useful marker of transition and also gives information on the streamwise extent of the transition region, which is identified by the vertical dashed lines corresponding to $\gamma_{peak}=0.1$ and $\gamma_{peak}=0.9$.
The binary segmentation can also be used to separate the laminar and turbulent contributions to the different statistical quantities by means of conditional averaging. The results for the two quantities characterizing the transition, the skin-friction coefficient and the shape factor, have been added to the subfigures of Fig. \ref{fig:intermittency}. At first glance, the laminar-conditioned statistics closely follow the overall statistics in the laminar region and extend continuously into the transition region. The turbulent-conditioned quantities emerge within the transition region before converging towards the overall statistics in the fully turbulent region. The evolution of the conditioned $C_f$ is approximately opposite to that of the shape factor.
For the ZPG T3B case, the evolution is consistent with previous studies \citep{Nolan13,Perrin23}. The transition region is longer for T3C5, owing to the lower $Tu$ (3\% instead of 6\% for T3B) and the presence of a favorable pressure gradient, but the merging of the laminar- and turbulent-conditioned contributions in the transition region is very similar. The T3C2 and T3C3 cases exhibit a somewhat different behaviour owing to the competition between the tendency of the laminar boundary layer to separate and the development of attached turbulence. In particular, for T3C2, a peak in the laminar-conditioned shape factor can be observed, which disappears when the turbulent-conditioned contribution is included. This peak in $H$ is a signature of the adverse pressure gradient acting on the laminar boundary layer and can emerge when $Tu$ is slightly reduced (see Appendix \ref{app:T3C_sens}). For T3C3, the peak in $H$ is also present in the overall average, but is more pronounced in the laminar-conditioned average. A similar distinction is observed in the $C_f$ curves, which, in contrast, exhibit a more pronounced minimum in the transition region, approaching zero, i.e., separation, for T3C3. Finally, for the separated T3C4 case, transition occurs very rapidly, and the transition between the laminar- and turbulent-conditioned curves takes place over a very short streamwise distance. The second negative minimum of $C_f$ is seen to be primarily a feature of the laminar separation, which also leaves a weaker signature in the overall $C_f$.

\begin{figure}
\centering
\includegraphics[width=0.9\linewidth]{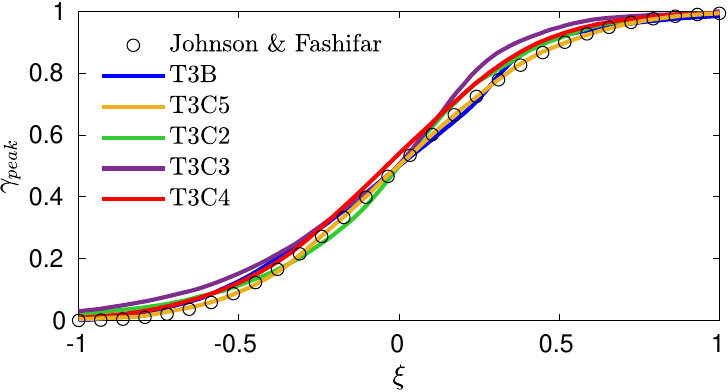}
\caption{Peak intermittency as a function of dimensionless streamwise coordinate $\xi$,
compared to the correlation of \cite{Johnson94}.}
\label{fig:intermittency_xi}
\end{figure}
  
Several authors have proposed semi-empirical correlations to describe the evolution of intermittency using a coordinate normalized by the transition length,
$\xi=(x-x_{tr})/\Delta x_{tr}$,
where $x_{tr}$ is the position at which $\gamma_{peak}=0.5$, and $\Delta x_{tr}$ is the distance between $x_{\gamma_{peak}=0.1}$ and $x_{\gamma_{peak}=0.9}$. For boundary layers in equilibrium, \cite{Nolan13} showed good similarity for different ZPG conditions and for constant FPG and APG with the correlation of \cite{Johnson94},
\begin{equation}
\gamma=1-\exp\left[-0.6(\xi+1.05)^3\right].
\label{gam_JF}
\end{equation}
A reasonable collapse is obtained (Fig. \ref{fig:intermittency_xi}) even for the abrupt transition associated with separation in T3C4.

\begin{figure*}
\centering
\includegraphics[width=0.75\linewidth]{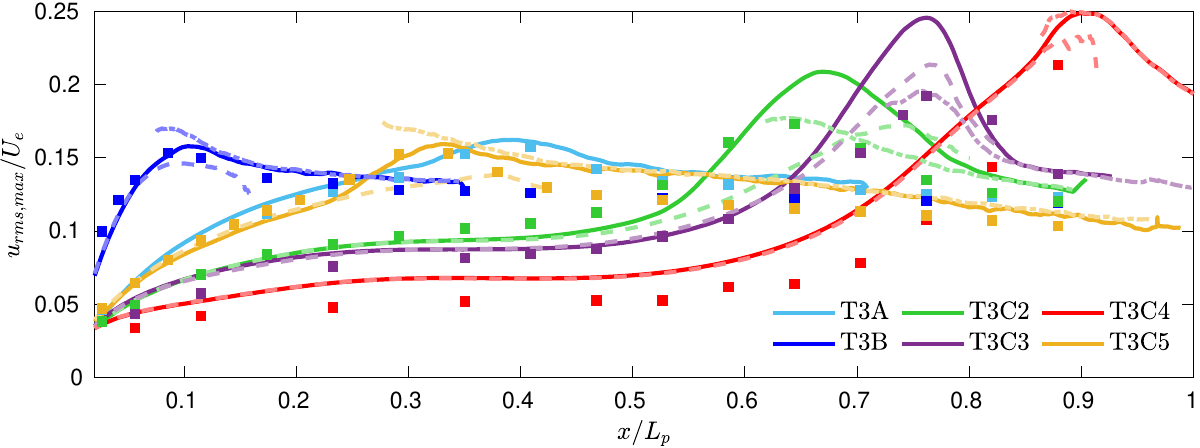}
\caption{Evolution of the maximum streamwise turbulence intensity (solid lines) compared to experiments (symbols).
Conditioned laminar (dashed) and turbulent (dash-dotted) data are superimposed.}
\label{fig:maxurms}
\end{figure*}

Finally, Fig. \ref{fig:maxurms} provides a comparison of all cases on a single graph. Here, we consider the evolution of the maximum value of $u_{rms}$, which is another indicator commonly used in the literature. If the transition process is considered to reach its maximum intensity when this peak is attained, the differences between the configurations become apparent. The different transition scenarios are well reproduced when compared with the experimental data, although some differences in magnitude can be observed, particularly in the peak values for T3C2--C4.
This representation clearly highlights two distinct transition scenarios. On the one hand, we have the ZPG cases T3A and T3B and the high-Reynolds-number T3C5, which undergoes transition relatively early while the pressure gradient is still favorable. Transition is more rapid for T3B, which has a $Tu$ twice as large as that of T3A and T3C5. The latter two cases exhibit very similar evolutions, particularly in terms of transition length. For these three cases, the maximum $u_{rms}$ is around 0.15, consistent with the results of \cite{Nolan13} for equilibrium boundary layers under ZPG, FPG, and APG conditions.
In contrast, the other three configurations, T3C2--C4, for which transition occurs under an adverse pressure gradient over the second half of the plate, exhibit an unusually higher peak, between 0.2 and 0.25. Interestingly, the superposition of the laminar- and turbulent-conditioned $u_{rms,max}$ shows that a significant part of this peak is associated with the laminar contribution. This contribution is related to the amplification of the velocity fluctuations as the boundary layer approaches separation. For example, in T3C3, the laminar contribution to the peak clearly exceeds the turbulent contribution. For the abrupt separation-induced transition in T3C4, the laminar contribution has already reached 0.23 when the turbulent contribution suddenly emerges, further increasing the peak to approximately 0.25.

\section{Analysis of Reynolds-stress budgets}\label{sec:budgets}

We now analyse the Reynolds-stress and turbulent-kinetic-energy (TKE) budgets, which can provide useful guidance for the development of turbulence models by elucidating the different mechanisms involved during laminar-to-turbulent transition, both under zero and non-zero external pressure gradients. The budgets of the individual Reynolds-stress components can, in particular, be used to assess nonlinear anisotropic Reynolds-stress-based models \citep{Lardeau04}. Comparison with data available in the literature for fully turbulent boundary layers subjected to zero or non-zero pressure gradients \citep{Voke95,Ovchinnikov08,Schlatter10,He13,Vinuesa17,Gungor22} can further help distinguish the effects associated with transition from those induced by the external pressure gradient.

\subsection{ZPG cases}

\begin{figure*}[t]
  \centering
  \includegraphics[width=0.9\textwidth]{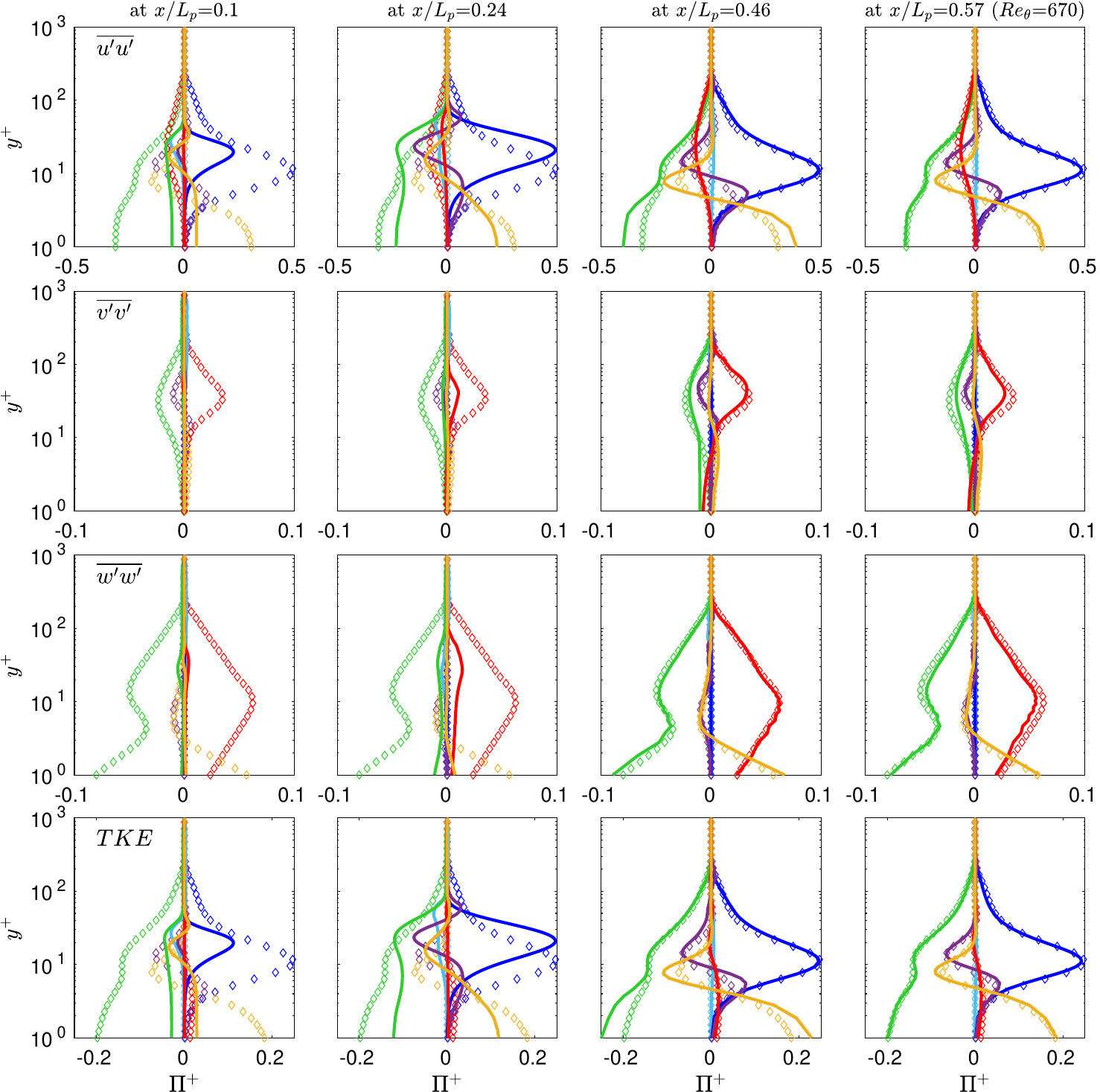}
  \caption{Reynolds-stress budgets normalized by inner scales for case T3A:
    production {\color{blue}\protect\traite};
    viscous dissipation {\color{forest_green}\protect\traite};
    turbulent transport {\color{dark_purple}\protect\traite},
    velocity-pressure gradient {\color{red}\protect\traite};
    viscous diffusion {\color{dark_yellow}\protect\traite};
    convection {\color{light_blue}\protect\traite}.
    Symbols with the same color are used for the corresponding terms in the
    DNS results of \cite{Schlatter10} at $Re_\theta=670$.}
  \label{bud_inner_T3A}  
\end{figure*}

We first consider the T3A and T3B cases, for which the external pressure gradient is zero. The different terms of the Reynolds-stress and turbulent-kinetic-energy budgets for the T3A case are presented in Fig.~\ref{bud_inner_T3A}.
The main terms of the budget (written using the repeated indices convention only for the j index) of the normal stresses $\frac{1}{2}\overline{\rho}\widetilde{u_i''u_i''}$ are
the production $\left(- \overline{\rho u_i''u_j''}\frac{\partial \widetilde{u}_i}{\partial x_j}\right)$,
the turbulent transport $\left(-\frac{1}{2}\frac{\partial}{\partial x_j}\overline{\rho u_i''^2u_j''}\right)$,
the pressure strain $\left(\overline{p'\frac{\partial u_i''}{\partial x_i}}\right)$,
the pressure transport $\left(-\frac{\partial}{\partial x_j} \overline{p'u_i'}\right)$,
the viscous diffusion $\left(\frac{\partial}{\partial x_j}\overline{u_i'\tau_{ij}}\right)$,
the viscous dissipation $\left(-\overline{\tau_{ij}\frac{\partial u_i'}{\partial x_j}}\right)$,
and the mean convection $\left(\frac{1}{2}\frac{\partial}{\partial x_j}\overline{\rho u''^2_i}\tilde{u}_j\right)$.
Here the compressible form is implemented with $\overline{(\cdot)}$ the Reynolds average and $\widetilde{(\cdot)}$ the Favre average;
prime quantities are Reynolds fluctuations and double prime Favre fluctuations.
Note that Reynolds and Favre averaging provide essentially the same results at the present Mach numbers (the
budgets divided by $\frac{1}{2}\overline{\rho}$ are simply noted budgets of $\overline{u'u'}$, $\overline{v'v'}$, $\overline{w'w'}$).
These budgets are normalized in Fig.~\ref{bud_inner_T3A} using inner variables, with the scaling $u_\tau^4/\nu$. Results are shown at four streamwise locations: in the pre-transitional region, at the location of the minimum $C_f$, shortly upstream of the maximum $C_f$, and in the fully turbulent region ($Re_\theta=670$). The incompressible DNS results of \cite{Schlatter10} at $Re_\theta=670$ are superimposed on each plot for $\overline{u'u'}$, $\overline{v'v'}$, $\overline{w'w'}$, and $k$, and are used as a fully turbulent reference. The pressure-strain and pressure-diffusion terms are combined into a single pressure-velocity interaction term in order to facilitate the comparison.

At the fully turbulent station, the first observation is the very good agreement between the DNS and the present T3A LES for all quantities. This provides an additional indication that the spatial and temporal resolution of the present LES is adequate. As expected, the turbulent-kinetic-energy budget exhibits features similar to those of the streamwise Reynolds-stress budget, $\overline{u'u'}$, which contains the dominant contribution from mean-shear production. In the near-wall region, this production is balanced primarily by viscous dissipation, while viscous diffusion becomes important close to the wall. Turbulent transport is also significant and contributes to the redistribution of turbulent kinetic energy within the boundary layer. For the wall-normal and spanwise components, $\overline{v'v'}$ and $\overline{w'w'}$, the dominant positive contribution is the pressure-velocity interaction, which is primarily balanced by viscous dissipation. For $\overline{w'w'}$, this pressure-velocity contribution combines with viscous diffusion near the wall to balance the dissipation. Thus, in the fully turbulent boundary layer, turbulent kinetic energy is produced predominantly through the streamwise component by the mean shear and subsequently redistributed among the three velocity components through the pressure-strain interaction.
Keeping the same abscissa limits for each row, at the first station, all budget terms are substantially smaller. In this pre-transitional region, production in the $\overline{u'u'}$ budget is the dominant contribution and is primarily balanced by dissipation. As also observed by \cite{He13} for temporally developing transition in a plane channel, the pressure-velocity interaction is almost negligible at this stage. Consequently, pressure-strain redistribution is not yet effective in transferring the energy associated with the streamwise fluctuations towards the wall-normal and spanwise components. The streamwise fluctuations produced in the boundary layer are consistent with the development of the Klebanoff streaks induced by the free-stream turbulence.
An interesting feature is the significant negative convection term in the $\overline{u'u'}$ budget near the edge of the boundary layer. Such a contribution is not reported in the analyses of \cite{He13} or \cite{Ovchinnikov08}. With the present sign convention, the convection term for $\overline{u'u'}$ is given by $-\overline{u}\frac{\partial\overline{u'u'}}{\partial x}-\overline{v}\frac{\partial\overline{u'u'}}{\partial y}$.
The first contribution dominates in the region considered, indicating a strong streamwise growth of the streamwise Reynolds stress. Thus, even before transition, the boundary layer exhibits a pronounced downstream development of the streamwise turbulent fluctuations resulting from its interaction with the free-stream turbulence. In contrast, the response of the other Reynolds-stress components remains weak at this stage. This strong anisotropy is consistent with the shear-sheltering effect, whereby the mean shear of the boundary layer filters the external turbulent fluctuations and limits their penetration into the near-wall region. The budgets therefore suggest that the pre-transitional response is not simply the passive transmission of the free-stream turbulence into the boundary layer, but rather an anisotropic response progressively modified by the boundary-layer shear.

\begin{figure*}
  \centering
  \includegraphics[width=0.9\textwidth]{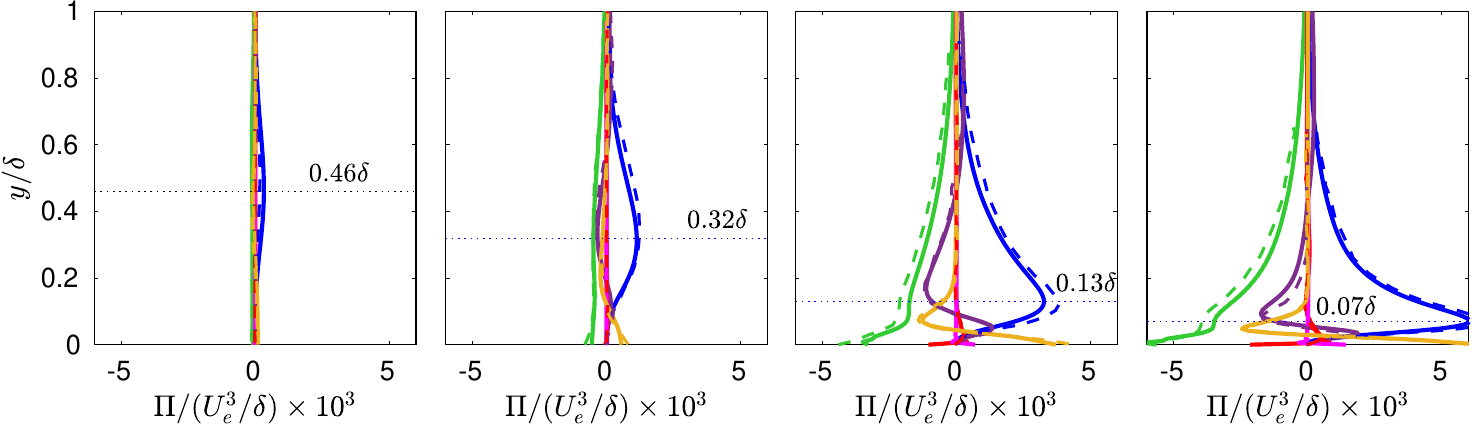}
  \caption{Comparison of turbulent-kinetic-energy budget of case T3B (thick solid lines)
    with the DNS of \cite{Ovchinnikov08} (dashed lines) at four
    streamwise locations where $Re_x$=$12\,000$, $50\,000$, $91\,500$ and $123\,250$.
    Same line colors as Fig. \ref{bud_inner_T3A}.}
  \label{bud_outer_T3B_OCP}  
\end{figure*}

At the location of the minimum $C_f$, which is considered here as the onset of transition, the peak production of $\overline{u'u'}$ has already reached a level comparable to that observed in the fully turbulent boundary layer, although it is located farther from the wall. In the $\overline{u'u'}$ budget, a second dissipation peak develops within the viscous sublayer and is balanced primarily by viscous diffusion. The terms in the $\overline{v'v'}$ budget remain weak, with the pressure-velocity redistribution term being the first contribution to become significant. This term reaches a similar magnitude in the $\overline{w'w'}$ budget, where it also begins to extend towards the wall and is balanced by dissipation.
As the flow approaches the maximum $C_f$ (third column), the budget terms reach their maximum amplitudes. The distributions of the different contributions become remarkably similar to those observed in the fully turbulent regime. The main difference is the enhanced dissipation near the wall, which is accompanied by a corresponding increase in viscous diffusion, thereby preserving the local budget balance.

The results obtained for T3B normal stresses are qualitatively very similar and are therefore not reproduced here for conciseness. Figure~\ref{bud_outer_T3B_OCP} presents the turbulent-kinetic-energy budgets at four streamwise locations, compared with the corresponding results from the DNS of the T3B case by \cite{Ovchinnikov08} (given in Fig.~24 of their supplementary material). The budgets are normalized using external scales, $U_e(x)$ and $\delta(x)$. When normalized using $u_\tau(x)$, which varies strongly as $C_f$ through the transition region, the relative evolution of the absolute budget levels is difficult to assess. The external normalization therefore provides a more appropriate representation of the relative magnitude of the different terms throughout transition.
In good agreement with the results of Ovchinnikov \textit{et al.}, all terms increase rapidly through the transition region, with the notable exception of the pressure-strain contribution. Their crosswise distributions, however, evolve differently among the four stations. In particular, the production peak is initially located around $\delta/2$ and progressively moves towards the wall, eventually reaching a position corresponding to approximately $y^+$=12 in the fully turbulent regime, as indicated by the dotted lines in the figure. Interestingly, transition models based on a transport equation for intermittency or for the laminar kinetic energy also aim to reproduce a production peak for these quantities at approximately $\delta/2$ during the pre-transitional stages \citep{Menter15}.

\begin{figure}
  \centering
  \includegraphics[width=0.8\linewidth]{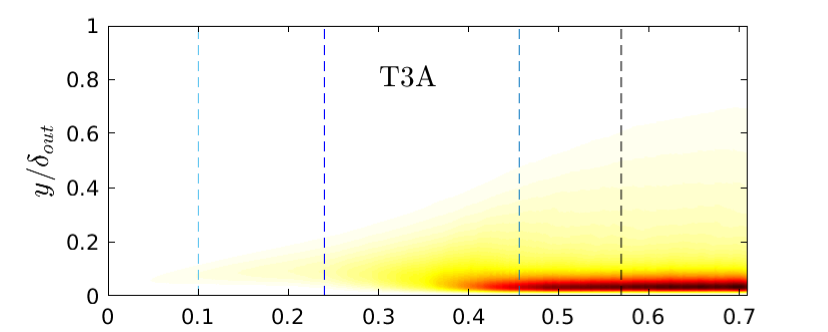}
  \includegraphics[width=0.8\linewidth]{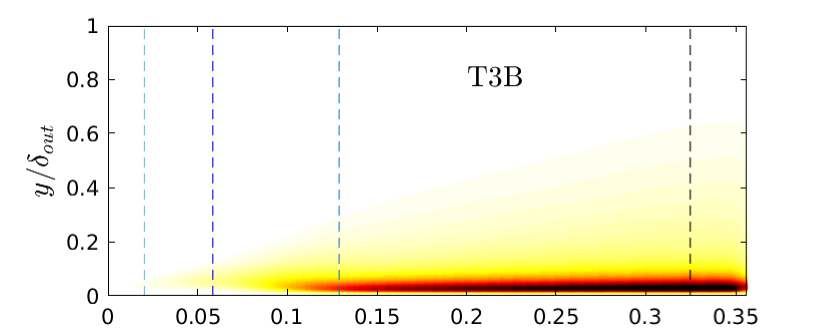}
  \includegraphics[width=0.8\linewidth]{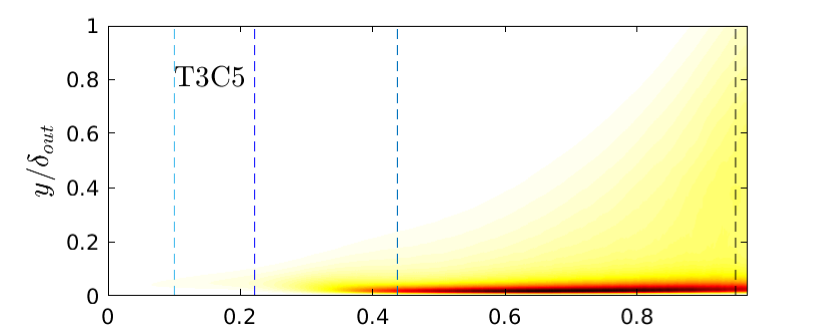}
  \includegraphics[width=0.8\linewidth]{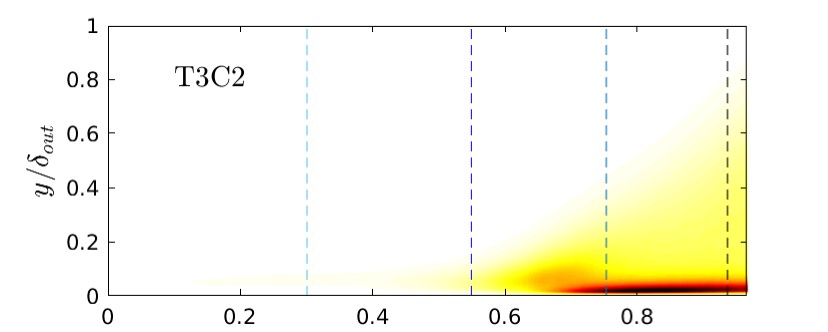}
  \includegraphics[width=0.8\linewidth]{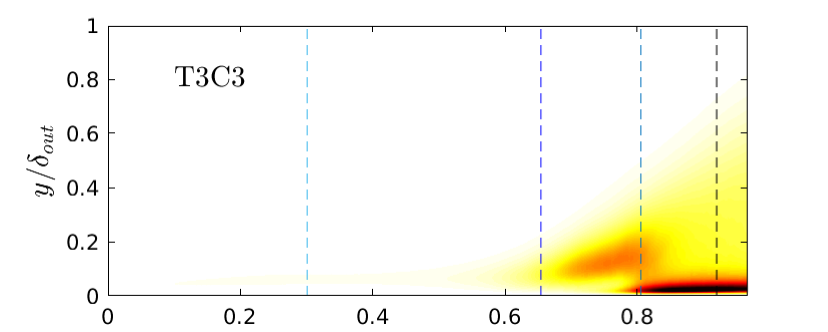}
  \includegraphics[width=0.8\linewidth]{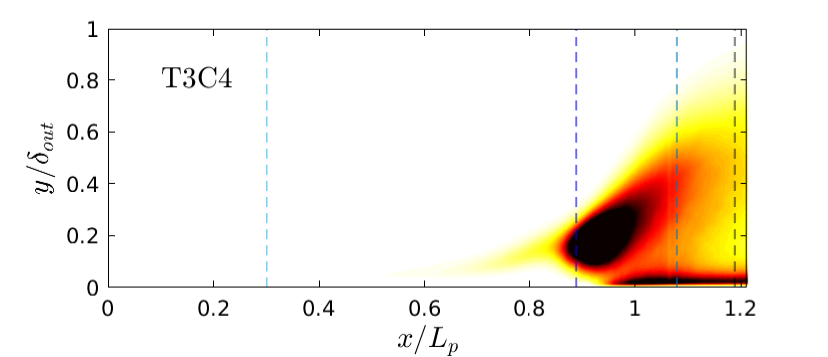}
  \caption{Maps of the production term in outer scaling for T3 cases (white-red-black color levels
    between 0 and 0.01). The vertical dashed lines denote the locations where the budget profiles are extracted, respectively in the
    pretransition region, at location of $C_{f,min}$, 50 points before $C_{f,max}$ is reached, and in the turbulent region.
    The $y$-axis is normalized by the boundary layer thickness at final location $\delta_{out}$.}
  \label{prod_T3}
\end{figure}
  

\subsection{Pressure gradient cases}

We now turn to the T3C cases, which are characterized by a streamwise-varying external pressure gradient. As pointed out, for example by \cite{Gungor22}, the friction velocity $u_\tau$ is not an appropriate scaling in the presence of flow separation, since its value approaches zero near the separation point and normalization using inner variables can consequently lead to unrealistically large budget levels.

\begin{figure*}
  \centering
  \includegraphics[width=0.9\textwidth]{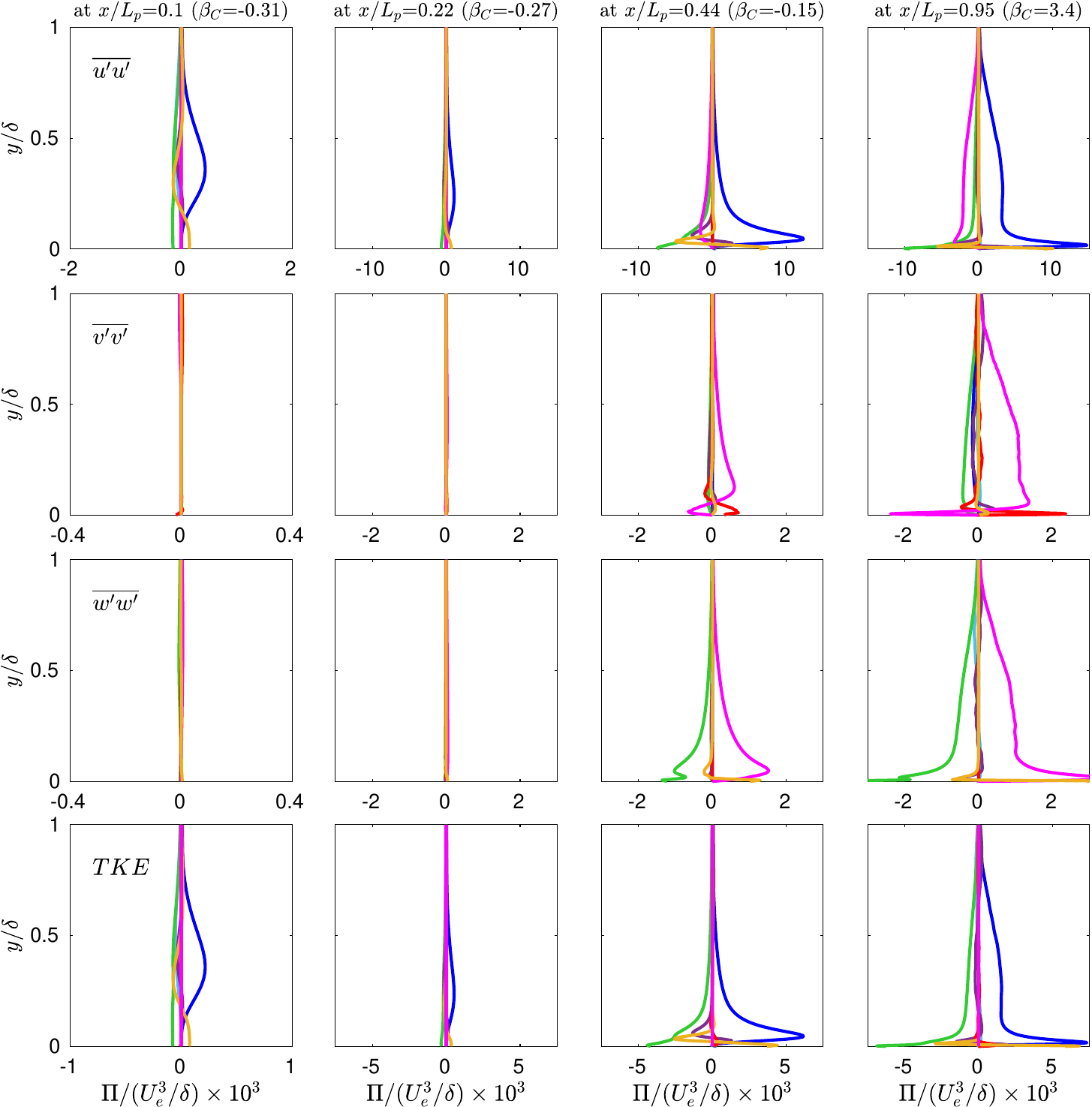}
  \caption{Reynolds-stress budgets in outer scaling for T3C5: production {\color{blue}\protect\traite}
viscous dissipation {\color{forest_green}\protect\traite}; turbulent transport {\color{dark_purple}\protect\traite};
pressure strain {\color{red}\protect\traite}; pressure transport {\color{magenta}\protect\traite};
viscous diffusion {\color{dark_yellow}\protect\traite}; convection {\color{light_blue}\protect\traite}.}
  \label{bud_outer_T3C5}  
\end{figure*}

\begin{figure*}
  \centering
  \includegraphics[width=0.9\textwidth]{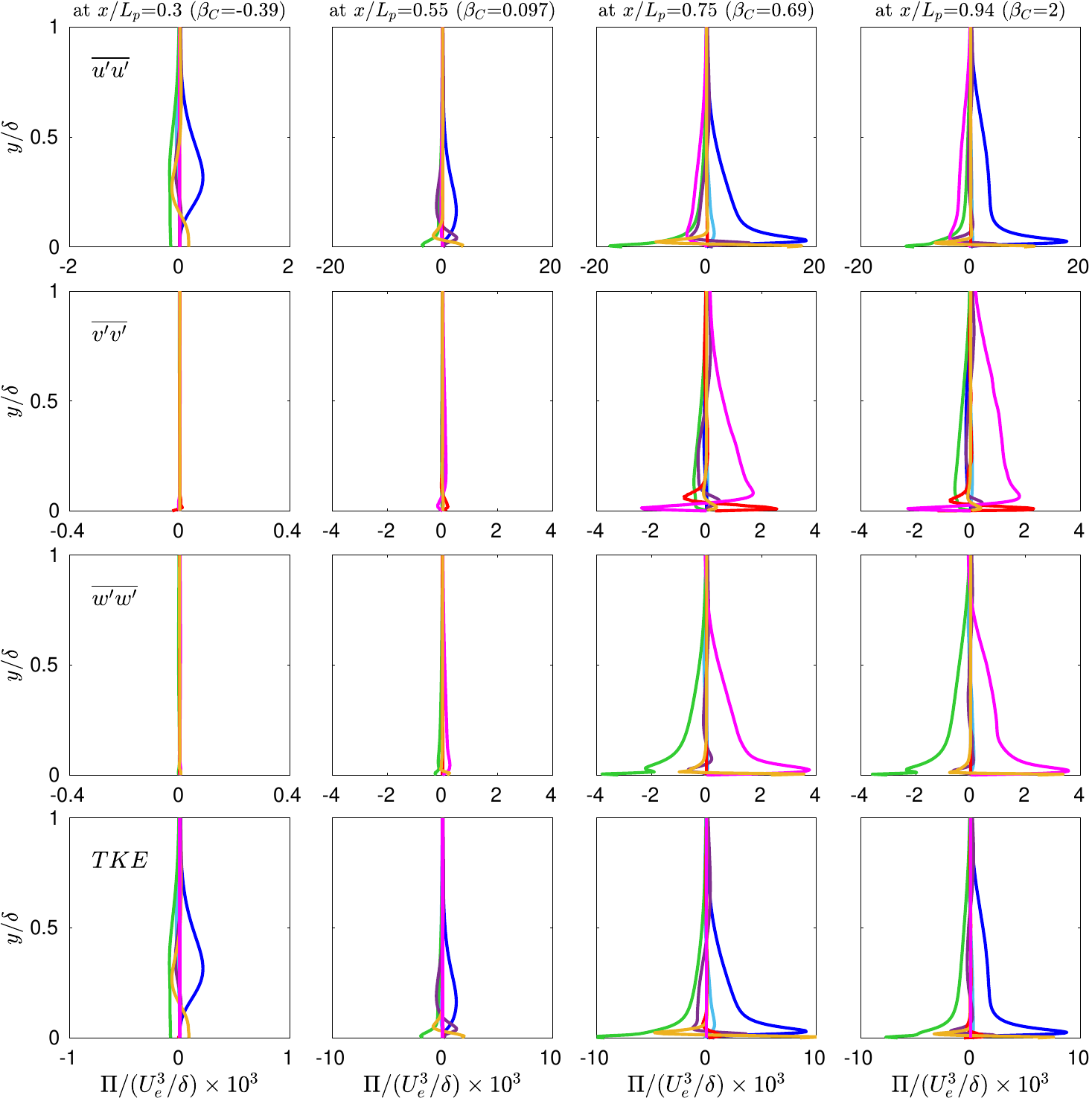}
  \caption{Reynolds-stress budgets in outer scaling for T3C2: production {\color{blue}\protect\traite}
viscous dissipation {\color{forest_green}\protect\traite}; turbulent transport {\color{dark_purple}\protect\traite};
pressure strain {\color{red}\protect\traite}; pressure transport {\color{magenta}\protect\traite};
viscous diffusion {\color{dark_yellow}\protect\traite}; convection {\color{light_blue}\protect\traite}.}
  \label{bud_outer_T3C2}  
\end{figure*}

\begin{figure*}
  \centering
  \includegraphics[width=0.9\textwidth]{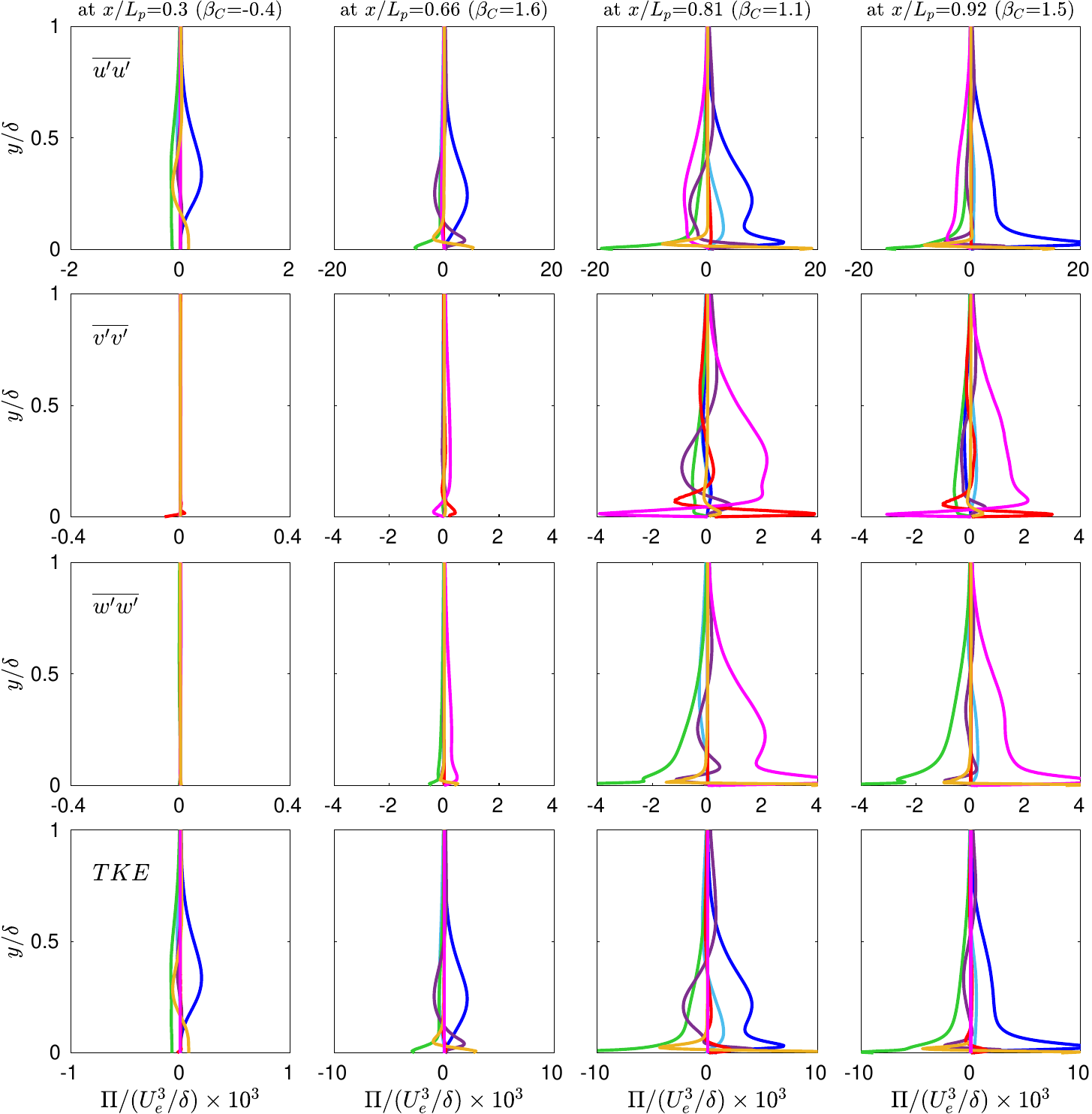}
  \caption{Reynolds-stress budgets in outer scaling for T3C3: production {\color{blue}\protect\traite}
viscous dissipation {\color{forest_green}\protect\traite}; turbulent transport {\color{dark_purple}\protect\traite};
pressure strain {\color{red}\protect\traite}; pressure transport {\color{magenta}\protect\traite};
viscous diffusion {\color{dark_yellow}\protect\traite}; convection {\color{light_blue}\protect\traite}.}
  \label{bud_outer_T3C3}  
\end{figure*}

\begin{figure*}
  \centering
  \includegraphics[width=0.9\textwidth]{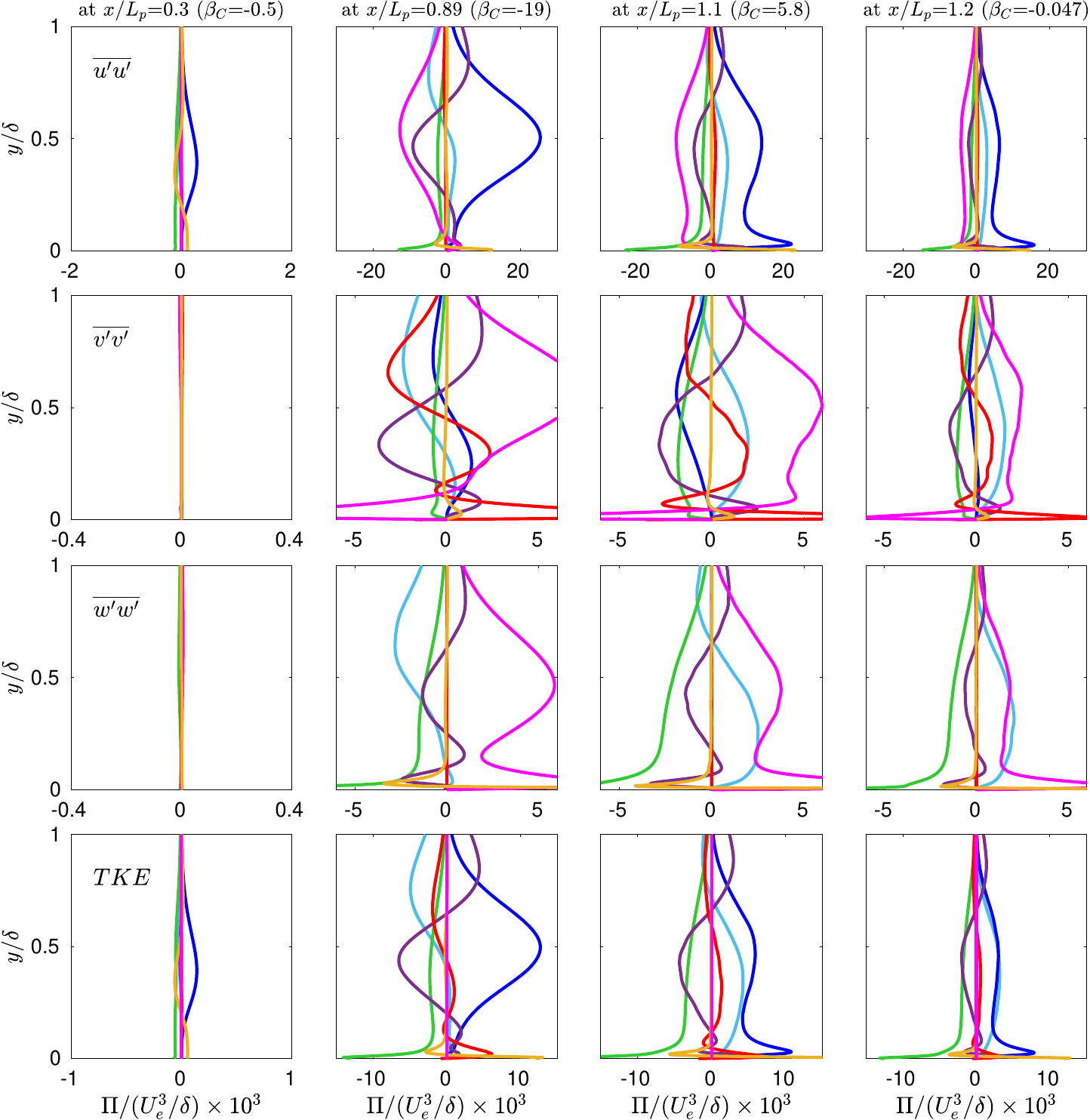}
  \caption{Reynolds-stress budgets in outer scaling for T3C4: production {\color{blue}\protect\traite}
viscous dissipation {\color{forest_green}\protect\traite}; turbulent transport {\color{dark_purple}\protect\traite};
pressure strain {\color{red}\protect\traite}; pressure transport {\color{magenta}\protect\traite};
viscous diffusion {\color{dark_yellow}\protect\traite}; convection {\color{light_blue}\protect\traite}.}
  \label{bud_outer_T3C4}  
\end{figure*}

Maps of the production term are reported in Fig.~\ref{prod_T3} for all cases. These production distributions provide a useful marker of the transition onset. In particular, the production peak is typically located at a higher wall-normal position in the transitional boundary layer, around $\delta/2$.
For T3C3 and T3C4, a secondary outer peak is clearly visible shortly before the emergence of the near-wall turbulent production within the inner layer. Its intensity appears to be related to the rapidity of the transition, which, in the present cases, is primarily associated with the increasing influence of the adverse pressure gradient. The T3C2 case again lies at the boundary between the two behaviours: the classical bypass-transition scenario observed for the ZPG cases and T3C5, and the increasingly dominant tendency of the laminar boundary layer to separate in the other cases, with separation ultimately driving the transition in T3C4.
The emergence of this secondary outer peak can already be observed for T3C2, suggesting that the transition from the conventional bypass mechanism to a separation-induced mechanism can be viewed as a continuous process. As the Reynolds number decreases, the adverse pressure gradient increasingly promotes the amplification of the laminar boundary layer and its tendency to separate, progressively modifying the transition mechanism from a conventional bypass process towards a separation-induced transition.
The profiles of the Reynolds-stress and turbulent-kinetic-energy budgets at four representative streamwise positions, identified by the dashed lines in Fig.~\ref{prod_T3}, are shown in Figs.~\ref{bud_outer_T3C5} to \ref{bud_outer_T3C4}, using the external scales $U_e(x)$ and $\delta(x)$.
For each station, the corresponding Clauser pressure-gradient parameter $\beta_C$ is indicated.
For the highest Reynolds-number case T3C5 (Fig.~\ref{bud_outer_T3C5}), the gradual bypass transition takes place primarily under the influence of a moderate favourable pressure gradient, and the budget balances remain qualitatively similar to those observed in the zero-pressure-gradient cases. At the first three stations, the pressure gradient is favourable. \cite{Vinuesa17} showed that, for mild favourable pressure gradients on the suction side of an airfoil, the turbulent budgets are only weakly affected. In the fully turbulent regime (last station), the dominant terms are similar to those observed in the T3A and T3B cases, but the production profile exhibits a secondary outer plateau associated with the strong adverse pressure gradient encountered downstream ($\beta_C=3.4$). This corresponds to the outer peak in $u_{rms}$ profiles (Fig. \ref{fig:APG}).

Similar observations can also be made for T3C2 in Fig.~\ref{bud_outer_T3C2}. At the first, pre-transitional station, the pressure gradient is favourable and the budgets are consistent with those observed in the previous cases. At the location of the minimum $C_f$, the Clauser parameter is approximately 1.6. The production peak is located closer to the wall, while the other budget terms also become significant in the near-wall region compared with the corresponding stations in the ZPG and T3C5 cases. These differences in the $\overline{u'u'}$ budget appear to be primarily associated with the proximity and rapidity of transition rather than with a direct effect of the adverse pressure gradient itself.
For the pressure-gradient cases, the pressure-strain and pressure-transport contributions are plotted separately. At the two downstream stations of T3C2, the pressure-strain contribution clearly dominates the pressure-related terms in the $\overline{v'v'}$ and $\overline{w'w'}$ budgets, whereas pressure transport remains a moderate sink, confined mainly to the near-wall region. This behaviour is consistent with the observations of \cite{Lardeau04,He13,Bhushan18} and does not support a dominant role of pressure transport in the growth of perturbations within the boundary layer, as proposed in \cite{Mayle97}. It is also interesting to note the non-negligible positive convection term in both the $\overline{u'u'}$ and TKE budgets, indicating a strong departure from local equilibrium in the transitional boundary layer. This contribution is particularly relevant in the present cases because the boundary layer is simultaneously subjected to the downstream development of the free-stream-turbulence-induced response and to a rapidly varying external pressure gradient.

For T3C3 (Fig.~\ref{bud_outer_T3C3}), transition occurs further downstream under an adverse pressure gradient, and the pre-transitional boundary layer is already close to its separation limit. As a consequence, the outer production peak identified in the production map of Fig.~\ref{prod_T3} is already developing at the second station and significantly modifies the budgets at the third one. At this location, a rapid redistribution of the turbulent energy produced in the outer region among the velocity components is particularly apparent through the pressure-strain term, which is the dominant contribution in both the $\overline{v'v'}$ and $\overline{w'w'}$ budgets. Turbulent transport and pressure transport in the near-wall region become more important than viscous dissipation in balancing the pressure-strain contribution in the $\overline{v'v'}$ budget.
A similar, but even more pronounced, situation is observed at the third station of T3C4 (Fig.~\ref{bud_outer_T3C4}). This case is separated, and the second station is located within the separation bubble, consistent with the value $\beta_C=-19$. All budget terms exhibit strong variations in the outer region of the boundary layer, and the outer production peak in the $\overline{u'u'}$ budget can exceed the near-wall peak. This pronounced non-equilibrium is reflected in the large convection, turbulent-transport, and pressure-transport contributions. The redistribution of turbulent kinetic energy from the streamwise component towards the wall-normal and spanwise fluctuations is also anisotropic, with the pressure-strain contribution being larger in the $\overline{v'v'}$ budget than in the $\overline{w'w'}$ budget. The delayed redistribution towards $v'$ compared with $w'$ may be attributed to the stronger constraint imposed by the wall on wall-normal fluctuations through the fluctuating pressure field, whereas spanwise fluctuations are constrained primarily by viscous effects and can therefore extend further into the viscous scales \citep{Voke95}. For the two latter cases, the fully turbulent stations (last column) also exhibit an outer production peak, or more generally a plateau-like distribution in the outer part of the boundary layer, which is characteristic of turbulent boundary layers subjected to adverse pressure gradients \citep{Vinuesa17,Bobke17}.

Overall, the Reynolds-stress budgets reveal a progressive departure from the local-equilibrium picture as the boundary layer approaches separation. The increasing importance of convection and turbulent transport, together with the rapid pressure-strain redistribution of the turbulent energy produced in the streamwise component, is particularly pronounced near separation and during the most rapid stages of transition. These observations suggest that transition models based solely on local-equilibrium assumptions, or on a purely local response to the instantaneous pressure gradient, may struggle in reproducing the strong non-equilibrium effects observed in the T3C cases. In particular, an adequate model should account for the coupled effects of free-stream turbulence, streamwise flow history, pressure-gradient-induced non-equilibrium, and the redistribution of turbulence among the Reynolds-stress components, which cannot be captured by standard linear eddy viscosity models.

\section{Conclusions} \label{sec:conclusion}

A high-fidelity numerical database of free-stream-turbu\-len\-ce-induced transition has been generated using wall-resolved iLES of the ERCOFTAC T3 flat-plate experiments, under zero (T3A, T3B) and varying (T3C2--C5) pressure gradients. All cases are computed within a unified framework, in which an upstream slip-wall block lets the free-stream turbulence develop and interact naturally with the leading-edge region. The database provides mean and fluctuating fields, intermittency, conditioned statistics, and Reynolds-stress budgets.

The agreement with the experiments is overall satisfactory. The decay of the streamwise free-stream turbulence, the skin friction, the integral boundary-layer quantities, and the streamwise fluctuation profiles are well reproduced for the ZPG cases and for T3C5, T3C2, and T3C3, including the shape-factor peak of T3C3. The main discrepancies concern T3C4, for which the simulations predict a separation bubble that is not reported experimentally, and the transverse fluctuations, which are systematically higher than measured. The re-analysis of the experimental database also identifies several limitations of these otherwise remarkable experiments: the reported turbulence intensity corresponds to the streamwise component only, which matters for the T3C cases where the free-stream turbulence becomes anisotropic; X-wire measurements underestimate the fluctuation levels; the laminar skin friction is underestimated; and the integral length scale, the contoured-wall geometry, and the leading-edge flow are insufficiently documented. Combined with the strong sensitivity of the T3C cases to the inflow turbulence, these uncertainties should be kept in mind when the T3 data are used for model calibration.

The simulations show that the varying pressure gradient progressively modifies the transition mechanism. Since the T3C cases share the same nominal turbulence intensity and pressure distribution, the Reynolds number controls where transition occurs relative to the pressure gradient. T3C5, like the ZPG cases, follows the classical bypass scenario under a favourable pressure gradient. As the Reynolds number decreases, transition moves into the adverse-pressure-gradient region and is increasingly governed by the tendency of the laminar boundary layer to separate. This shows up in a peak of the shape factor, unusually high streamwise fluctuations largely carried by the laminar contribution, and an outer production peak preceding near-wall turbulence production. T3C2 lies at the boundary between the two behaviours, T3C3 is close to separation, and T3C4 undergoes separation-induced transition. The change of scenario is therefore continuous rather than abrupt, with both mechanisms competing within the same flow. The intermittency nevertheless follows a similar distribution in all cases when expressed in terms of the transition length, whereas the budgets reveal strong anisotropy before transition and a growing departure from local equilibrium near separation.

These results highlight several challenges for RANS transition modelling. Correlation-based closures rely on a free-stream turbulence intensity that is not uniquely defined under a varying pressure gradient, and on local pressure-gradient parameters that cannot capture the strong history and non-equilibrium effects observed here. Intermittency or laminar-kinetic-energy transport models partly account for these effects, but the continuous evolution from bypass to separation-induced transition requires both mechanisms, and their competition, to be represented within a single model rather than through threshold-based corrections. Finally, experimental biases may have been partly absorbed into existing model calibrations, a risk that is even greater for data-driven approaches. The present internally consistent, full-field database offers a suitable basis for deriving and assessing interpretable model corrections, and future work will extend it to a wider range of turbulence and pressure-gradient conditions.

\section*{Acknowledgments}
This work has been granted access to the HPC resources of CINES and TGCC under the allocation A0182A13457
and A0192A01736 made by GENCI (Grand Equipement National de Calcul Intensif).
PC acknowledges support by the French National Research Agency (ANR) under the France 2030
program, within the framework of the IA Cluster initiative (grant ANR-23-IACL-0007).

\section*{Data availability statement}
The data that support the findings of this study are openly available in Zenodo at
\href{https://zenodo.org/records/17166216}{zenodo.org/records/17166216}, reference number 17166216.
Time resolved data can be made available upon request, subject to storage and transfer constraints.

\appendix

\section{Analysis of experimental database} \label{app:expe}

\begin{figure}
\centering
\includegraphics[width=\linewidth]{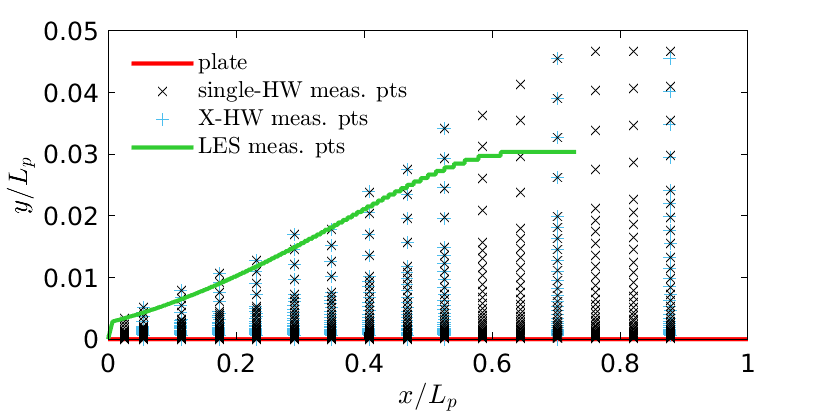}\
\hspace{-0.4cm}\includegraphics[width=0.9\linewidth]{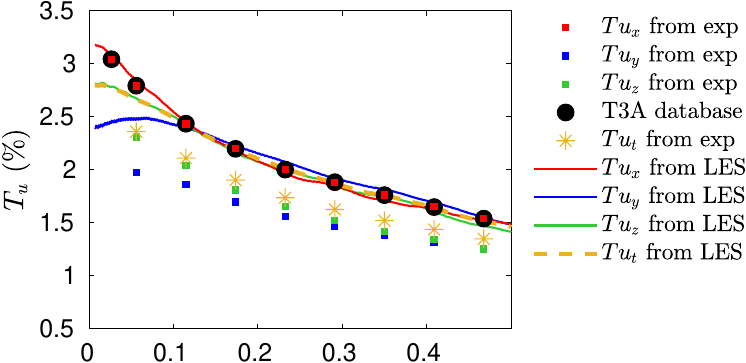}\
\includegraphics[width=0.85\linewidth]{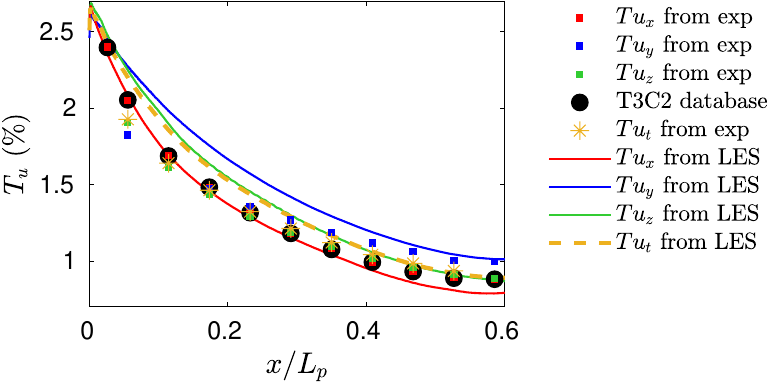}
\caption{Determination of FST intensity in T3 experiments.}
\label{fig:Tu}
\end{figure}

The experimental database available online from the ERCOFTAC (European Research Community on Flow, Turbulence and Combustion) committee is very comprehensive and has already been extensively post-processed. Nevertheless, for comparison with scale-resolving simulations, it is useful to re-analyse some quantities in order to reproduce these configurations as accurately as possible and provide a numerical version of the database for turbulence modelling purposes. This re-analysis also helps to explain some of the discrepancies between simulations and experiments that have already been reported in previous studies. In light of this additional post-processing, it is possible to distinguish between uncertainties related to the imperfect reproduction of the experimental setup and those arising from measurement errors. Three aspects are considered in this appendix: the determination of the freestream turbulence intensity $Tu$, the bias affecting the measurements of the transverse velocity components obtained with X-hotwires, and the direct or indirect determination of the skin-friction coefficient, which remains a challenging quantity to measure experimentally. In particular, the first point clarifies the difference between the initial simulations \citep{Zemmour25}, which constitute version v1 of our database available on Zenodo, and the updated simulations presented in this article, which constitute version v2 of the Zenodo database.

We first attempt to recalculate $Tu$ from the hotwire (HW) data available online. The top panel of Fig. \ref{fig:Tu} shows the location of the hotwire measurements for the T3A case. The experimental measurement points are concentrated close to the wall, particularly for the single-hotwire measurements, and some traverses are not available for the X-hotwire measurements. To recalculate the freestream turbulence intensity, we consider the measurement point furthest from the wall and calculate $Tu$ for each component as
\begin{equation}
Tu_x=\frac{u_{rms}}{U_e},\qquad
Tu_y=\frac{v_{rms}}{U_e},\qquad
Tu_z=\frac{w_{rms}}{U_e},
\end{equation}
and the overall turbulence intensity as
\begin{equation}
Tu_t=\frac{(u_{rms}+v_{rms}+w_{rms})/3}{U_e}.
\end{equation}

The resulting values are reported in Fig. \ref{fig:Tu} for the T3A and T3C2 cases, using coloured symbols, while the large black circles represent the data reported in the ERCOFTAC T3 database. For all configurations, we observe that the values reported in the database actually correspond to $Tu_x$ at the last available measurement point. This can be explained by the greater accuracy of the single-wire measurements and by the absence of X-wire traverses at some streamwise locations. Assuming perfect isotropy, $Tu_x=Tu_t$. In this case, for the ZPG T3A and T3B configurations, the measurement of $Tu_x$ is sufficient to specify the FST level. However, we observe, on the one hand, a strong anisotropy in the measurements downstream of $x$=0.1$L_p$ compared with the LES, most likely due to the uncertainty of the X-wire measurements, and, on the other hand, anisotropy between $x$=0 and $0.1L_p$ in both the simulations and experiments due to the presence of the leading edge. The experimental measurements do not exhibit exactly the same LE anisotropy, again most likely because of the uncertainty in the $v$ and $w$ measurements (see also Figs.\ref{fig:Tu_t3a} and \ref{fig:Tu_t3b}).

For the T3C configurations, even if the grid-generated turbulence is assumed to be initially isotropic, the turbulence above the plate is not. This is analogous to the flow through a turbine passage, since the contoured upper wall is intended to reproduce the conditions experienced by an aft-loaded turbine blade. This anisotropy is visible in the LES results of Fig. \ref{fig:Tu_t3c}, with a systematic decrease in the $Tu_x$ component and an increase in $Tu_y$. The spanwise component $Tu_z$, in between, provides a reasonable estimate of the overall turbulence level. This behaviour is observed for all cases in the transition region, with some variability near the leading edge and at the end of the plate in the fully turbulent regime due to the proximity of the measurement location to the boundary layer. The anisotropy is also fairly apparent in the experimental data for T3C2, shown in the bottom panel of Fig. \ref{fig:Tu}, at least near the centre of the channel ($0.3\!<\!x/L_p\!<\!0.6$), although it is less pronounced. Comparing the LES $Tu_t$ (yellow dashed line) shows that it underestimates the turbulence levels that would be present in the experiment.

In summary, the fact that $Tu_{exp}=Tu_x$ implies that, for the cases with a varying pressure gradient, the target turbulence level for the simulations should be specified in terms of the longitudinal component $Tu_x$ in order to avoid underestimating the freestream turbulence present above the transition region. Furthermore, the evolutions of $Tu_y$ and $Tu_z$ indicate that, in all cases, the transverse fluctuation levels are substantially underestimated by the X-hotwires, although the discrepancy is somewhat smaller for T3C2.

\begin{figure}
\centering
\includegraphics[width=0.32\linewidth]{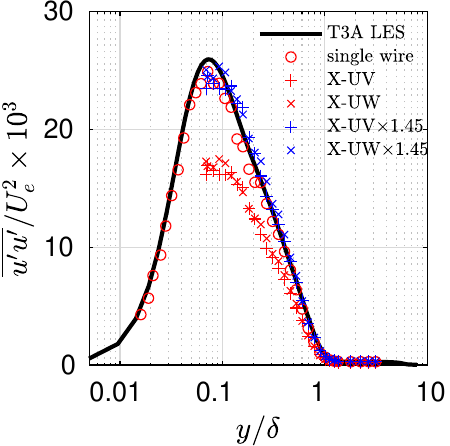}
\includegraphics[width=0.32\linewidth]{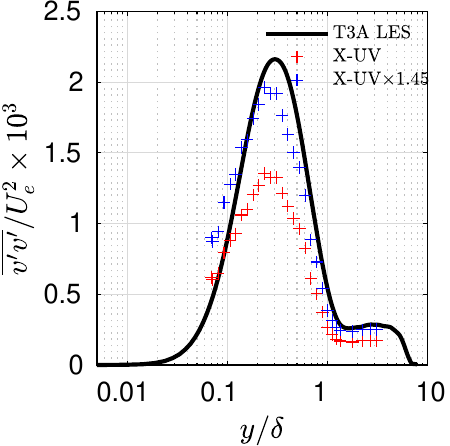}
\includegraphics[width=0.32\linewidth]{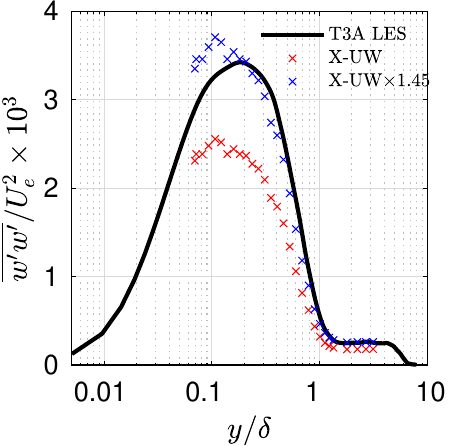}
\caption{Comparison of single-HW and X-HW for three components of Reynolds stress at $x/L_p=0.4$ in the T3A case.}
\label{fig:HW}
\end{figure}

To further assess the systematic underestimation of the transverse and vertical turbulent fluctuations, we replot several profiles obtained from the X-wire measurements. These measurements provide access to two velocity components, either $u$ and $v$ or $u$ and $w$, which makes it possible to compare the $\overline{u'u'}$ measurements obtained with single- and X-hotwires, as shown in the left panel of Fig. \ref{fig:HW}. The $\overline{u'u'}$ profile obtained with the single-hotwire is in good agreement with the simulation, whereas the profiles obtained with the X-hotwires for the simultaneous $uv$ and $uw$ measurements show a clear underestimation of the fluctuation levels. For both measurements, a correction factor of 1.45 brings the results into agreement with the levels obtained from the single-wire measurements. It should also be noted that X-wires do not allow measurements very close to the wall, which may also explain the difficulties encountered near the leading edge, where the boundary layer is very thin.
The two other panels of Fig. \ref{fig:HW} compare the X-wire measurements of the vertical and transverse components, respectively, with and without the correction factor. Applying the same correction as for $\overline{u'u'}$ yields levels compatible with the LES. The figure illustrates one location for T3A, but the same observation is made at other locations and for other configurations, supporting the conclusion that the fluctuation levels obtained from the X-wires are underestimated. In the main body of the paper, we retain the original measurements without attempting to apply an ad-hoc correction.

\begin{figure}
\centering
\includegraphics[width=0.77\linewidth]{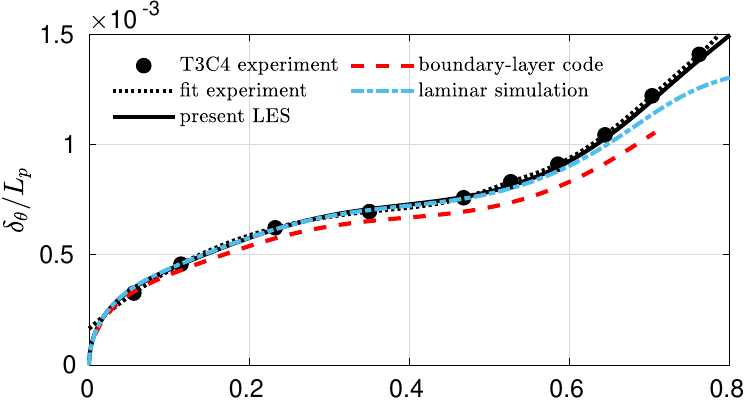}
\includegraphics[width=0.75\linewidth]{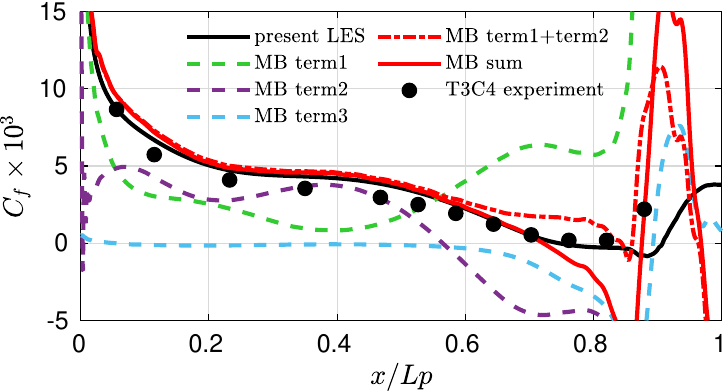}
\caption{Evolution of momentum thickness (top) and momentum balance (MB) equation (\ref{eq:MB}) applied to
  LES results (bottom) for T3C4 case.}
\label{fig:MB1}
\end{figure}

Finally, we recalculate the experimental skin-friction coefficients. Four techniques are used in the experiments: either a direct estimate from single-hotwire profiles extending very close to the wall, within the viscous sublayer ($y^+\approx3$), an indirect momentum-balance method based on the von K\'arm\'an integral equation, the use of a Preston tube or, for profiles identified as fully turbulent, the Clauser-chart method, which consists of fitting the standard logarithmic law $u^+\!=\!(1/\kappa)\log(y^+)\!+\!C$, with $\kappa$=0.41 for the von K\'arm\'an constant and $C$=5.2 for the intercept. In the literature, RANS calculations generally predict higher values of $C_f$ in the laminar region. We observe a similar discrepancy, which motivates a closer examination of the possible sources of error. Another suspicious feature is the discontinuity at the transition location for the T3C5 case, which is not observed in the LES results. This additional post-processing was carried out for all cases, but only selected results are presented here for conciseness.
The largest discrepancy in the laminar $C_f$ is observed for the lowest-Reynolds-number case, T3C4. However, Fig. \ref{fig:MB1}(top) reveals very good agreement for the evolution of the momentum thickness in the laminar region. In this figure, we also compare the results with those from a boundary-layer code accounting for the external pressure gradient, as well as with a simulation on the same grid as the T3C4 LES, but performed in two dimensions without incoming turbulence. We find that this laminar calculation is in very good agreement with the three-dimensional simulation, in which Klebanoff streaks develop from the leading edge of the plate.
Despite the good agreement for $\delta_\theta$, the $C_f$ reported in the ERCOFTAC database appears to be underestimated. We therefore re-evaluate $C_f$ from the momentum balance:
\begin{equation}
  \frac{C_f}{2}=\underbrace{\frac{d\delta_\theta}{dx}}_{\text{term1}}
  +\underbrace{\frac{1}{U_e}\frac{d U_e}{dx}}_{\text{term2}}
  -\underbrace{\frac{1}{U_e^2}\int_0^{\delta}\left(\overline{u'^2}-\overline{v'^2}\right)dy}_{\text{term3}}
  \label{eq:MB}
\end{equation}
The procedure is first applied to the LES data in Fig. \ref{fig:MB1} (bottom), where the three terms on the right-hand side of Eq.~(\ref{eq:MB}) are plotted to quantify their relative contributions. In particular, the pressure-gradient term is seen to play an important role for this boundary layer and produces the inflection in the $C_f$ evolution. As transition is approached, the non-equilibrium term involving the difference between the streamwise and wall-normal velocity fluctuations, $\overline{u'^2}-\overline{v'^2}$, becomes non-negligible. The reconstructed $C_f$ is in very good agreement with the direct LES result up to separation, beyond which the boundary-layer approximation breaks down.

\begin{figure}
\centering
\includegraphics[width=0.76\linewidth]{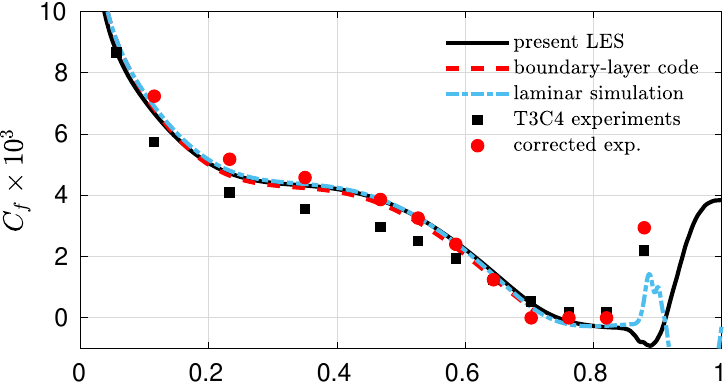}
\includegraphics[width=0.75\linewidth]{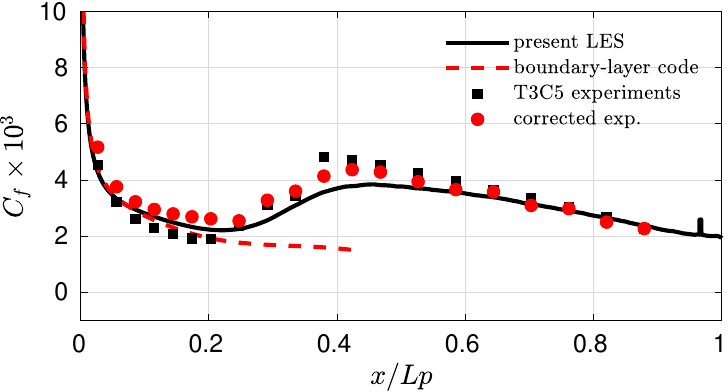}
\caption{Re-evaluation of friction distribution for T3C4 (top) and T35 (bottom) configurations.}
\label{fig:MB2}
\end{figure}

For the experimental data, a reliable estimate of $\overline{u'^2}-\overline{v'^2}$ is not available, as this quantity is too strongly affected by measurement noise. We therefore retain only the first two terms of Eq.~(\ref{eq:MB}) up to the location $x_c$, where the contribution of the non-equilibrium term is identified as becoming significant from the LES. A least-squares smoothing procedure is used to mitigate the numerical noise associated with differentiating the sparse experimental data. Beyond $x_c$, we re-evaluate $C_f$ for the T3C cases using an estimate of the wall-normal velocity gradient from the velocity profiles (direct method). The re-evaluated $C_f$ is reported in Fig. \ref{fig:MB2} (top) for T3C4 configuration and shows better agreement with the LES data and the laminar calculations. In the bottom panel, the re-evaluated $C_f$ for T3C5 produces a slight increase of laminar values for $x\!<\!xc\!=\!0.2L_p$. Beyond this value, the direct evaluation of $C_f$ yields lower values near the peak compared to the original data, which were most probably estimated from Clauser-chart method even if the first profiles are still transitional.

\section{Sensitivity of T3C cases to inlet turbulence} \label{app:T3C_sens}

\begin{figure}[htbp]
\centering
\includegraphics[width=0.9\linewidth]{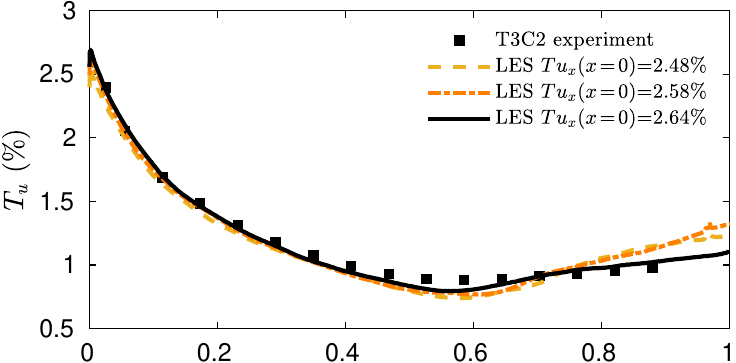}\\
\hspace{0.2cm}\includegraphics[width=0.78\linewidth]{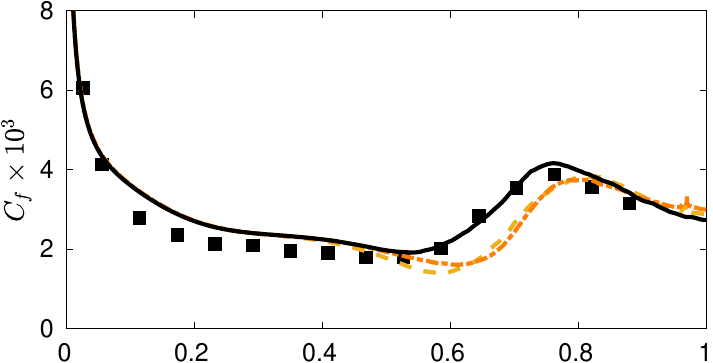}\\
\includegraphics[width=0.9\linewidth]{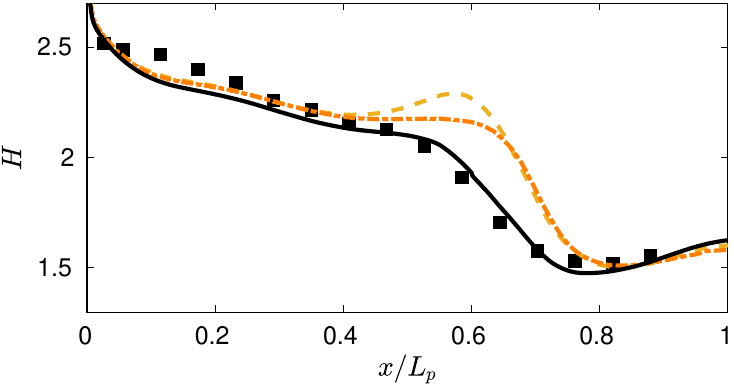}
\caption{Sensitivity of T3C2 case to initial turbulence definition.}
\label{fig:conv_t3c2}
\end{figure}

\begin{figure}[htbp]
\centering
\includegraphics[width=0.9\linewidth]{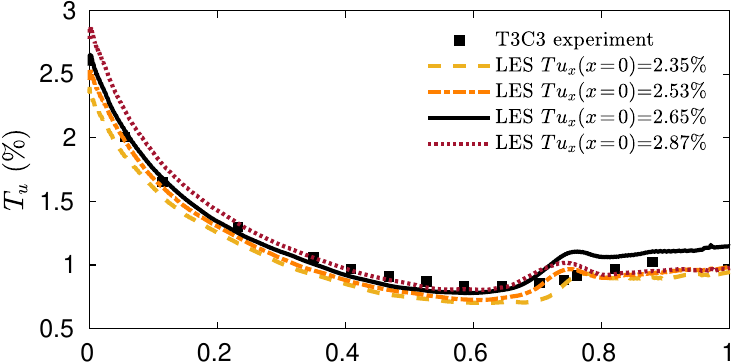}\\
\hspace{0.2cm}\includegraphics[width=0.78\linewidth]{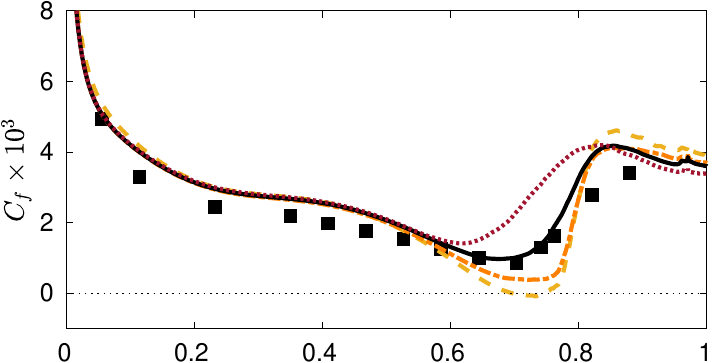}\\
\includegraphics[width=0.9\linewidth]{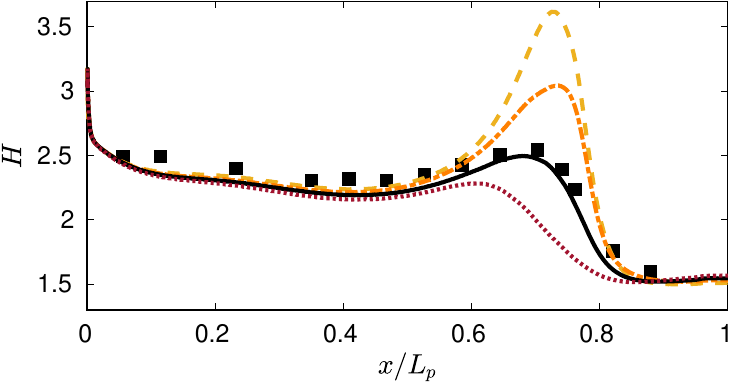}
\caption{Sensitivity of T3C3 case to initial turbulence definition.}
\label{fig:conv_t3c3}
\end{figure}

The analysis of the experimental database revealed that the reported $Tu$ corresponds only to the streamwise component. Consequently, the simulations of the T3C configurations presented in \cite{Zemmour25} actually imposed a $Tu$ that is too low compared with the experimental setup. New LES were therefore performed, revealing a strong sensitivity of transition under a varying pressure gradient to the precise level of incoming turbulence. The most useful indicator is the shape factor, as it highlights the competition between the tendency of the laminar boundary layer to separate and the reattachment associated with the transition to a turbulent state. For example, a variation of less than 0.2\% in the turbulence level at the leading edge can determine whether or not a bump appears in the T3C2 shape-factor evolution, as shown in Fig. \ref{fig:conv_t3c2}. Similarly, for T3C3 (Fig. \ref{fig:conv_t3c3}), which has a lower Reynolds number and is therefore closer to separation, a variation of 0.5\% can lead from a strong overestimation of $H$ (excessive separation tendency) to an underestimation. Underestimating $Tu$ (for example, by defining it as the average of the three components rather than using only the streamwise component) delays transition, as shown by the evolution of $C_f$. The lowest $Tu$ of 2.35\% considered here for T3C3 even leads to boundary-layer separation. The adjustment of the T3C2 and T3C3 cases was performed without modifying the integral length scale $L_f$ selected by trial and error in \cite{Zemmour25}. This characteristic length scale, which is related to the size of the largest energetic structures in the incoming turbulence, controls its decay rate. It is therefore possible that another combination of ($Tu$, $L_f$) could produce a similar disturbance level at the transition location and consequently a similar transition process.

\begin{figure}[htbp]
\centering
\includegraphics[width=0.9\linewidth]{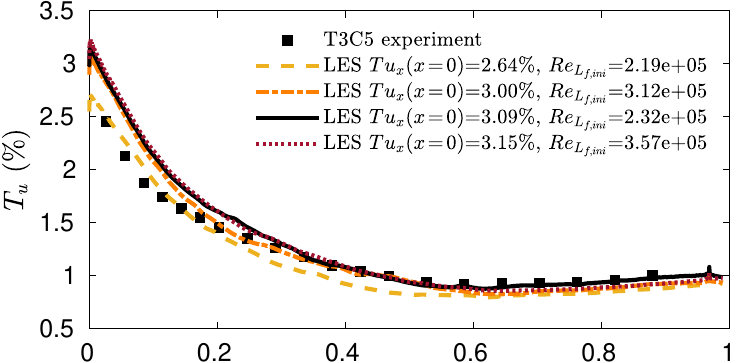}\\
\hspace{0.2cm}\includegraphics[width=0.78\linewidth]{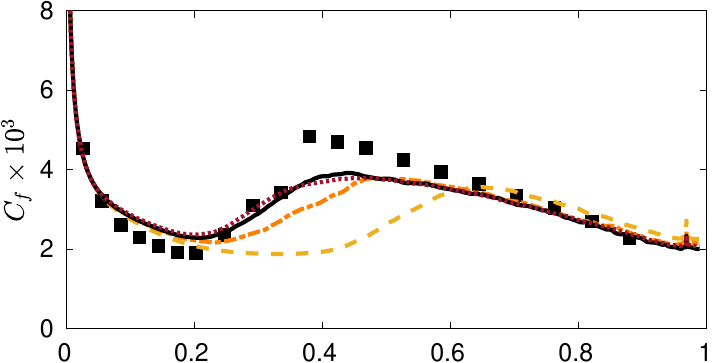}\\
\includegraphics[width=0.9\linewidth]{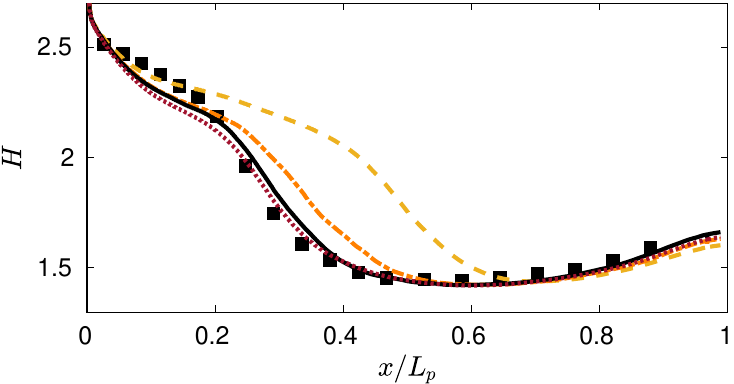}
\caption{Sensitivity of T3C5 case to initial turbulence definition.}
\label{fig:conv_t3c5}
\end{figure}

For the T3C5 case, different values of $L_f$ were also tested to control the turbulence decay and obtain levels similar to the experimental data in the region where transition is initiated. Since transition starts while the favorable pressure gradient is still strong, a small variation in the initial conditions, which also controls the $Tu$ level at $x=0$, can significantly delay the transition to turbulence. Figure \ref{fig:conv_t3c5} shows that the same turbulence level, and therefore the same transition location, is obtained either with $Tu_{x=0}=3.09\%$ and a length scale corresponding to $Re_{L_f,in}=2.32\times10^5$, or with the slightly higher $Tu_{x=0}=3.15\%$ and $Re_{L_f,in}=3.57\times10^5$, corresponding to a slower decay. We retained the simulation with a $Tu$ closer to 3\% in the main body of the paper.

\begin{figure}[htbp]
\centering
\includegraphics[width=0.9\linewidth]{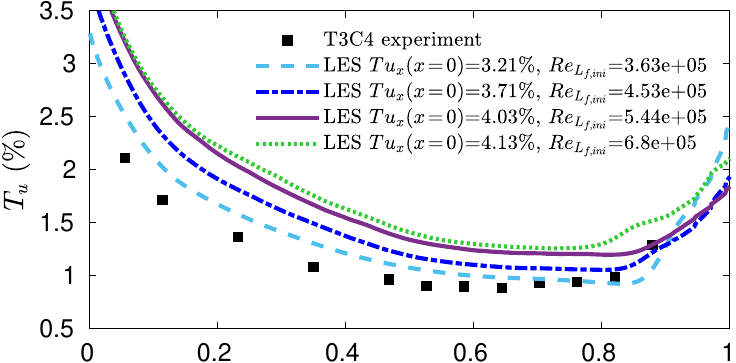}\\
\hspace{0.1cm}\includegraphics[width=0.79\linewidth]{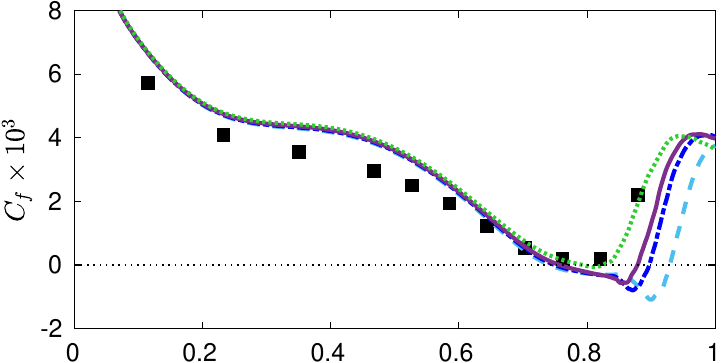}\\
\hspace{0.2cm}\includegraphics[width=0.77\linewidth]{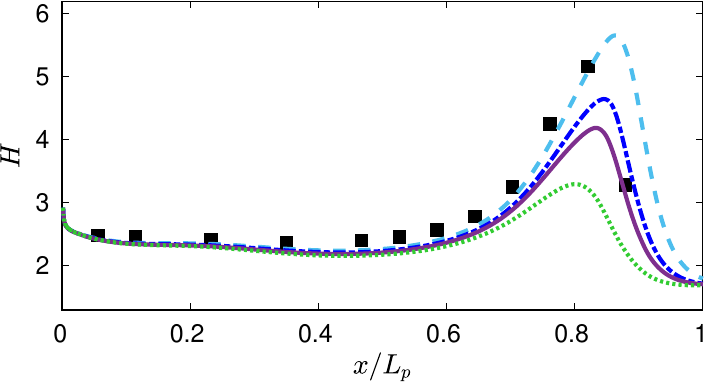}
\caption{Sensitivity of T3C4 case to initial turbulence definition.}
\label{fig:conv2_t3c4}
\end{figure}

\begin{figure}[htbp]
\centering
\includegraphics[width=0.9\linewidth]{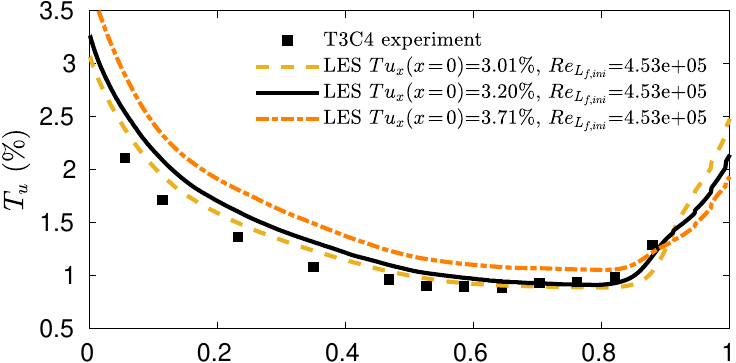}\\
\hspace{0.1cm}\includegraphics[width=0.79\linewidth]{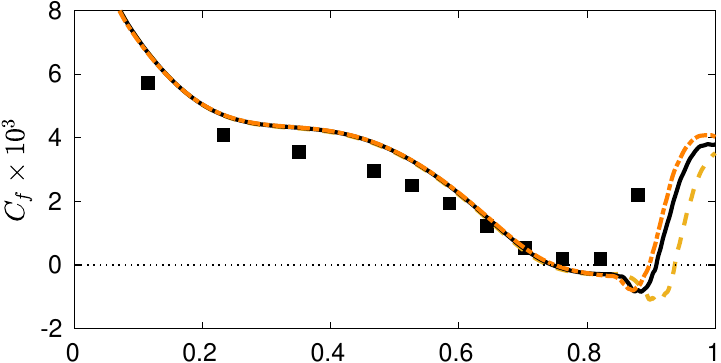}\\
\hspace{0.2cm}\includegraphics[width=0.77\linewidth]{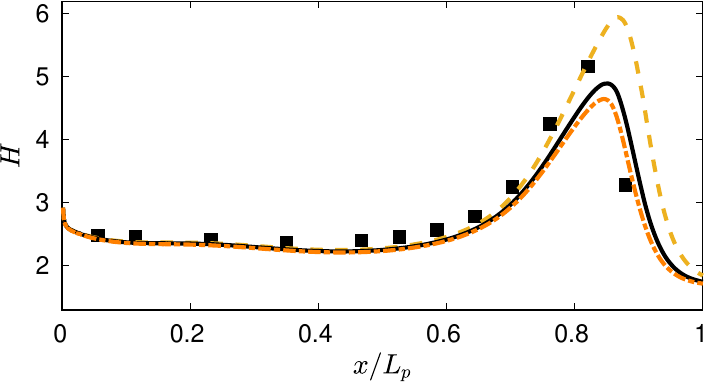}
\caption{Sensitivity of T3C4 case to initial turbulence definition.}
\label{fig:conv_t3c4}
\end{figure}

Different values of $Tu$ and $L_f$ were also tested for the T3C4 case. This case is not reported as separated based on the experimental $C_f$, but our simulations exhibit separation near the end of the plate owing to the low Reynolds number. Several attempts were made to limit the separation, which may potentially be present in the experiment but sufficiently mitigated by transition not to be identified as such. Figure \ref{fig:conv2_t3c4} gathers several simulations with the same $Tu$ at inlet but different values of $L_f$. Increasing $L_f$ results in a slower turbulence decay and therefore higher turbulence levels at the leading edge location and in the transition region, slightly advancing the transition onset. Only the last case with an exaggerated $Tu$=4.13\% is able to reduce significantly the separation and to provide a $C_f$ rise in agreement with experiments. Similarly, a simple increase in $Tu$ at constant $L_f$ (Fig. \ref{fig:conv_t3c4}) limits the increase in $H$, indicating that an earlier transition takes over from the separation process. In all cases, negative $C_f$ values mark the presence of a separation bubble. We selected for the main text the case with $Tu$=3.2\% and $Re_{L_f,in}$=$4.53\times10^5$, which reasonably captures the balance with the external pressure gradient.

\section{Algorithm for laminar-turbulent discrimination} \label{app:lam_turb}

The algorithm for the laminar-turbulent discrimination has been developed in \cite{Bienner24c,Bienner24d}
and is adapted from \cite{Durovic22}. We first interpolate the field on a coarser mesh, the interpolation acting as a low-pass filter, in addition to reducing the computational cost of the procedure.
For each wall-normal plane of a volume snapshot, the 2D scalar field of the spanwise velocity fluctuations $w'$, noted $s_{w'}(\mathbf{x})$, is first used to define $s_{extr,w'}$ containing the local extrema of $w'$. The same extraction is performed for the wall-normal velocity fluctuations $v'$ to obtain $s_{extr,v'}$. A 3D scalar field $s_{extr,3D}$ is then reconstructed from the 2D extractions, $s_{extr}=s_{extr,v'}+s_{extr,w'}$, and a smoothing $S$ is applied to it. The original Laplace kernel has been replaced by a Gaussian kernel,
\begin{equation}
S(\mathbf{r}) =
\exp\left(-\frac{|\mathbf{r}|^2}{2r_s^2}\right),
\end{equation}
where $r_s$ controls the smoothing radius. In the present implementation, the Gaussian kernel is separated into a two-dimensional contribution in the $(x,y)$ plane and a one-dimensional contribution in the spanwise direction. This factorization considerably reduces the computational cost and memory requirements compared with the original three-dimensional kernel.
As the distance from the extrema increases, $S(\mathbf{r})$ decreases rapidly. The smoothing is therefore applied over a cubic domain $D^3$ restricted to the locations where $S(\mathbf{r})\leq 0.025$, limiting the computational cost. The extrema density scalar field $f_{extr,3D}$, weighted by the extrema values, is computed as
\begin{equation}
f_{extr,3D}(\mathbf{x}) =
\sum_{\mathbf{x}_s\in D^3}
S\left(\frac{\mathbf{x}_s-\mathbf{x}}{r_s}\right)
s_{extr,3D}(\mathbf{x}_s),
\end{equation}
where $r_s$ is determined by trial and error and is taken equal to $5\Delta x_{interp}$, with $\Delta x_{interp}$ denoting the streamwise spacing on the interpolated grid. The density scalar field $f_{extr,3D}$ is then interpolated back onto the computational grid, and a binary segmentation is realized using the method of \cite{Otsu79}. A threshold is obtained at each wall-normal location
independently, and levels are smoothed in the wall-normal direction to ensure continuity
in the discrimination. The discrimination is applied for each case using 2000 to 3000 sub-volumes over approximately two flow through times.

\bibliographystyle{cas-model2-names}
\bibliography{bibli}

\end{document}